\documentclass[draft]{agujournal2019} 
\usepackage{url} 
\usepackage{soul}
\usepackage{tikz}
\usetikzlibrary{shapes.geometric, arrows}
\tikzstyle{input} = [rectangle, minimum width=3cm, minimum height=1cm, text centered, draw=black, fill=yellow!30]
\tikzstyle{process} = [rectangle, minimum width=3cm, minimum height=1cm, text centered, draw=black, fill=blue!30]
\tikzstyle{output} = [ellipse, minimum width=3cm, minimum height=1cm, text centered, draw=black, fill=green!30]
\tikzstyle{arrow} = [thick,->,>=stealth]
\input{LatexFiles/preamble}
\draftfalse

\journalname{NA}

\begin{document}

\title{Exploring climate change effects on concurrent floods and concurrent droughts via statistical deep learning}

%
%




\authors{C. J. R. Murphy-Barltrop\affil{1,2}, J. Richards\affil{3}, B. Poschlod\affil{4,5}, A. Sasse\affil{6}, J. Zscheischler\affil{2,7,8} }

\affiliation{1}{TUD Dresden University of Technology, Institut Für Mathematische Stochastik, Helmholtzstraße 10, 01069 Dresden, Germany}
\affiliation{2}{Center for Scalable Data Analytics and Artificial Intelligence (ScaDS.AI) Dresden/Leipzig, Germany}
\affiliation{3}{School of Mathematics and Maxwell Institute for Mathematical Sciences, University of Edinburgh, UK}
\affiliation{4}{Institute of Global Water Security, Hamburg University of Technology, Hamburg, Germany}
\affiliation{5}{Earth and Society Research Hub (ESRAH), Universität Hamburg, Hamburg, Germany}
\affiliation{6}{Department of Geography, Ludwig-Maximilians-Universität München, Munich, Germany}
\affiliation{7}{Department of Compound Environmental Risks, Helmholtz Centre for Environmental Research–UFZ, Leipzig, Germany}
\affiliation{8}{Department of Hydro Sciences, TUD Dresden University of Technology, Dresden, Germany}

\correspondingauthor{Callum Murphy-Barltrop}{callum.murphy-barltrop@tu-dresden.de}



\begin{keypoints}
\item Contemporary extreme value modelling techniques accurately capture marginal and joint extremes for simulated discharge.
\item Compound flooding and drought-like events becoming more likely in the Alpine Foreland region of the Upper Danube catchment. 
\item Deep learning offers practical advantages for modelling compound extreme events.
\end{keypoints}

%
%

%
%


\begin{abstract}
Concurrent floods and concurrent droughts in nearby catchments pose challenges to risk assessment and water management. Climate change is affecting extremely high and low discharge, but the complex interplay between changes in individual catchments and in the dependence across catchments make it difficult to provide accurate assessments of the occurrence probabilities of concurrent extremes. In this work, we use a contemporary statistical deep learning model to capture concurrent river floods and droughts in four catchments in the Upper Danube basin, based on discharge simulated by a hydrological model driven with large ensemble climate model output. The statistical model is able to accurately capture the tail behaviour of the simulated discharge, which we assess by making use of the large available sample size. We subsequently use our statistical model to study changes in joint tail behaviour of discharge over time, finding that both compound flooding and drought-like conditions are becoming increasingly likely towards the end of the 21st century under a high-emission scenario. In particular, our results highlight that changes in the dependence structure of extremes strongly contribute to the detected changes, an aspect that would be difficult to capture with traditional approaches. This work paves the way for highly flexible, general inference on compound extremes in hydrological applications, and demonstrates key advantages of using statistical deep learning in this setting.
\end{abstract}

\section*{Plain Language Summary}
Understanding how concurrent droughts or floods in multiple catchments change with climate change is important for risk assessment but very challenging. There exist a wide range of statistical techniques for modelling joint extremes, many of which lack the flexibility to represent observed physical behaviour and/or suffer from various practical drawbacks. To overcome such limitations, we use a contemporary modelling approach, whereby statistical deep learning is used to model compound extremes. We show that this model can accurately capture marginal and joint extremes, while simultaneously respecting physical conditions. We find that concurrent floods and concurrent drought-like conditions become more likely with climate change for four catchments in the Upper Danube basin. The method provides an exciting new paradigm for modelling and understanding compound extremes and their changes in hydrological applications.

\section{Introduction} \label{sec:intro}

River floods and droughts can endanger lives, damage properties, and seriously disrupt societies. Due to large-scale weather conditions, such as widespread precipitation extremes \cite{Bevacqua2021}, nearby rivers may experience floods simultaneously \cite{Berghuijs2019}. Such concurrent river floods can cause overall impacts that surpass the sum of individual localised flood events \cite{Zscheischler2018}. Widespread concurrent droughts can also be highly impactful, as they can affect river transport and energy security \cite{Bevacqua2024}. Both concurrent floods and droughts are of particular concern when limits for emergency response, disaster rescue, and insurance payouts are exceeded \cite{Kemter2020,Jongman2014}. In Europe, the spatial extent of floods has increased over the last 70 years \cite{Fang2024}. At the same time, drought frequency and severity have significantly increased over Central and Southern Europe during the last decades, driven primarily by rising temperatures and enhanced evapotranspiration \cite{Ionita2021}.

There is growing recognition that climate change is affecting the frequency and intensity of floods and droughts. Changes in spatial dependence due to regionally different intensity changes or changes in large-scale weather patterns can make risk assessments particularly challenging \cite{Zscheischler2017,Zscheischler2020}. For instance, extremely large floods are projected to increase further in the future \cite{Fang2025}. This underscores the crucial importance of developing novel techniques to study the evolving risk posed by such events over time, enabling better adaptation and mitigation strategies to reduce societal vulnerability.

Historically, most modelling approaches for extreme river discharge have focused on univariate techniques, where a statistical model is fitted to observations at an individual location \cite{Zhang2018}. Practitioners opt for a \textit{peaks over threshold} (POT) approach, which consists of fitting a \textit{generalised Pareto distribution} (GPD) to the excesses of data above some high threshold \cite{Coles2001}. Given an independent and identically distributed (IID) sequence of random variables $X_1,\dots,X_n,$ and a sufficiently large threshold $u$, the positive threshold excesses $Y_t := (X_t - u \mid X_t > u)$, {$t=1,\dots,n,$} can be modelled using a GPD, with distribution function
\begin{equation} \label{eqn:gpd}
    \Pr(Y_t \leq y) = H(y; \sigma,\xi) = 1 - \left(1 + \frac{\xi y}{\sigma}\right)_+^{-1/\xi}, \; \; \; y>0, 
\end{equation} 
where $z_+ = \max(0,z)$, $\sigma > 0$, and $\xi \in \mathbb{R}$ \cite{Balkema1974,Pickands1975}. We refer to $\sigma$ and $\xi$ as the \textit{scale} and \textit{shape} parameters, respectively. The shape parameter dictates the support of the tail distribution, with negative shape values omitting a finite upper bound \cite{Davison1990}. The probability density function associated with equation~\eqref{eqn:gpd} is given by
\begin{equation} \label{eqn:gpd_dens}
    h(y; \sigma,\xi) = \frac{1}{\sigma} \left(1 + \frac{\xi y}{\sigma}\right)_+^{-1/\xi-1}, \; \; \; y>0.
\end{equation}
A wide range of approaches have been proposed for fitting the GPD and selecting the threshold $u$; see, e.g., discussions by \citeA{Davison2015,Murphy2024}. We note that POT approaches are not restricted to data in the upper tail; observing that $(X_t - u  \mid X_t < u) \,{\buildrel d \over =}\, {( u - X_t \mid -X_t > -u)}$, one can equivalently use a GPD to model observations in the lower tail by taking increments below $u$.  

In practice, POT approaches are often used to quantify risk associated with extreme events \cite{Rootzen2013,Darcy2023a}. A common metric used to summarise risk is the \textit{return level}; given some return period of $N$ years, we define the $N$-year upper tail return level, $x^{1-1/N}$, to be the value one would expect $X_t$ to exceed on average once every $N$ years. Equivalently, one can also define the $N$-year lower tail return level, $x^{1/N}$, as the value one would expect $X_t$ to fall below on average once every $N$ years. 

Univariate extreme value analysis techniques are useful for quantifying and summarising risk at an individual location or for a single random variable, but provide no information about the interdependence of extreme events between different locations or variables. From a practical perspective, taking data across multiple locations or variables also increases the amount of information available for modelling. This can result in more robust, intricate, and detailed risk assessments \cite{Boulaguiem2022} -- particularly if there exist strong interdependencies between different locations, as is almost always the case on river networks.  

Techniques for modelling concurrent floods have been developed for many applications; for example, flood risk mitigation \cite{Keef2013a}, actuarial science \cite{Quinn2019}, climate change adaptation \cite{Rojas2011}, and catastrophe modelling \cite{Wing2020}. 
Assessment of concurrent extremes requires a model for capturing the dependence between extremes at different locations; this dependence is termed the extremal dependence structure, and multivariate extreme value theory is the branch of statistics concerned with the modelling of such structure \cite{Beirlant2004}. 
In the context of river discharge, an `extreme' can correspond to both extremely large or small values at a given location, i.e., flooding and drought-like conditions, respectively. 
 
Traditional models for multivariate extremes rely on the assumption of multivariate regular variation \cite{Resnick2002,Beirlant2004}, but such an assumption is untenable for modelling environmental processes, where variables often do not exhibit extreme values at the same times  \cite{Heffernan2004,Opitz2016,Huser2019,Huser2025}. Classical modelling of multivariate extreme events is further complicated by the fact there is no unique definition of a multivariate extreme event due to the lack of ordering in multivariate data \cite{Barnett1976}. Focusing only on regions where all variables are extremely large (or small) simultaneously is too simplistic since many different combinations can result in high-impact events, including cases when a subset of variables are not extreme at all \cite{Zscheischler2020,Murphy-Barltrop2023}. As the number of variables increases, the number of possible combinations for which a subset of variables are extreme simultaneously grows exponentially \cite{Simpson2020}. Developing robust modelling frameworks that can flexibly capture many combinations of extreme events represents an active area of research.   

In this work, we use a contemporary approach that allows flexible modelling of extremal dependence without strong parametric assumptions. In particular, we apply the recently proposed \textit{semi-parametric angular-radial} (SPAR) framework \cite{Mackay2023}, via a deep extremal regression framework \cite{Richards2024}, and demonstrate that this technique captures well the joint discharge extremes on a river network located in the Alpine Foreland region of the Upper Danube catchment. We use simulated discharge from a hydrological model forced with climate simulations from a regional 50-member single model initial-condition large ensemble (SMILE) resulting in a hydro-SMILE. The use of a hydro-SMILE allows us to evaluate the robustness and sensitivity of the SPAR model. 
Furthermore, using climate models allows one to model and infer extremes in future time periods, enabling the study of changes in complex extremal dependence structures over time. 

\section{Data and Methods}
\subsection{Study area and hydrological simulations} \label{subsec:data_set}
\begin{figure}[h]
    \centering
        \includegraphics[width=\textwidth]{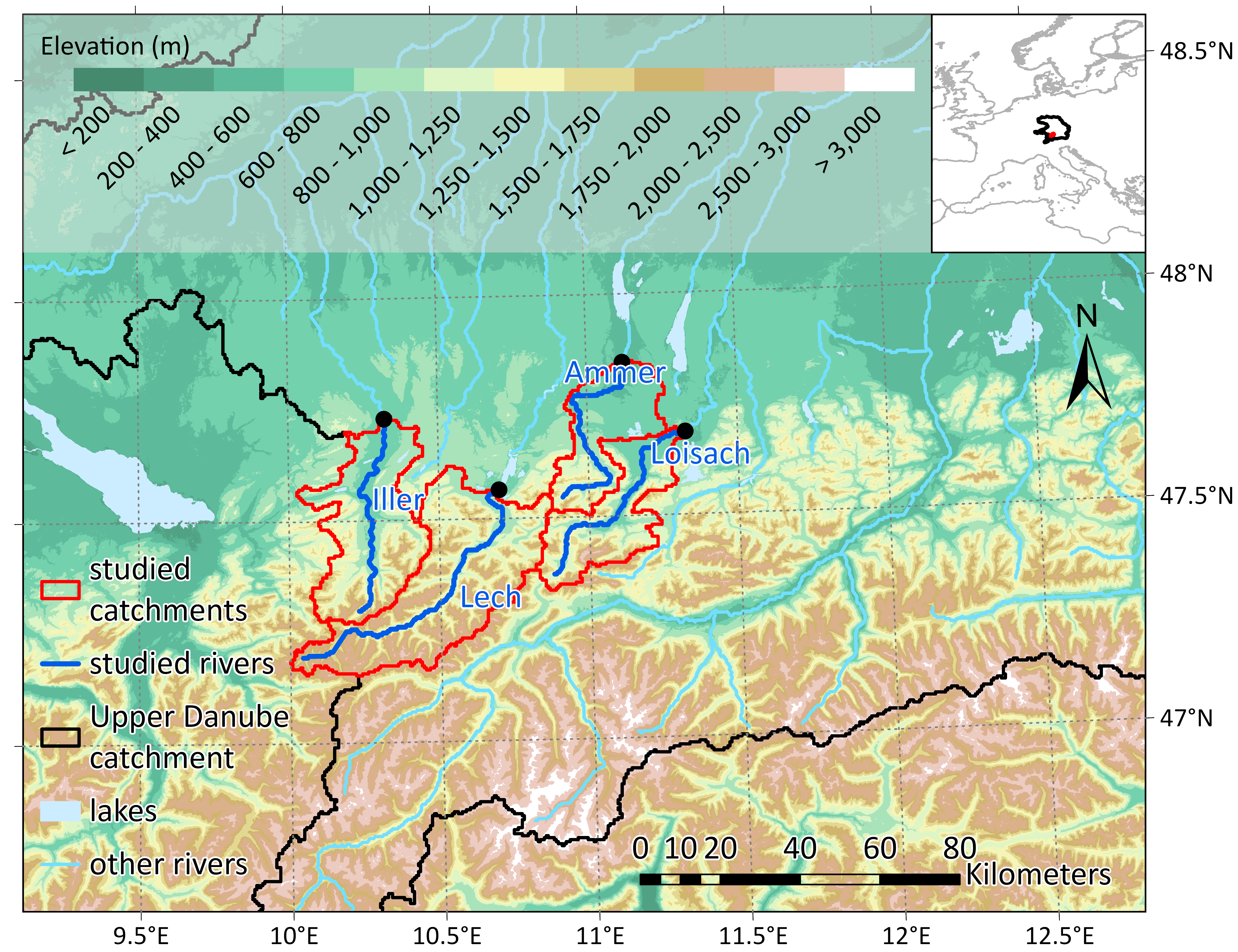}   
    \caption{Topography of the four neighbouring catchments under study. Elevation is taken from the GLO30 digital elevation model \protect\cite{copernicus_glo30}.}
    \label{fig:spar_map}
\end{figure}

We explore concurrent extremes in the four neighbouring head catchments of the rivers Ammer, Iller, Lech, and Loisach in the Alpine Foreland region. The river catchments range in size from approximately 600 to 1,400 km$^{2}$ and span an elevation gradient from about 550 m to over 3,000 m at the highest peak of the Lech Alps (Figure~\ref{fig:spar_map}). The southern catchment areas are predominantly forested, whereas the northern parts are mainly covered by pastures and natural grasslands. Driven by an annual precipitation of approximately 1,500 mm, characterized by a wet summer season, pronounced snow accumulation in winter and melt dynamics in spring, the rivers exhibit a pluvio-nival to nivo-pluvial flow regime, with peak mean discharges occurring in April for the Ammer and Loisach, and in June for the Lech and Iller \cite{poschlod2020}.

We base our analysis on a hydro-SMILE, i.e., hydrological model discharge simulations driven by a SMILE of climate model simulations. The forcing data stems from the Canadian Earth System Model version 2 large ensemble (CanESM2-LE; \citeA{fyfe2017}) under the high-emission scenario RCP8.5, which was downscaled with the Canadian Regional Climate Model version 5 (CRCM5) at 12 km spatial resolution over a European domain \cite{leduc2019}. This dynamical downscaling covers the time period 1955--2099. After a bias adjustment and further statistical downscaling to a resolution of 500 metres \cite{willkofer2018}, the climate simulations were used to drive the process-oriented spatially distributed Water balance Simulation Model (WaSiM; \citeA{schulla_wasim_2021}) at 3-hourly time steps. The WaSiM setup was calibrated and validated against observational discharge \cite{willkofer2020}. The 50 members of the resulting hydro-SMILE cover 1961--2099, and thus only differ due to internal climate variability, enabling one to disentangle climate change and climate variability. The hydro-SMILE has been applied to study floods \cite{willkofer2024, brunner2021} and river droughts \cite{poschlod2026} under climate change. 
Within the WaSiM validation \cite{willkofer2020}, the model setup yields Nash-Sutcliffe efficiency (NSE) values of 0.76, 0.76, 0.76, and 0.67, and logarithmic NSE values of 0.79, 0.75, 0.80, and 0.74 for the Ammer, Iller, Lech, and Loisach catchments, respectively. We therefore assume that the hydrological simulations reproduce the characteristics of floods and droughts sufficiently well to serve as a testbed for the SPAR framework. 

To investigate climate change effects on concurrent extremes in the four catchments, we divide the transient simulations into four 30-year periods: 1980--2009, 2010--2039, 2040--2069, and 2070--2099. We assume stationarity within each period and take the second period to represent the modern day. From these periods, we sample weekly maxima to account for temporal dependence and study upper and lower tail behaviour of the resulting distribution. The resulting series samples the upper‐tail behaviours across rivers and therefore characterises univariate and concurrent flood magnitudes. However, we acknowledge that weekly maxima will not always be appropriate for characterising droughts as a window size of 7 days will not be large enough to de-cluster the data (i.e., drought events lasting several weeks or months), resulting in residual temporal dependence within the series \cite{poschlod2026}. Nevertheless, we consider 7-day maxima a suitable compromise for assessing concurrent floods and droughts in a single coherent modelling framework.


\subsection{POT modelling for non-identically distributed data} \label{subsec:ns_pot}

There exist many extensions of the standard univariate POT approach that account for cases where data are not identically distributed. For example, many environmental datasets exhibit temporal non-stationarity, where the statistical properties of variables change over time. This can result in both seasonal and long-term trends. To capture such trends with the POT approach, it is common for covariate-dependent functional forms to be specified for the threshold and GPD parameters, $u$ and $(\sigma,\xi)$, respectively. 

Let $\{X_t\}_{t=1}^n$ denote a non-stationary time-series indexed by $t\in\{1,\dots,n\}$ (typically representing time) and $\{\boldsymbol{Z}_t=(Z_{1,t},\dots,Z_{p,t})\}^n_{t=1}$ a covariate process (with $p$ predictor variables) that influences the marginal tail behaviour of $X_t$. Furthermore, given some exceedance level $\alpha \in (0,1)$ close to 0, let $u(\cdot)$ be the function solving the equation $\alpha = \Pr (X_t>u(\boldsymbol{z}_t) | \boldsymbol{Z}_t = \boldsymbol{z}_t)$ for all $\boldsymbol{z}_t$. Applying the POT paradigm, a natural model for the conditional excesses variable $Y_t := (X_t - u(\boldsymbol{z}_t)) \mid (\boldsymbol{Z}_t = \boldsymbol{z}_t, X_t > u(\boldsymbol{z}_t))$ is given by 
\begin{equation} \label{eqn:ns_gpd}
    H(y; \sigma(\boldsymbol{z}_t),\xi(\boldsymbol{z}_t)) = 1 - \left(1 + \frac{\xi(\boldsymbol{z}_t) y}{\sigma(\boldsymbol{z}_t)}\right)_+^{-1/\xi(\boldsymbol{z}_t)}, \; \; \; y>0, 
\end{equation}
where $u(\cdot)$, $\sigma(\cdot),$ and $\xi(\cdot)$ now denote covariate-dependent threshold, scale, and shape parameter functions, respectively. A wide range of methodologies, which we term \textit{GPD regression techniques}, have been proposed for modelling the functions $u(\cdot)$, $\sigma(\cdot),$ and $\xi(\cdot)$ \cite{Davison1990,Chavez-Demoulin2005,Randell2016,Youngman2019}. In particular, approaches based on deep learning \cite{Pasche2024,Richards2023a} offer a high degree of flexibility for capturing complex covariate interactions, and allow a detailed description of marginal tail behaviour. As we show, such techniques can also be adapted to model multivariate extremes. 

\subsection{The SPAR framework} \label{sec:SPAR}

Many recent modelling approaches for multivariate extremes decompose data into radial and angular components (defined relative to some centroid), and fit an asymptotically motivated distribution to the tail of the radial variable conditional on the angle. Compared to classical modelling approaches, these contemporary techniques offer a more complete and detailed picture of extremal dependence and can be used for a broader range of applications \cite{Nolde2022, Murphy-Barltrop2024c, Papastathopoulos2025}.

In this work, we use the SPAR model for multivariate extremes, which offers a high degree of flexibility and practical utility. \citeA{Mackay2023} demonstrated theoretically that this model accurately approximates joint tail behaviour for a wide range of popular copulas and marginal scales. The framework has been successfully applied in many practical applications \cite{Murphy-Barltrop2024,Simpson2024,Mackay2025,Mackay2025b}. Furthermore, unlike many approaches to modelling multivariate extremes, SPAR can be applied without the need for explicit marginal modelling, resulting in decreased estimation uncertainty and fewer modelling assumptions \cite{Towe2023,Kakampakou2024}. Remarkably, the SPAR framework can also be used to approximate the tail distributions of marginal variables, although these distributions are not modelled explicitly. 

\begin{figure}[h]
    \centering
     \begin{minipage}{0.48\textwidth} 
        \centering
        \includegraphics[width=\textwidth]{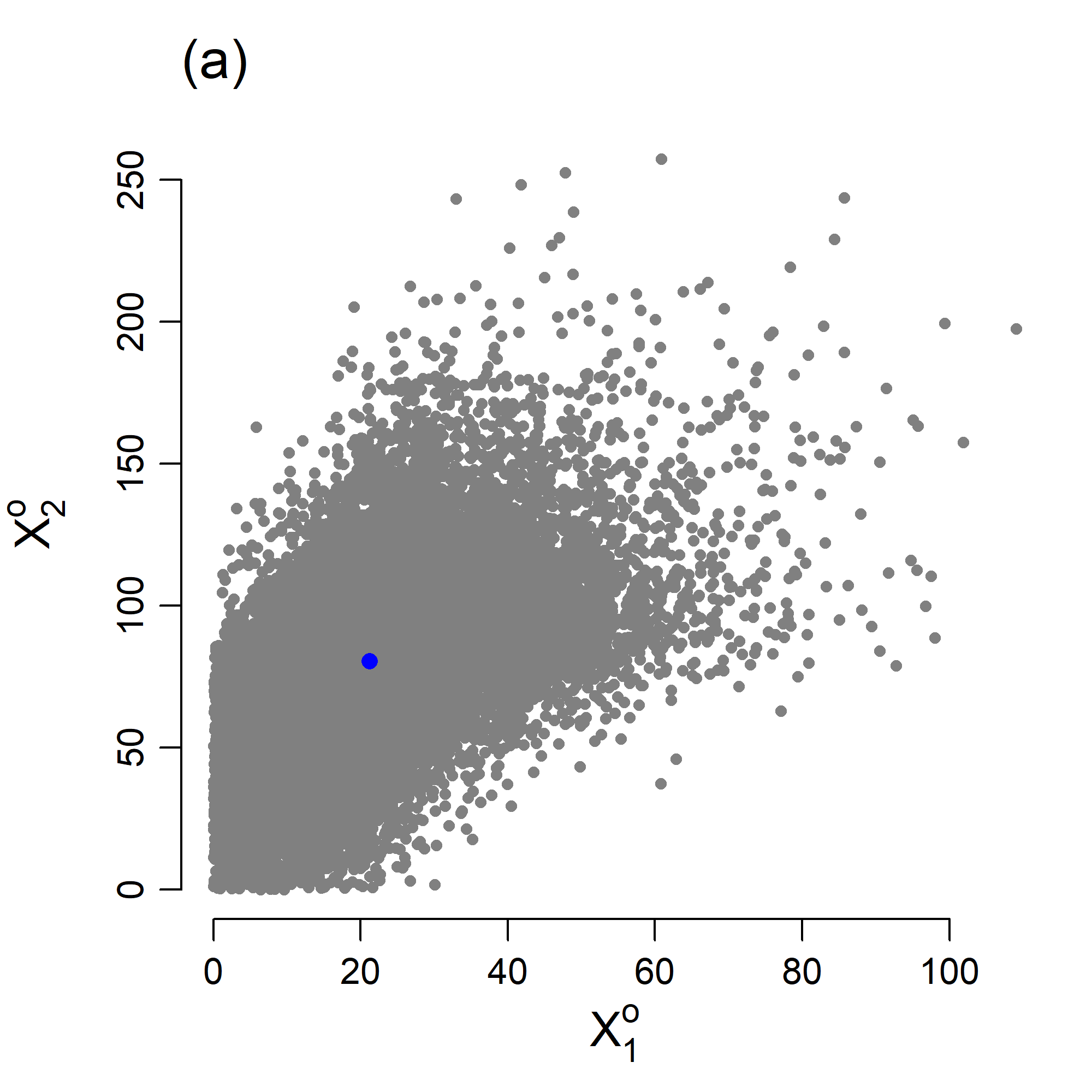}
    \end{minipage}
    \hfill
    \begin{minipage}{0.48\textwidth} 
        \centering
        \includegraphics[width=\textwidth]{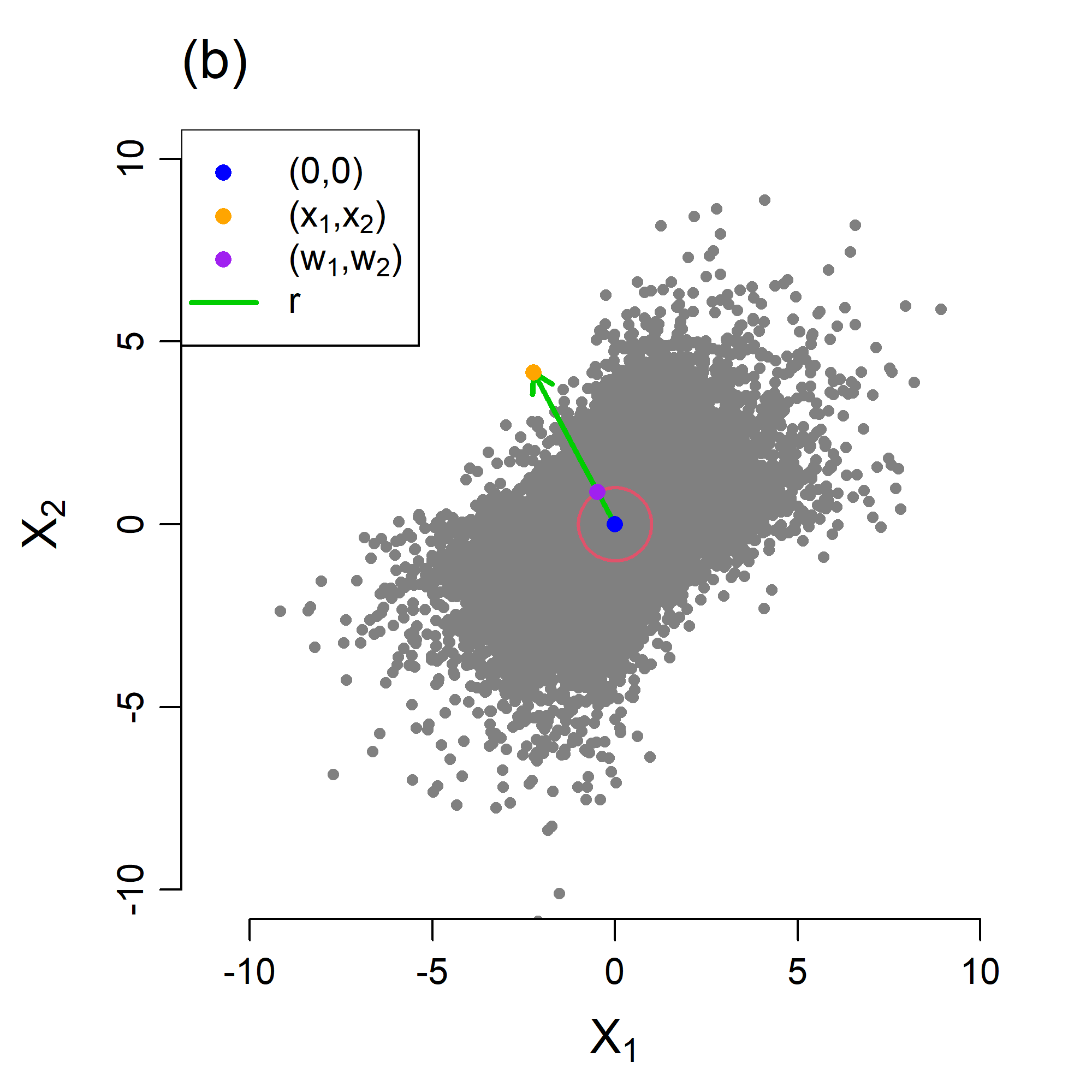}
    \end{minipage}
    \hfill
        \begin{minipage}{0.48\textwidth} 
        \centering
        \includegraphics[width=\textwidth]{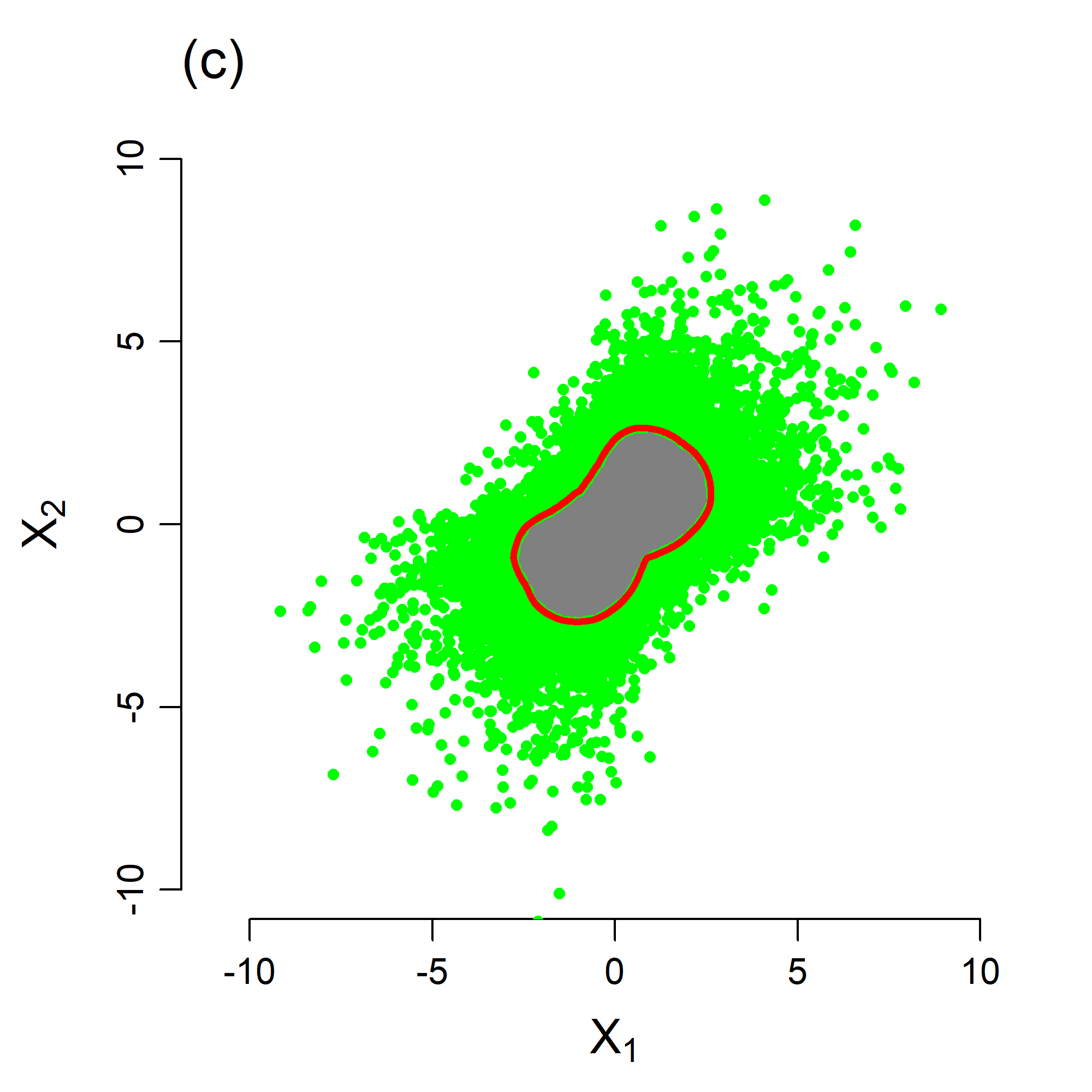}
    \end{minipage}
    \hfill
        \begin{minipage}{0.48\textwidth} 
        \centering
        \includegraphics[width=\textwidth]{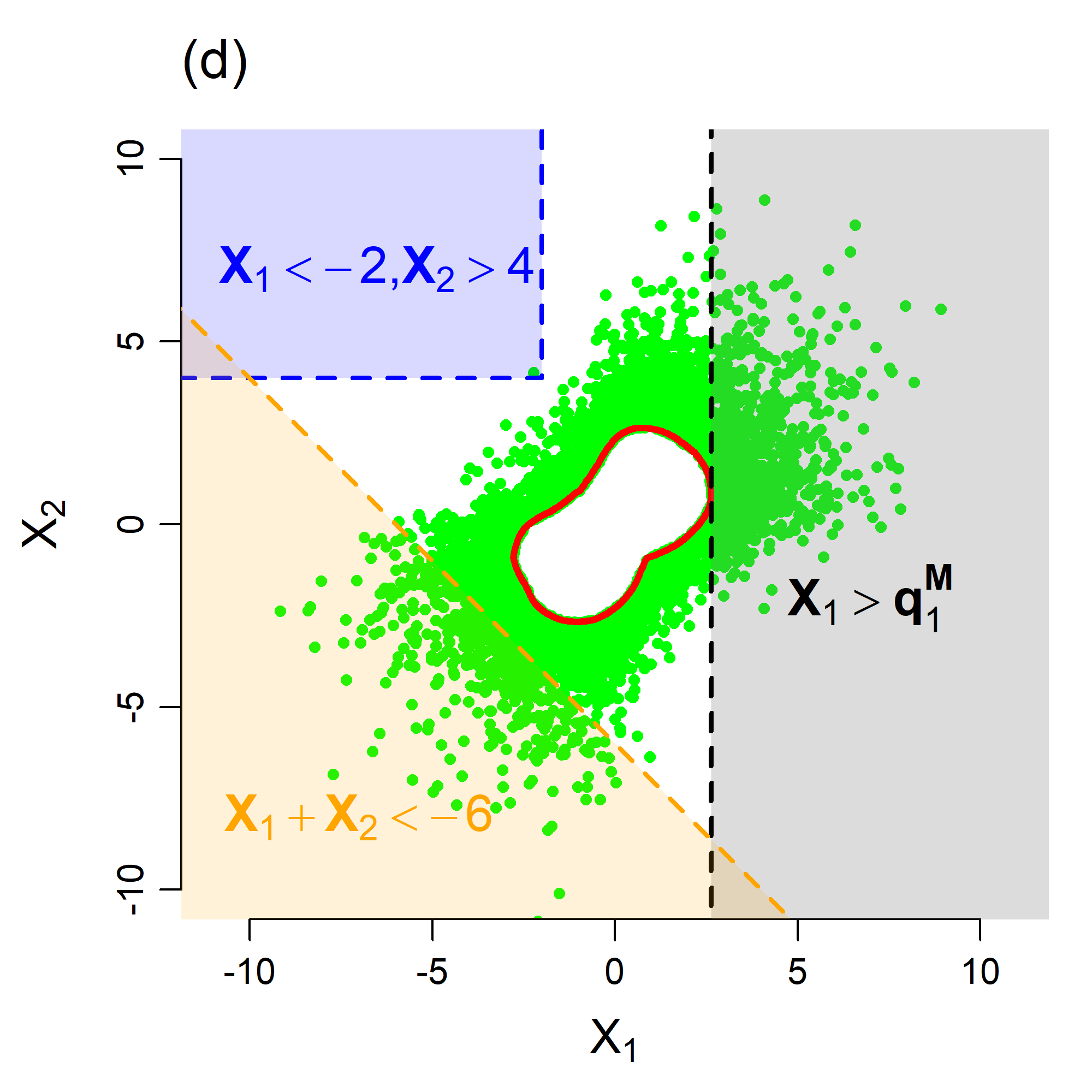}
    \end{minipage}
    
    \caption{Visual illustration of the pre-processing and SPAR modelling procedures for $d=2$. (a) Bivariate data prior to pre-processing, with a star-centre at $(21.3,80.4)$ (blue). (b) Bivariate data post-processing with star-centre $(0,0)$ (blue), alongside example angular (purple) and radial (green) components for a single observation (orange) and the unit circle (red). (c) The boundary of the quantile set $\mathbb{Q}_{u}$ for $\alpha=0.2$ (red) and threshold exceeding observations (green). (d) Example probability regions in $\RR^2$ for which the SPAR model can be used for inference, with the minimal threshold $q_1^M$ defined in Appendix~\ref{appen:sim_prob}. }
    \label{fig:SPAR_illustrative}
\end{figure}

Let $\boldsymbol{X} = (X_1,\hdots,X_d)$ denote a $d$-dimensional IID random vector defined on a subset $\mathcal{R} \subset \mathbb{R}^d \setminus \boldsymbol{0}$ with continuous density function $f_{\boldsymbol{X}}$ and $\boldsymbol{0} = (0,\hdots,0)$. We assume that $\mathcal{R}$ is star-shaped, with its star-centre at $\boldsymbol{0}$; this implies that for every point $\boldsymbol{x} \in \mathcal{R}$, the line segment from $\boldsymbol{0}$ to $\boldsymbol{x}$ is also in $\mathcal{R}$, i.e., $\{t\boldsymbol{x} \mid t \in (0,1] \} \subset \mathcal{R}$ \cite{demonte2025generative}. For the simulated discharge, this assumption holds under a simple pre-processing, illustrated in Figure~\ref{fig:SPAR_illustrative}a--b, which transforms the discharge magnitudes from $\mathbb{R}_+$ to $\mathbb{R}$ and centres the random vector around $\boldsymbol{0}$; see Appendix~\ref{appen:marg_stand} for further details. Note that for our discharge data we have $d=4$.

Under the SPAR modelling approach, the random vector $\boldsymbol{X}$ is decomposed into a radial component $R$, and a vector of (pseudo-)angles $\boldsymbol{W}$, defined as
\begin{equation*}
R=\|\boldsymbol{X}\|\in\mathbb{R}_+, \qquad \boldsymbol{W} = \boldsymbol{X}/\|\boldsymbol{X}\| \in \mathbb{S}^{d-1},
\label{eqn:RWdecomposition}
\end{equation*}
where $|| \cdot||$ denotes the Euclidean norm and $\mathbb{S}^{d-1} := \{ \boldsymbol{x} \in \mathbb{R}^d \mid \| \boldsymbol{x}\| = 1 \}$ is the unit $(d-1)$-hypersphere; see Figure~\ref{fig:SPAR_illustrative}b, for example. The joint density function of $R$ and $\boldsymbol{W}$, denoted here by $f_{R,\boldsymbol{W}}$, is related to $f_{\boldsymbol{X}}$ via the equation ${f_{R,\boldsymbol{W}}(r,\boldsymbol{w}) = r^{d-1} \, f_{\boldsymbol{X}}(r \boldsymbol{w})}$, $r>0, \boldsymbol{w}\in\mathbb{S}^{d-1}$, where $r^{d-1}$ is the Jacobian determinant for the transformation. Note that the mapping $\boldsymbol{X}\mapsto (R,\boldsymbol{W})$ is one-to-one, implying no information about the stochastic behaviour of $\boldsymbol{X}$ is lost via this decomposition.

Observe that extremely large (absolute) magnitudes in at least one component of $\boldsymbol{X}$ cause large extremes of the radial variable $R$; this observation motivates the SPAR framework \cite{Mackay2023}. Applying Bayes' theorem, we decompose the joint density of $(R,\boldsymbol{W})$ into the conditional form: $f_{R,\boldsymbol{W}}(r,\boldsymbol{w}) = f_{\boldsymbol{W}}(\boldsymbol{w}) f_{R|\boldsymbol{W}}(r|\boldsymbol{w}).$ In this way, the problem of modelling multivariate extremes is transformed to that of separately modelling an angular density, $f_{\boldsymbol{W}}$, and the tail of the univariate conditional radial density, $f_{R|\boldsymbol{W}}$. 
As noted in the Introduction, the GPD is a natural choice for modelling the tail of a univariate random variable. The SPAR framework applies this principle to $R|(\boldsymbol{W} = \boldsymbol{w})$; treating the angle $\boldsymbol{W}$ as a covariate, one can adapt the existing frameworks for GPD regression, introduced in Section~\ref{subsec:ns_pot}, to model the tail of the radial component. In this way, the SPAR framework intuitively links univariate and multivariate paradigms for modelling extreme events. 

First, define the `threshold' function $u : \mathbb{S}^{d-1} \mapsto \mathbb{R}_+$ as the conditional $(1-\alpha)$-quantile of $R|(\boldsymbol{W} = \boldsymbol{w})$, with $\alpha\in(0,1)$ close to 0, i.e., the solution $u(\cdot)$ of the equation $\alpha = \Pr (R>u(\boldsymbol{w}) | \boldsymbol{W} = \boldsymbol{w})$. In direct analogy to equation~\eqref{eqn:ns_gpd}, we assume the conditional excesses $(R - u(\boldsymbol{w})) \mid (\boldsymbol{W} = \boldsymbol{w}, R > u(\boldsymbol{w}))$ follow a GPD, implying that for large values of $r>u(\boldsymbol{w})$,
\begin{equation*} 
    f_{R,\boldsymbol{W}} (r,\boldsymbol{w}) = f_{\boldsymbol{W}}(\boldsymbol{w}) \,f_{R|\boldsymbol{W}}(r|\boldsymbol{w}) \approx \alpha f_{\boldsymbol{W}} (\boldsymbol{w}) h(r - u(\boldsymbol{w}); \sigma (\boldsymbol{w}), \xi(\boldsymbol{w})),
\end{equation*} 
where $h$ is the GPD density in equation~\eqref{eqn:gpd_dens}, and {$\sigma (\boldsymbol{w})>0$} and $\xi(\boldsymbol{w}) \in \mathbb{R}$ denote angle-dependent scale and shape parameters, respectively. 

Combined with an appropriate model for the angular density $f_{\boldsymbol{W}}$, the SPAR model can be used to perform inference on the joint extremes of $\boldsymbol{X}$. In particular, one can evaluate the distribution of $\boldsymbol{X}$ in regions where the radial component is (conditionally) extreme. To see this, first note that restricting attention to the threshold-exceeding radial observations does not alter the resulting angular density, i.e., $(\boldsymbol{W} \mid R > u(\boldsymbol{W})) \,{\buildrel d \over =}\, \boldsymbol{W},$ implying standard statistical techniques can be used to model $f_{\boldsymbol{W}}$ \cite{Mackay2023,Papastathopoulos2025}. Letting $\mathbb{Q}_{u} := \{ \boldsymbol{x} \in \mathbb{R}^{d} \mid \| \boldsymbol{x}\| \leq u(\boldsymbol{x}/\| \boldsymbol{x}\|) \}$ be the set of non-threshold exceeding observations, the SPAR model is valid on the region $\mathbb{R}^d \setminus \mathbb{Q}_{u} = \mathbb{Q}^c_{u}$, i.e., the set of points that are (radially) far from the centre of $\mathcal{R}$. Under our modelling assumptions, the joint density of $\boldsymbol{X}$ on $\mathbb{Q}^c_{u}$ is given by 
\begin{equation*}
    f_{\boldsymbol{X}}(\boldsymbol{x}) = (\|\boldsymbol{x}\|)^{1-d} \alpha f_{\boldsymbol{W}} (\boldsymbol{x}/\|\boldsymbol{x}\|) h\left[\|\boldsymbol{x}\| - u(\boldsymbol{x}/\|\boldsymbol{x}\|); \sigma(\boldsymbol{x}/\|\boldsymbol{x}\|), \xi(\boldsymbol{x}/\|\boldsymbol{x}\|)\right], \quad \boldsymbol{x} \in \mathbb{Q}^c_{u},
\end{equation*}
where $(\|\boldsymbol{x}\|)^{1-d}$ denotes a Jacobian term. Consequently, the SPAR framework fully describes the tail behaviour of $\boldsymbol{X}$ on $\mathbb{Q}^c_{u}$. We henceforth refer to $\mathbb{Q}^c_{u}$ as the joint tail of $\boldsymbol{X}$. 

An illustrative schematic for the SPAR framework is provided in Figure~\ref{fig:SPAR_illustrative}. One can observe the joint tail region $\mathbb{Q}^c_{u}$ for which the SPAR model is valid, and note that this region encompasses values in both the upper and lower marginal tails. Figure~\ref{fig:SPAR_illustrative}d illustrates example tail regions where the SPAR framework is valid, and can thus be used to there make probabilistic statements. Further details about how the SPAR model can be used for simulation and probability estimation can be found in Appendix~\ref{appen:sim_prob}. 

    

\subsection{Model Fitting} \label{sec:model_fit}

\subsubsection{The deep SPAR framework} \label{subsec:deepSPAR}

In this work, we use deep learning to estimate our SPAR model: the so-called \textit{deep SPAR} framework \cite{Mackay2025b} represents the GPD threshold and parameter functions via flexible neural networks that are optimised using standard deep learning methods \cite{Kingma2017}. The deep SPAR model builds upon recent work using deep learning to perform GPD regression, which has been applied to study complex conditional tail behaviours in a variety of applications, including wildfire modelling and flood risk management \cite{Richards2023a, Cisneros2023, Pasche2024, Majumder2025}; see overview by \citeA{ Richards2024,Richards2022a}. 

We briefly introduce the deep SPAR setup in this section, and refer the reader to \citeA{Mackay2025b} for further details. The exceedance threshold $u(\cdot)$, scale $\sigma(\cdot)$, and shape $\xi(\cdot)$ parameter functions are represented as multilayer perceptrons (MLPs), which are a class of fully-connected, feed-forward neural networks \cite{Goodfellow2016}. Neural networks offer models a high degree of flexibility and scalability, owing to established approximation theorems \cite{Hornik1989,Schmidt-Hieber2020,Farrell2021}. This therefore allows one to fit the SPAR model in a highly flexible and general manner without imposing strong constraints on the threshold and parameter functions. Moreover, deep learning permits the use of the SPAR framework in higher dimensional settings than competing approaches.  

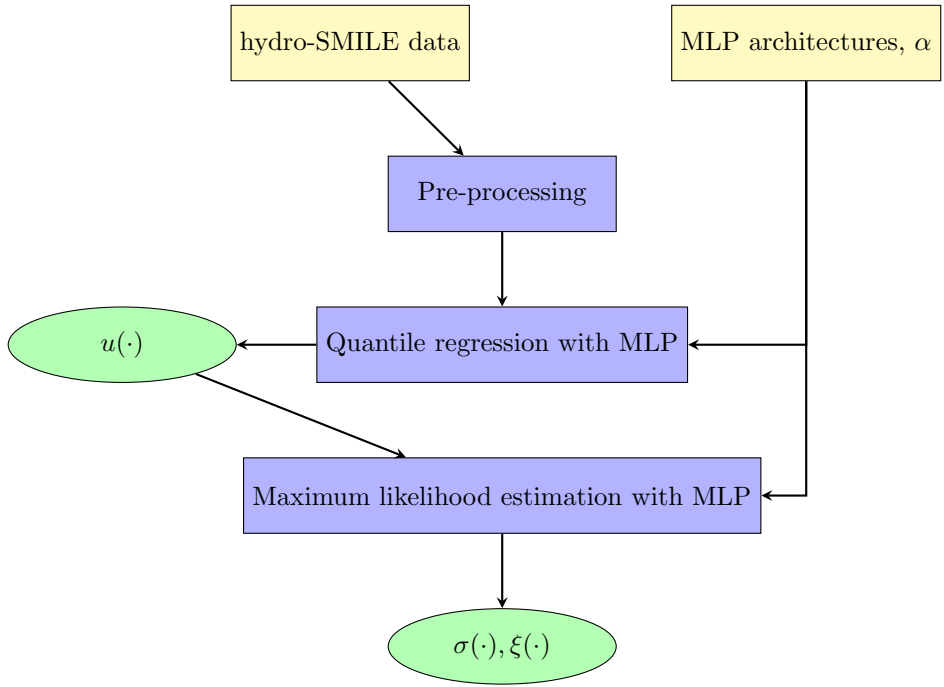
\begin{figure}[h!]
    \centering

    \begin{tikzpicture}[node distance=2cm]
    
    \node (data) [input] {hydro-SMILE data};
    \node (tuning) [input, right of=data, xshift=4cm] {MLP architectures, $\alpha$};
    
    \node (preprocess) [process, below of=data, xshift=2cm] {Pre-processing};
    \node (quantile) [process, below of=preprocess] {Quantile regression with MLP};
    \node (mle) [process, below of=quantile] {Maximum likelihood estimation with MLP};
    
    \node (threshold) [output, left of=quantile, xshift=-3cm] {$u(\cdot)$};
    \node (parameters) [output, below of=mle] {$\sigma(\cdot),\xi(\cdot)$};
    
    \draw [arrow] (data) -- (preprocess);
    \draw [arrow] (tuning) |- (quantile);
    \draw [arrow] (tuning) |- (mle);
    \draw [arrow] (preprocess) -- (quantile);
    \draw [arrow] (quantile) -- (threshold);
    \draw [arrow] (mle) -- (parameters);
    \draw [arrow] (threshold) -- (mle);
    \end{tikzpicture}

    \caption{Flowchart illustrating our implementation of the deep SPAR modelling framework. We take the hydro-SMILE data, threshold exceedance probability $\alpha$, and MLP architectures as inputs, and output estimates of the threshold $u(\cdot)$ and GPD parameter functions, $(\sigma(\cdot),\xi(\cdot))$.}
    \label{fig:threshold_gpd_flowchart}

\end{figure}

Similarly to standard inference for GPD regression models, inference for the deep SPAR model proceeds in two stages. First, we estimate the threshold function, $u(\boldsymbol{w})$, via standard quantile regression techniques \cite{Koenker2017}. Conditional on $u(\boldsymbol{w})$, we subsequently estimate the GPD parameter functions, $\sigma(\boldsymbol{w})$ and $\xi(\boldsymbol{w})$, via maximum likelihood estimation with hard parameter sharing \cite{Ruder2017,Rothfuss2019}. This requires two neural network models; one for the threshold $u(\boldsymbol{w})$ and another for the parameter vector $( \sigma(\boldsymbol{w}), \xi(\boldsymbol{w}))$. Both $u(\boldsymbol{w})$ and $(\sigma(\boldsymbol{w}),\xi(\boldsymbol{w}))$ are modelled by MLPs, which are composed of multiple hidden layers of `neurons'. Each neuron passes a linear transformation of input variables through a nonlinear `activation function', and the output is then passed to the subsequent layer. In our setting, the MLP defining $u(\boldsymbol{w})$ takes in as input $\boldsymbol{w}\in \mathcal{S}^{d-1}$ and provides a single output $u(\boldsymbol{w})\in \mathbb{R}_+$. For the second stage of the modelling procedure, we fit a deep GPD regression model to excesses of $R$ above $u(\boldsymbol{w})$. This requires a single MLP which outputs a vector $(\sigma(\boldsymbol{w}),\xi(\boldsymbol{w}))$. In this case, we use an orthogonal reparametrisation of the GPD parameters \cite{Pasche2024}, but we omit the details for brevity. 
A flowchart illustrating our modelling procedure is given in Figure~\ref{fig:threshold_gpd_flowchart}. Further details regarding optimisation and parameter estimation can be found in the Supporting Information. 

\subsubsection{Architecture selection} \label{subsec:arch_design}

Selecting an appropriate architecture is a well-known challenge in the deep learning literature, as the performance of a model can vary significantly depending on its configuration and the specific dataset being analysed. In this work, we select architectures by comparing diagnostics (Section~\ref{subsec:diag}) across different setups and visually identifying those that demonstrate the best model fit to the discharge data. For discussion on architecture choice for deep SPAR models, see also \citeA{Mackay2026diags}.

For both MLPs, we take all activation functions to be the rectified linear unit, defined by $\text{ReLU}(\mathbf{x})=(\max\{x_1,0\},\max\{x_2,0\},\dots)$.  This is a popular choice well-studied in theoretical contexts \cite{Goodfellow2016,Schmidt-Hieber2020} that has shown good performance in extremal settings \cite{Richards2024}. Alternative activation functions (sigmoid, tanh) were considered, but found to provide worse model fits. 

The final layers of the MLPs are used to constrain the range of the angular functions: $u(\boldsymbol{w})$, $\sigma(\boldsymbol{w})$, and $\xi(\boldsymbol{w})$. We use exponential outputs to ensure $u(\boldsymbol{w})$ and $\sigma(\boldsymbol{w})$ are strictly positive for any input angle $\boldsymbol{w}$. For the shape parameter $\xi(\boldsymbol{w})$, we tested a range of activation functions which place finite bounds on $\xi(\boldsymbol{w})$. These include: $0.5\tanh(x)$, $0.75\tanh(x) + 0.25$, $\tan(x)/\pi$, $(3/2)\times (\tan(x)/\pi+0.5) - 0.5$, which  bound the support of $\xi(\boldsymbol{w})$ to either the interval $[-0.5,0.5]$ or $[-0.5,1]$. Note that $\xi$ must satisfy $\xi > -0.5$ to ensure regularity of the GPD likelihood \cite{Smith1985}. We found no advantage to allowing $\xi(\boldsymbol{w}) \in [-0.5,1]$ as opposed to $\xi(\boldsymbol{w}) \in [-0.5,0.5]$, since the outputs for the former never gave estimates of the shape parameter above $0.5$; and that the model fits with the $\tan$-based constraint were marginally better than those constructed using the $\tanh$ function. Consequently, we opt to use $\tan(x)/\pi$ to constrain the range of $\xi(\boldsymbol{w})$. 

Regarding depth and width of the MLPs, we found $3$ hidden layers and $32$ neurons per layer to be sufficient for our data. Adding additional flexibility did not significantly improve model fits, while using a shallower or narrower architecture reduced the quality of model fits.

Finally, our deep SPAR model also requires one to specify an exceedance level, $\alpha$, for estimating the threshold function $u(\cdot)$. This selection is critical for obtaining reliable model fits, representing a bias-variance trade-off \cite{Murphy2024}. We tested a range of levels, ultimately finding that setting $\alpha = 0.15$ was sufficient for obtaining reliable and accurate model fits that represent observed tail behaviour.  

We stress here that while our selected model architecture works well for our data, we do not advocate the general use of the selected architecture. We instead recommend that practitioners wishing to apply this framework consider a range of possible architectures and use model diagnostics (Section~\ref{subsec:diag}) to decide upon the most appropriate model for a given dataset. 

\subsubsection{Modelling the angular density function} \label{subsec:ang_dens}
For modelling of the angular density $f_{\boldsymbol{W}}$ (the secondary aspect of the SPAR framework), we opt to use an empirical estimator. In practice, deep learning models could also be used \cite{demonte2025generative, Wessel2025, Lhaut2025}. However, best practices for applying such techniques remain unclear, and we find an empirical estimator is sufficient. In practice, we resample (with replacement) from the angular observations when simulating from the SPAR model. Given our data application (Section~\ref{subsec:data_set}) is relatively low dimensional, empirical sampling suffices for this work; this is backed up by our strong model performance (Section~\ref{sec:results}).  

\subsubsection{Uncertainty assessment} \label{subsec:uncertainty}

We use non-parametric bootstrapping (i.e., resampling with replacement) to estimate uncertainty in our model fits. Bootstrap techniques are well established and offer desirable theoretical guarantees in terms of uncertainty quantification under mild conditions \cite{Efron1994}. Moreover, similar studies considering deep learning of multivariate extremes have found non-parametric bootstraps to be sufficient for quantifying uncertainty and providing good coverage \cite{demonte2025generative,Murphy-Barltrop2024c}. For a single time window, model fitting takes less than one hour to run on a machine equipped with an Intel(R) Core(TM) i5-8250U processor (4 cores, 1.60GHz) and 16 GB RAM running Windows 11. As such, repeated resampling and fitting of several models is not infeasible or an overly expensive computational challenge. For our analysis, we use $100$ bootstrap samples for each model setup. For all diagnostics and computed statistics, we report 95\% approximate confidence intervals, computed empirically across the bootstrapped model fits. 

\subsection{Model evaluation diagnostics} \label{subsec:diag}
We consider a range of goodness-of-fit diagnostics. When relevant, all diagnostics are computed on the scale of the data, i.e., after inverting the pre-processing step described in Appendix~\ref{appen:marg_stand}. 

\textbf{Quantile-quantile (QQ) plots:} In the case of the SPAR model, the model parameters for the GPD are angle dependent, and therefore deriving stationary QQ plots is not possible. Consequently, we transform all threshold exceeding observations to the standard exponential scale via the probability integral transform and equation~\eqref{eqn:ns_gpd}, and compare these values with quantiles from the standard exponential distribution; see, e.g., \citeA{Coles2001}. We henceforth refer to this diagnostic as the \textit{GPD QQ plot}. 

We further compute QQ plots comparing the observed and model quantiles in each margin for the tail region $\mathbb{Q}^c_{u}$. This allows us to observe how well the SPAR model recreates the marginal distribution, including the tails, of each variable. 

\textbf{Pairwise extremal dependence coefficients:} In the multivariate extremes literature, it is common to summarise extremal dependence using pairwise summary measures \cite{Coles1999}. Given two random variables $X_1$ and $X_2$, the most common extremal dependence summary is the upper-tail index $\chi = \lim _{u \rightarrow 1^-} \chi(u) \in [0,1]$, where
\begin{equation*}
    \chi(u) := \Pr(F_2(X_2)>u \mid F_1(X_1)>u) \in [0,1],
\end{equation*}
and $F_1,F_2$ denote the marginal distribution functions of $X_1$ and $X_2$, respectively \cite{Joe1997}. Higher values of $\chi$ correspond to stronger positive dependence in the upper tails of $X_1$ and $X_2$. 

In practice, we cannot evaluate the limit $\lim _{u \rightarrow 1^-} \chi(u)$ unless we know both the marginal and joint distributions of $(X_1,X_2)$. We thus estimate $\chi(u)$ empirically for values of $u$ close to 1. In our case, we take hydro-SMILE and SPAR-simulated data in the tail region $\mathbb{Q}^c_{u}$, and consider each pair of variables (resulting in $\binom{4}{2} = 6$ pairs). We then compute empirical estimates of $\chi(u)$ for $u \in \{0.8,0.81,\hdots,0.99 \}$. These plots allow us to assess how well the fitted model captures the extremal dependence in the joint tail of each pair of variables. 


\textbf{Return level plots:} We estimate return levels empirically by simulating a large number of observations from the SPAR model in both the joint body $\mathbb{Q}_{u}$ and tail $\mathbb{Q}^c_{u}$ regions, using the sampling schemes discussed in Appendix~\ref{appen:sim_prob}. In particular, we simulate $[ n_{tail}/(1-\alpha) ]$ and $n_{tail}$ observations in the regions $\mathbb{Q}_{u}$ and $\mathbb{Q}^c_{u}$, respectively, with $n_{tail} = 2\times 10^6$. We found such sample sizes to be sufficient for obtaining smooth estimates of return levels up to the large return periods considered in Section~\ref{sec:results}. We compute return level plots for both the upper and lower marginal tails.

\subsection{Joint tail probabilities} \label{subsec:tail_probs_runoff}

For assessing extremal dependence beyond the pairwise summary $\chi$, we consider a range of tail risk regions relevant in the context of flood and drought assessment. Specifically, we define the structure (response) variable $S := \sum_{i}X_i$, as the total sum of weekly maxima across all four catchments. The extremal behaviour of $S$ will be driven by the marginal extremal behaviour of each $X_i$, as well as the joint extremal dependence structure \cite{richards2022tail}. Extremes in both the upper and lower tail of $S$ can correspond to flood or drought-like conditions, respectively, at one or more locations.

We consider fixed thresholds in the tails of $S$: for the upper and lower tails, we consider $s=1,000m^3/s$ and $s=30m^3/s$, respectively, noting that these are relatively arbitrary choices. We then investigate temporal changes in the probability that the sum of the discharge rates $S$ over all four catchments is above or below the given thresholds; see, for example, the yellow area in Figure~\ref{fig:SPAR_illustrative}d. We note that the magnitudes of discharges vary across catchments, and thus larger rivers (e.g., Lech) will contribute more to the upper tail of $S$ compare to smaller rivers; however, large values of $S$ can be driven by a wide variety of joint tail events, including cases where extremes are only observed on smaller rivers. The corresponding tail probabilities take the forms $\Pr(S > 1,000(m^3/s))$ and $\Pr(S < 30(m^3/s))$.

We also estimate probabilities for concurrent extremes. In particular, we evaluate joint distribution and survivor functions in the tail, as given by $\Pr(X_i > x_i^{1 - 1/N}, \forall i \in \{1,\hdots,4 \} )$ and $\Pr(X_i < x_i^{1/N}, \forall i \in \{1,\hdots,4 \} )$, respectively, where $x_i^{1 - 1/N}$ and $x_i^{1/N}$ denote the marginal upper and lower tail, respectively, $N$-year return levels of $X_i,i=1,\dots,d$. These tail probabilities correspond to cases where all weekly maxima are simultaneously extreme, similar to the blue area in Figure~\ref{fig:SPAR_illustrative}d. Moreover, these probabilities are defined such that if the extremal dependence structure remains constant across time windows, the probabilities will not vary. For our application, we set $N = 10$ years, finding this to be sufficient for defining extreme probability regions in both the upper and lower joint tails. 

\section{Results} \label{sec:results}

\subsection{Model accuracy} \label{subsec:main_results}

\begin{figure}[t] 
    \centering
    \includegraphics[width=.45\textwidth]{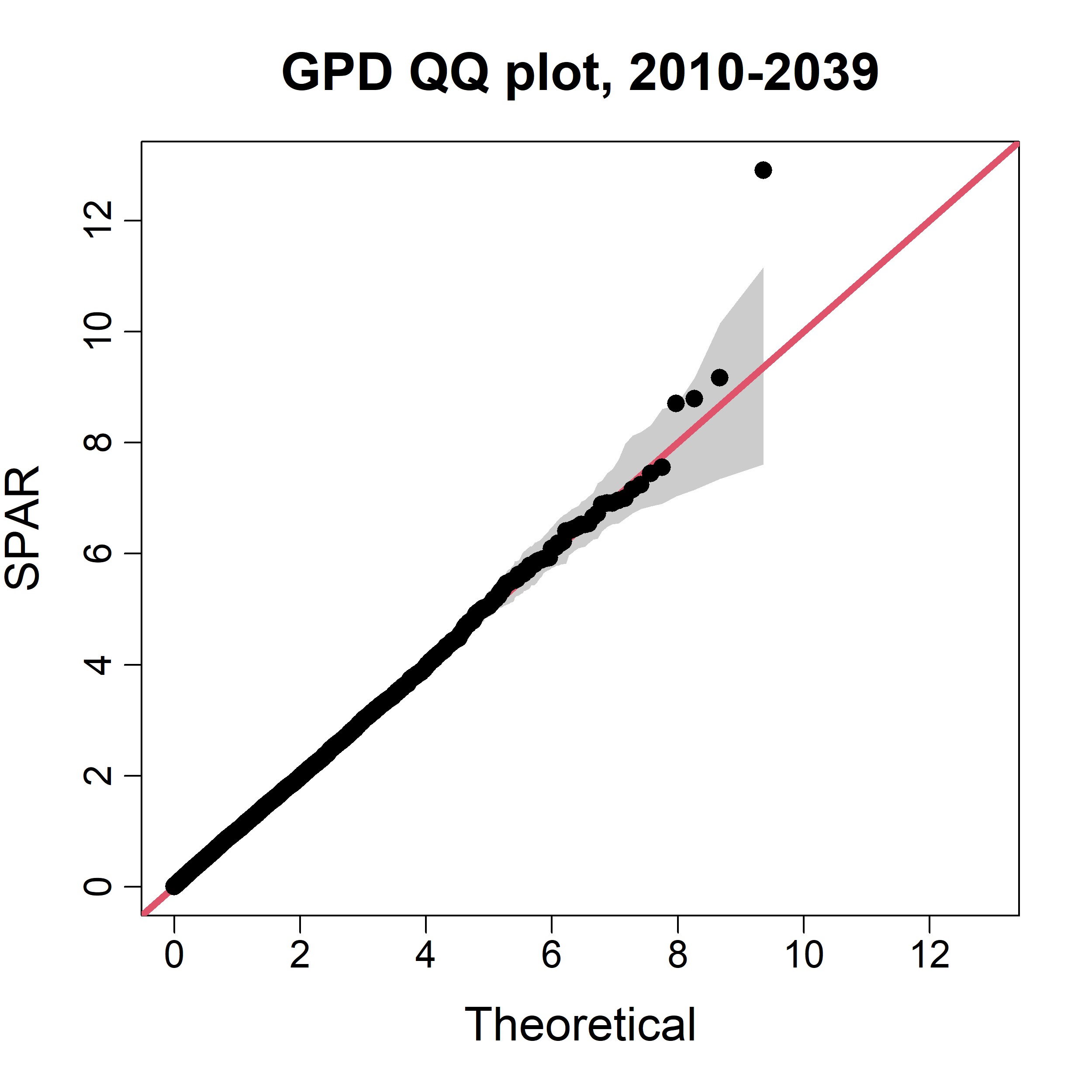}

    \caption{GPD QQ plot for the 2010--2039 time window. Shading represents 95\% confidence intervals computed via bootstrapping. }
    \label{fig:gpd_qq_plot_2}
\end{figure}

\begin{figure}[t!] 
    \centering
    \includegraphics[width=1\textwidth]{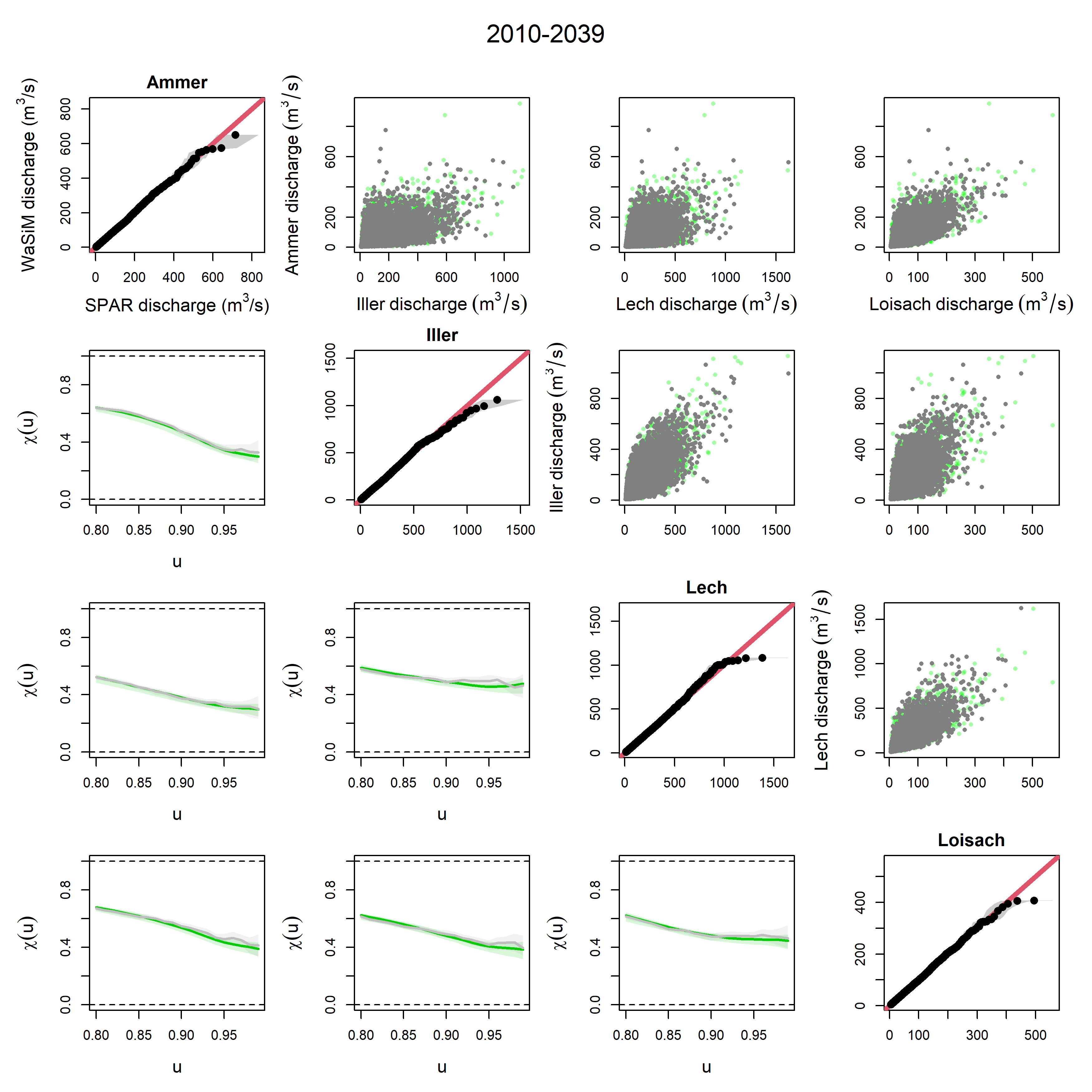}
    \caption{Model performance in the joint tail region $\mathbb{Q}^c_{u}$ for the 2010--2039 time period. Marginal QQ plots are shown on the diagonal, the upper diagonal illustrates pairwise scatterplots and the lower diagonal illustrates empirical estimates of $\chi(u)$ for a sequence of $u$ values approaching one. For the former and the latter, we also plot 95\% confidence intervals computed via bootstrapping with grey shading. For the upper and lower diagonals, the green and grey represent the data samples generated by SPAR and simulated by WaSiM, respectively.}
    \label{fig:scatter_2}
\end{figure}

We fit the deep SPAR model across each time window. Overall, the SPAR model represents the four-dimensional ($d=4$) dataset, including the multivariate extremes, very well. Here and throughout the remainder of Section~\ref{sec:results}, we present diagnostics only for the time period encompassing the modern day (2010--2039), with the plots for the remaining time periods given in the Supporting Information. Diagnostics for all time periods are qualitatively similar. The GPD QQ plots show very good agreement in all cases, indicating validity in the GPD assumption for each time window (Figure~\ref{fig:gpd_qq_plot_2}). 
Furthermore, we find very good agreement between the generated and observed marginal distributions and pairwise relationships (Figure~\ref{fig:scatter_2}). The model appears able to recreate well the physical structure of each pairwise combination. There exists some discrepancy between marginal quantiles at the most extremes observations, but in general there is good agreement. 

\begin{figure}[ht] 
    \centering
    \begin{minipage}{0.48\textwidth} 
        \centering
        \includegraphics[width=\textwidth]{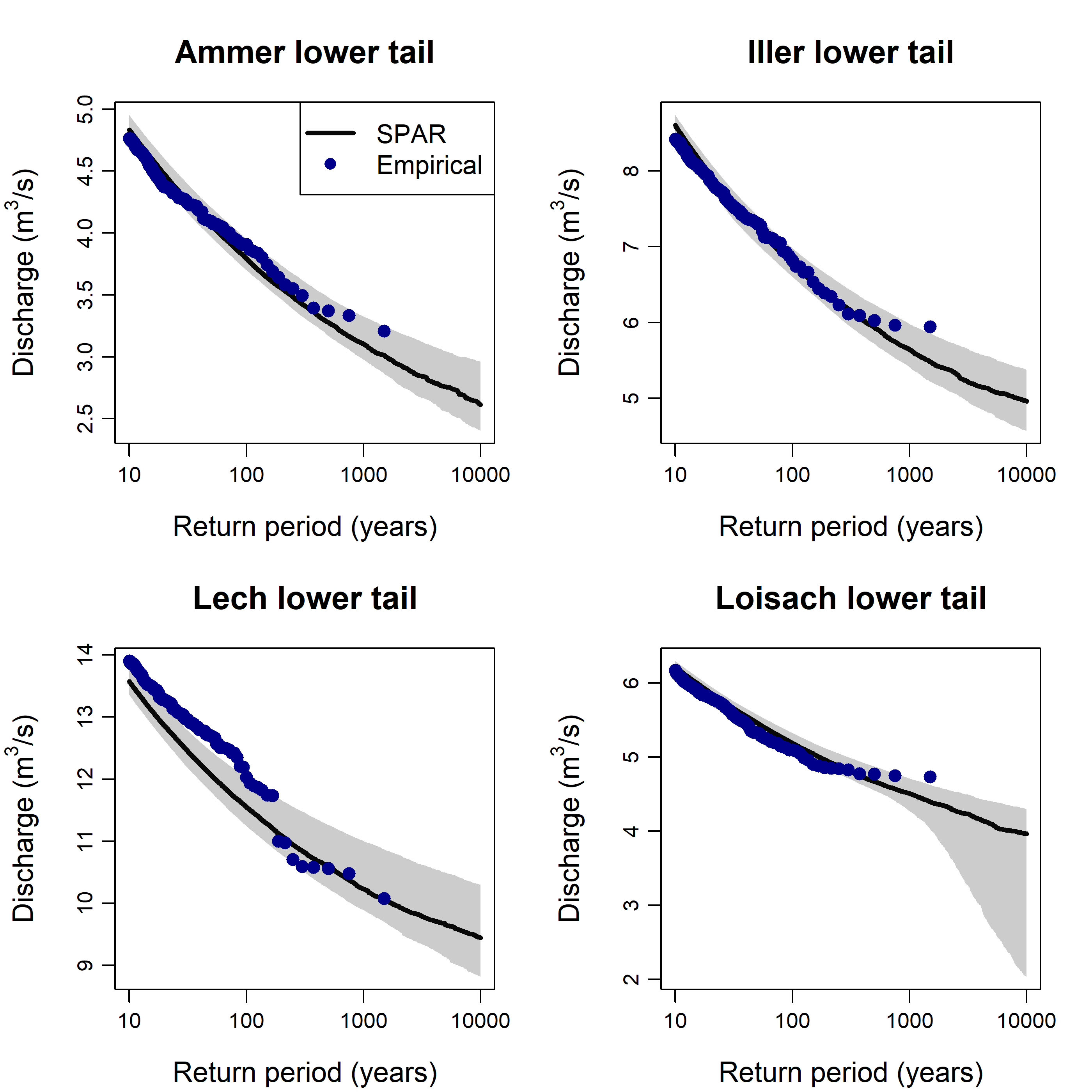}
    \end{minipage}%
    \hfill
   \begin{minipage}{0.48\textwidth} 
        \centering
        \includegraphics[width=\textwidth]{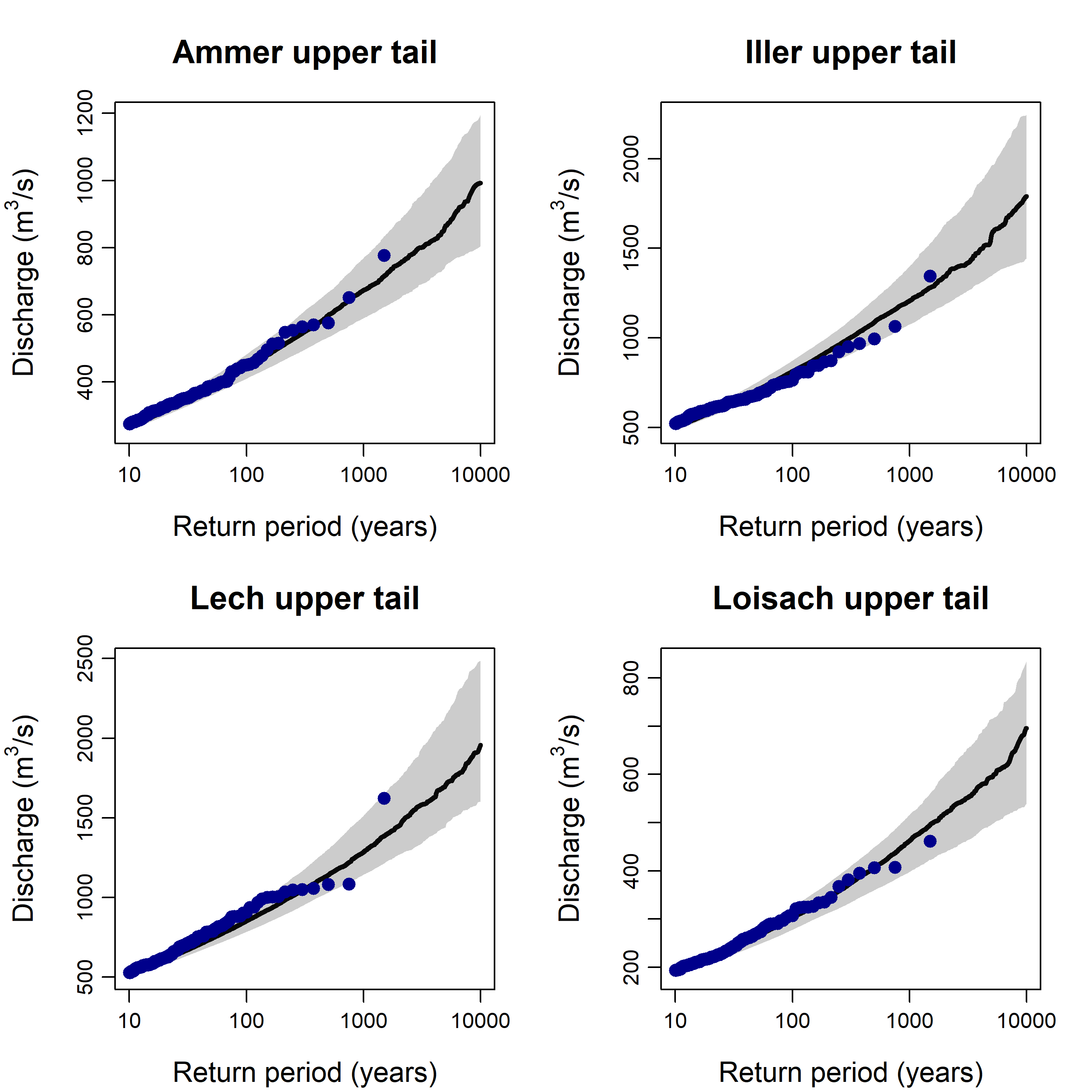}
    \end{minipage}%
    \caption{Return level plots in the lower (left) and upper (right) tails of each marginal variable for the 2010--2039 time period. In each panel, the black lines give the return levels estimated by SPAR (with 95\% confidence intervals based on bootstrapping), while the blue points give the corresponding empirical estimates from WaSim. Return periods are given on the log scale.}
    \label{fig:ret_level_2}
\end{figure}

We also observe very good agreement between the model and empirical estimates of the return levels in both the upper and lower tails, suggesting the SPAR fits capture well the marginal tails (Figure~\ref{fig:ret_level_2}). There are some cases where we observe slight discrepancies at the most extreme values, in particular for the lower tails, but such discrepancies are typically small relative to the observation scale. Moreover, we observe that the estimates for the original sample do not lie within the computed confidence intervals in a few cases; this arises due to the fact such confidence intervals are only approximate and the fact our model assumes data are independent, which may not be the case in the lower tails (see Section~\ref{sec:discuss}). 

Overall, these diagnostics indicate that our model fits well for each time window, and can accurately capture both the marginal and joint extremes of each catchment. This provides justification for the architectural choices and modelling assumptions. 

\subsection{Sensitivity analysis}\label{subsec:ss_comp}
The dataset introduced in Section~\ref{subsec:data_set} comprises data combined from 50 transient hydrological model simulations, resulting in a large sample size ($n=78,250$ data points per time window). However, in practice most environmental datasets will be much smaller. Yet this large ensemble allows us to test the robustness of the deep SPAR framework in small sample environments. We take subsamples of the full hydro-SMILE; in particular, we consider just the first $1$, $2$, $3$, $5$, and $10$ ensemble members, corresponding to $2\%$, $4\%$, $6\%$, $10\%$ and $20\%$ of the full data sample, respectively, and repeat our analysis on these smaller sample sizes. The first three subsamples ($2\%$, $4\%$, $6\%$) represent realistic observational periods for discharge datasets ($40$, $80$, and $120$ years, respectively).

For each subsample, we re-apply the marginal pre-processing procedure described in Appendix~\ref{appen:marg_stand}. Considering subsamples of the full sample significantly alters the marginal and joint behaviour, which requires the values associated with the pre-processing step to be recomputed. We re-fit the deep SPAR model to each subsample using the same setup and techniques introduced in Section~\ref{sec:model_fit}.

\begin{figure}[h]
    \centering
    \includegraphics[width=1\textwidth]{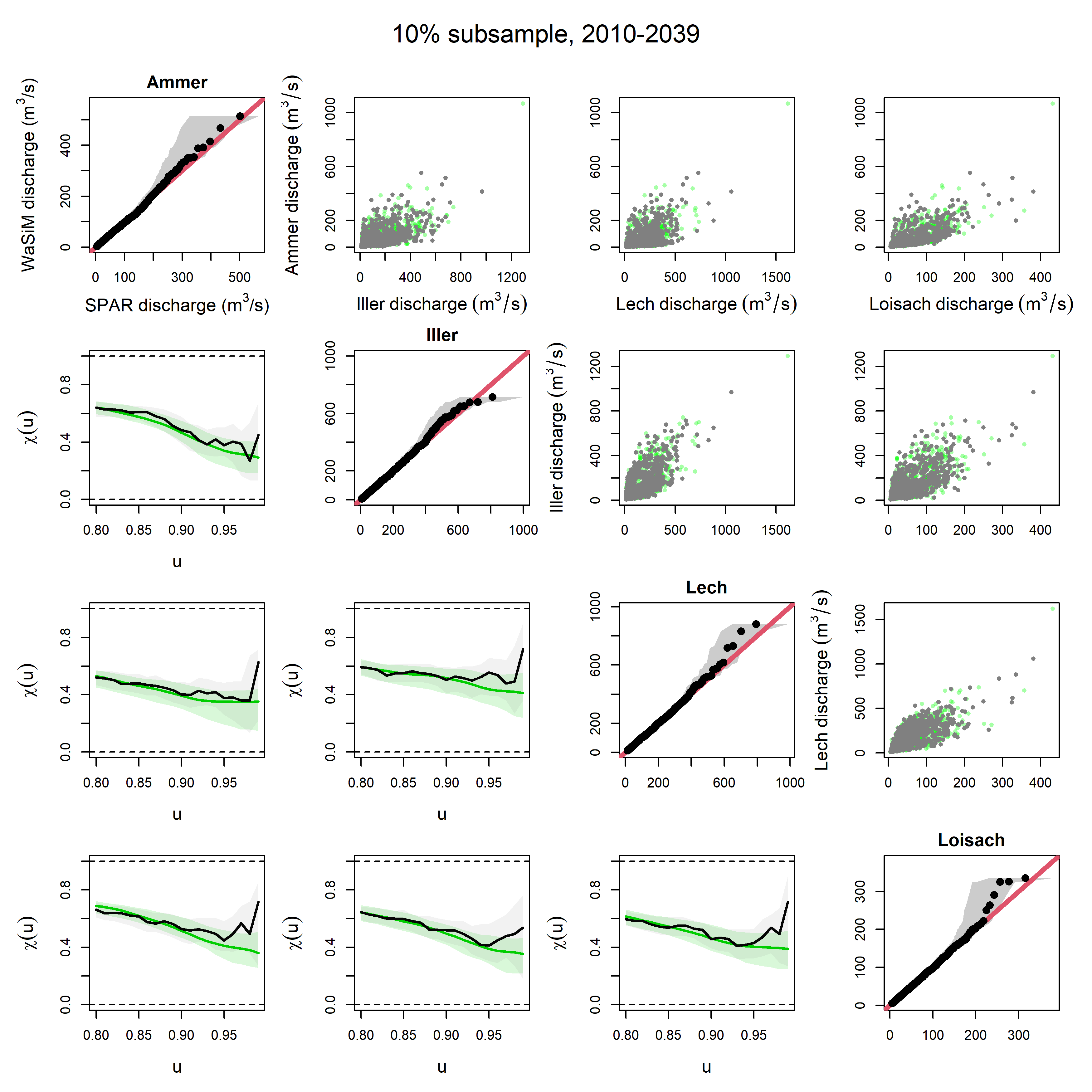}
    \caption{Model performance in the joint tail region $\mathbb{Q}^c_{u}$ for the 2010--2039 time period with the 10\% subsample equalling 150 years of data for model fitting. See Figure~\ref{fig:scatter_2} for details on the interpretation of individual panels.}
    \label{fig:scatter_tw2_nmodel_5}
\end{figure}

Reducing the sample size increases estimation uncertainty, resulting in much larger confidence intervals. For example, see Figure~\ref{fig:scatter_tw2_nmodel_5}, which illustrates the generated and observed marginal distributions and pairwise relationships for the 2010--2039 time period with the first 5 ensemble members; figures for other time periods and subsamples are given in the Supporting Information. Yet the estimated values are generally close to the hydro-SMILE quantities, especially for the $10\%$ and $20\%$ subsamples, indicating deep SPAR is still accurate for much smaller sample sizes. Moreover, the computed confidence intervals generally capture the corresponding hydro-SMILE values. 

\begin{figure}[h] 
    \centering
    \begin{minipage}{0.48\textwidth} 
        \centering
        \includegraphics[width=\textwidth]{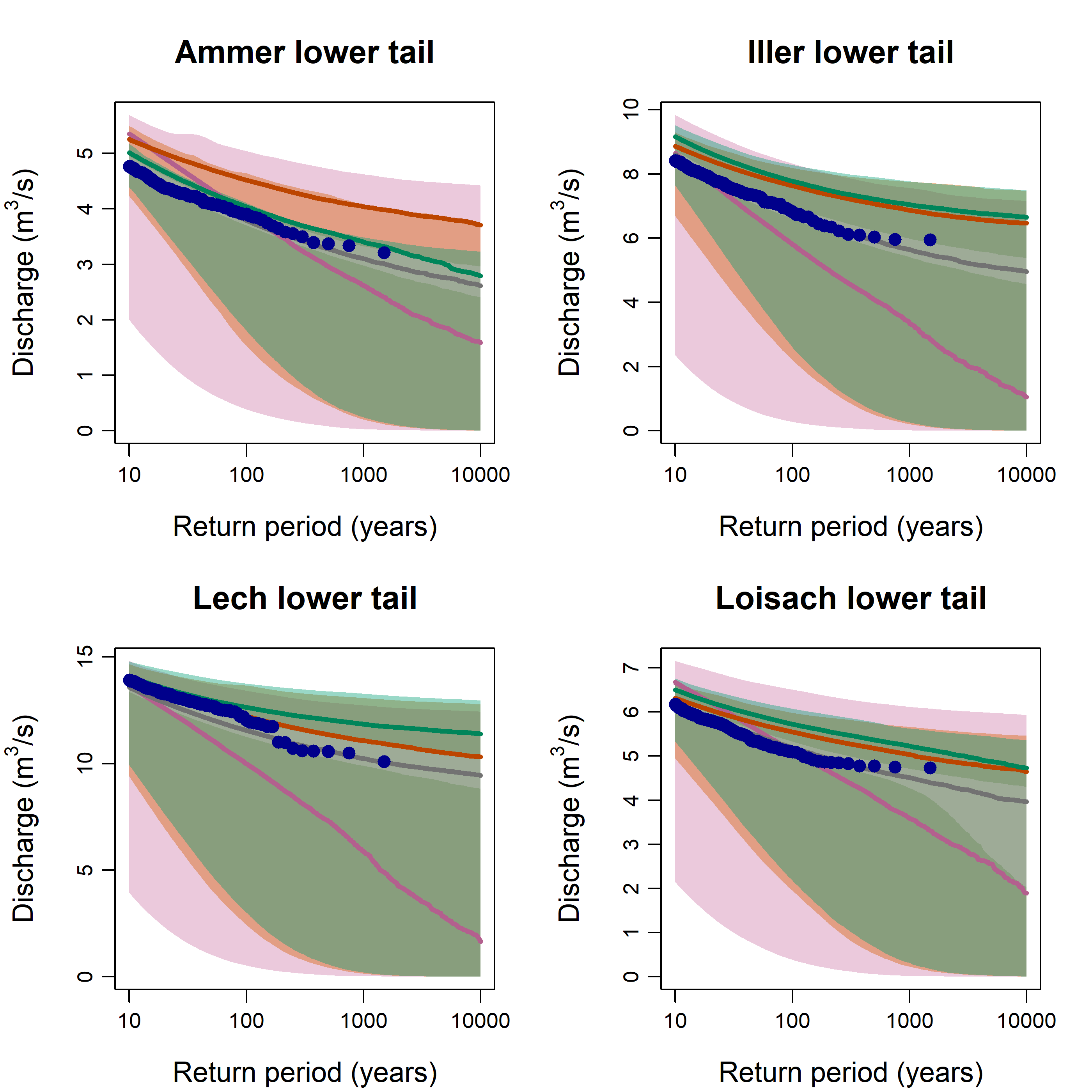}
    \end{minipage}%
    \hfill
   \begin{minipage}{0.48\textwidth} 
        \centering
        \includegraphics[width=\textwidth]{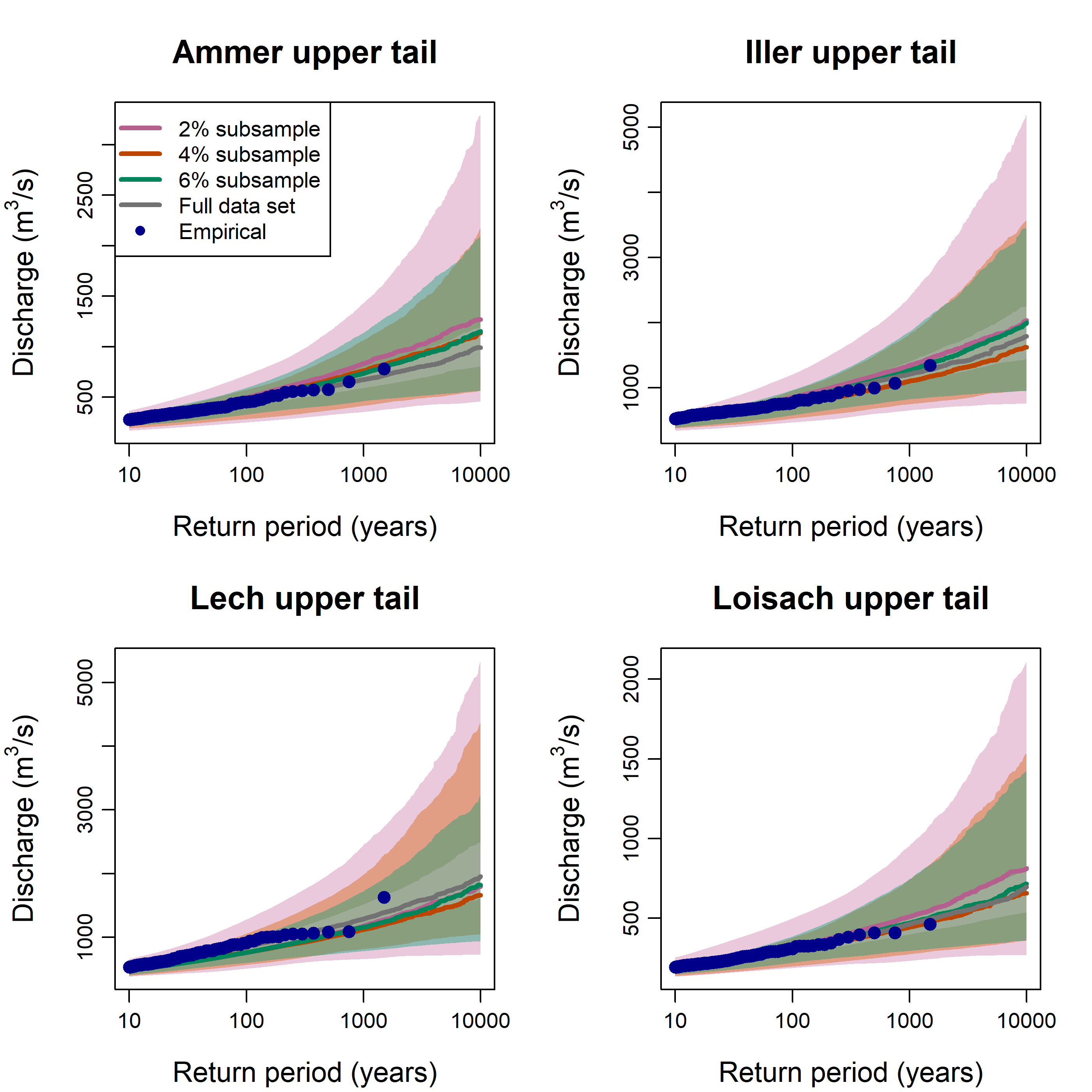}
    \end{minipage}%
    \hfill
    \begin{minipage}{0.48\textwidth} 
        \centering
        \includegraphics[width=\textwidth]{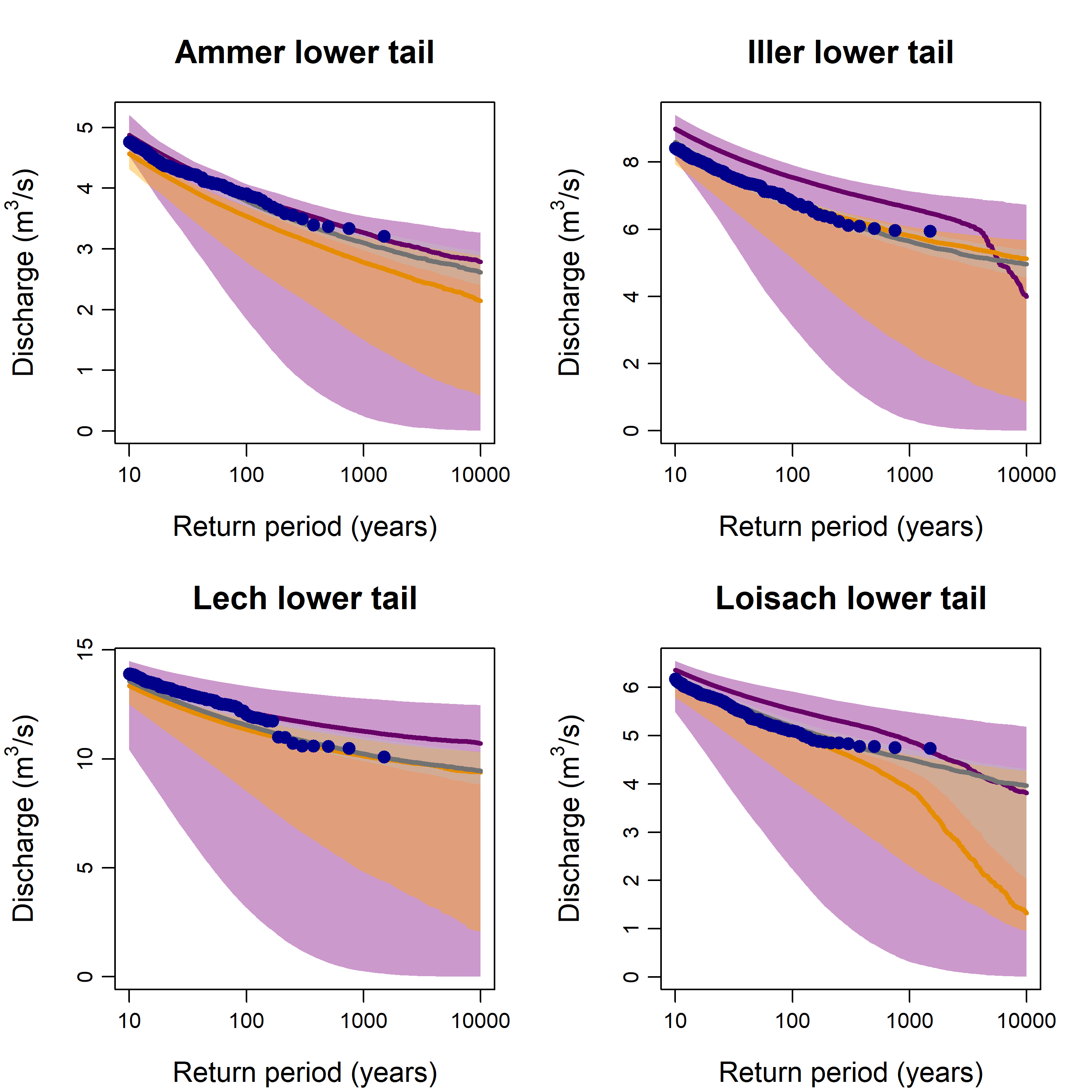}
    \end{minipage}%
    \hfill
   \begin{minipage}{0.48\textwidth} 
        \centering
        \includegraphics[width=\textwidth]{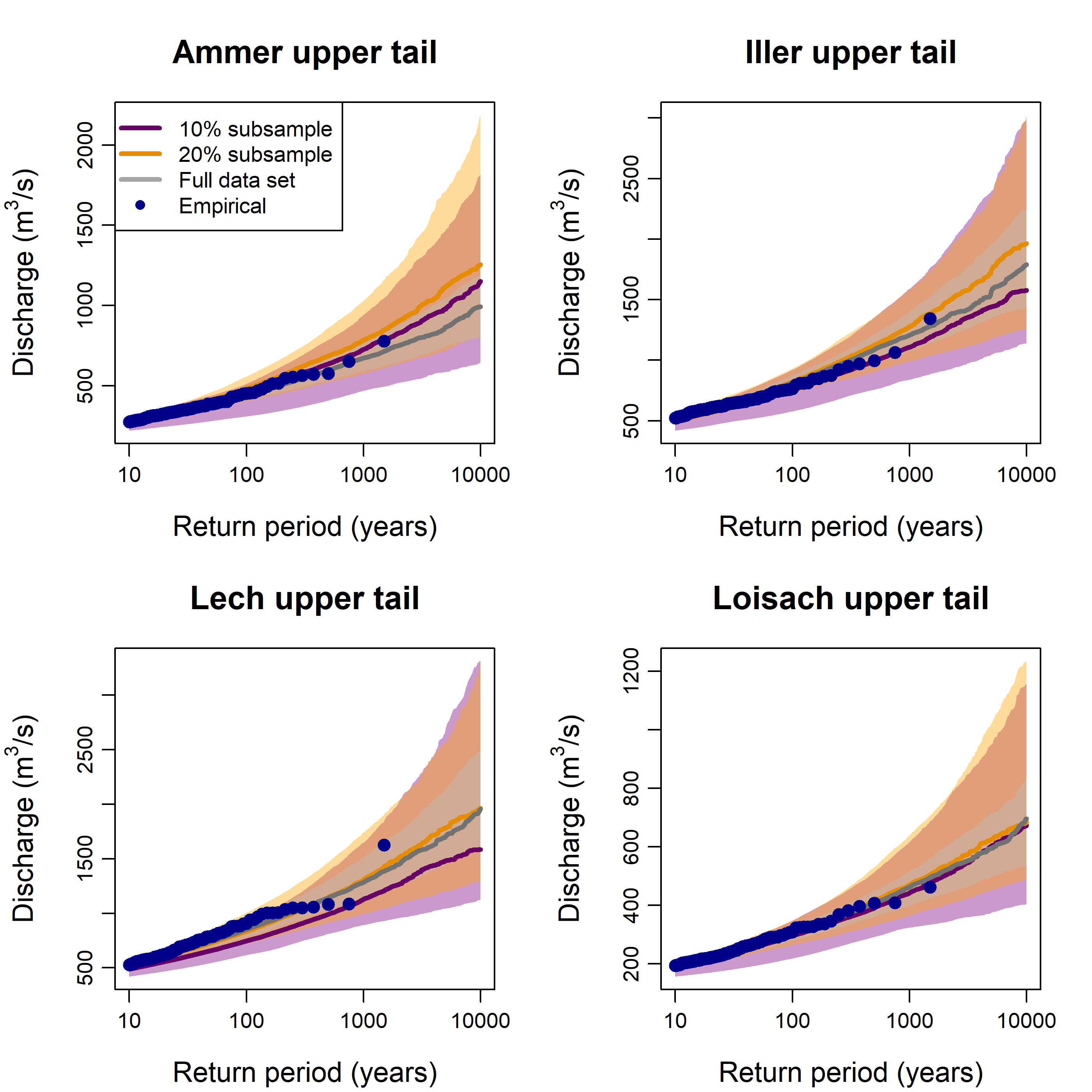}
    \end{minipage}%
    \caption{Comparison of return level estimates and uncertainty intervals (shaded; pointwise 95\% confidence intervals) from deep SPAR models trained with different sample sizes in the lower (left two columns) and upper (right two columns) tails of each marginal variable for the 2010--2039 time period. The upper and lower rows represent the smaller ($2\%$, $4\%$, $6\%$) and larger ($5\%$, $10\%$) subsamples, respectively. Blue points denote empirical return level estimates derived from the full dataset.}
    \label{fig:ret_level_comp_2}
\end{figure}

As an additional comparison and check, we compute return level estimates from the deep SPAR model fits across the range of considered subsample sizes; these are then compared against the empirical return levels obtained for the full hydro-SMILE, which allows us to investigate degradation of the quality of fits with decreasing sample size. Figure~\ref{fig:ret_level_comp_2} shows clearly how the width of the confidence intervals reduces as the sample size increases. Furthermore, the deep SPAR return levels are generally in better agreement with the empirical estimates as the sample size increases, yet the empirical estimates are mostly captured within the 95\% confidence intervals for every subsample. 
These results provide evidence that the deep SPAR setup still provides a useful inference framework for lower sample sizes. Of course, one needs to consider what level of uncertainty is tolerable, which will likely vary depending on a given application.  

\subsection{Changes in tail probabilities over time} 

\begin{figure}[h] 
    \centering
    \begin{minipage}{0.49\textwidth} 
        \centering
        \includegraphics[width=\textwidth]{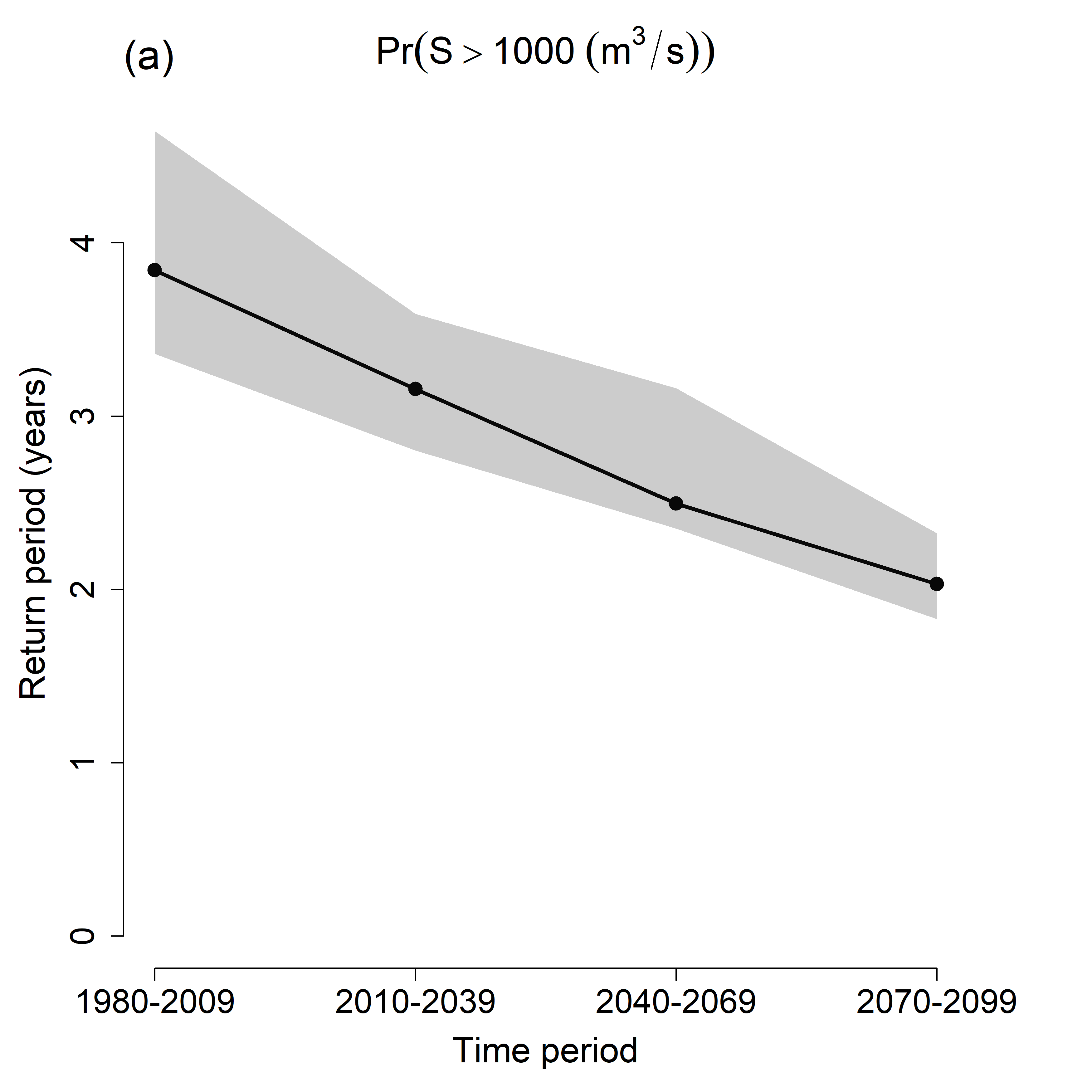}
    \end{minipage}%
    \hfill
       \begin{minipage}{0.49\textwidth} 
        \centering
        \includegraphics[width=\textwidth]{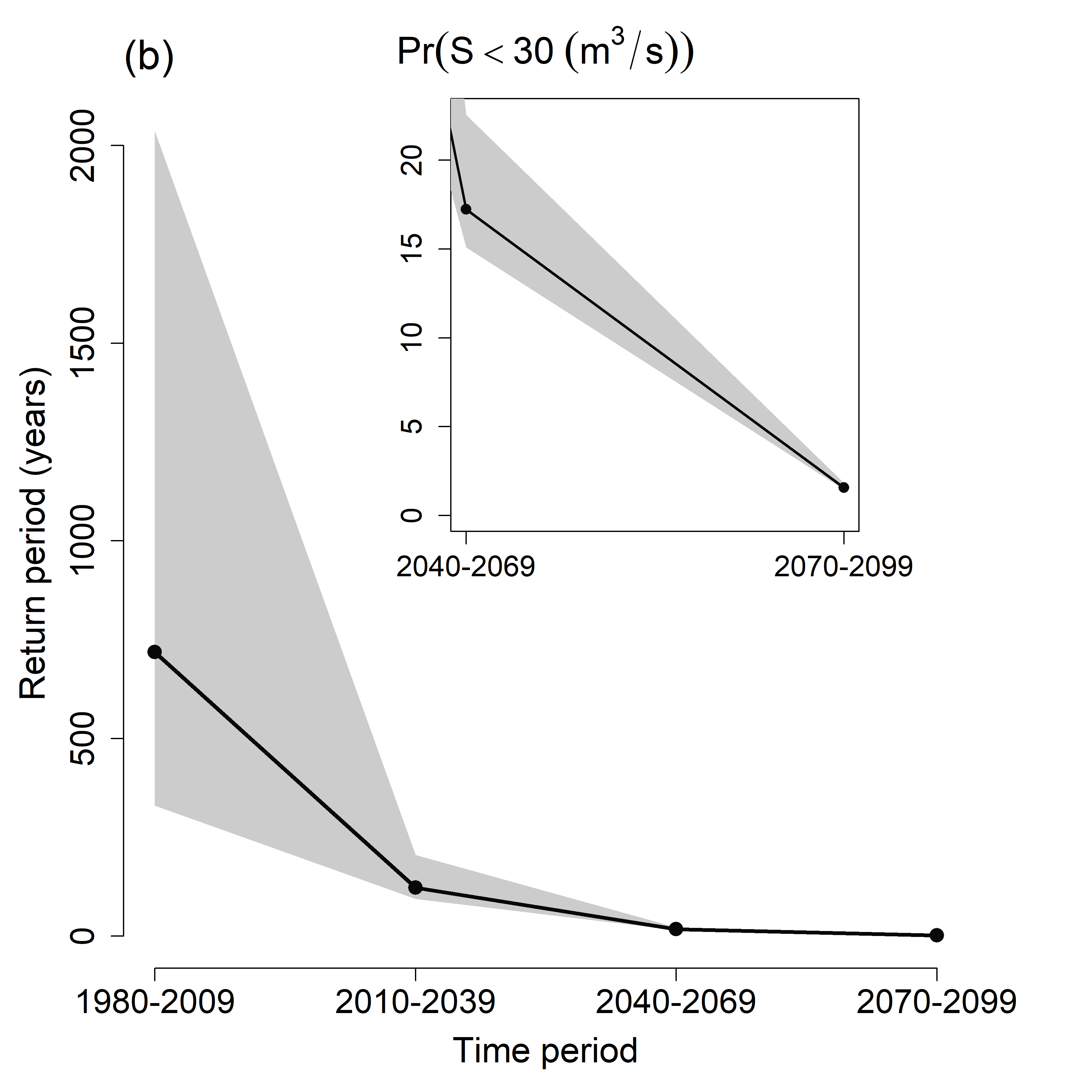}
    \end{minipage}%
     \hfill
       \begin{minipage}{0.49\textwidth} 
        \centering
        \includegraphics[width=\textwidth]{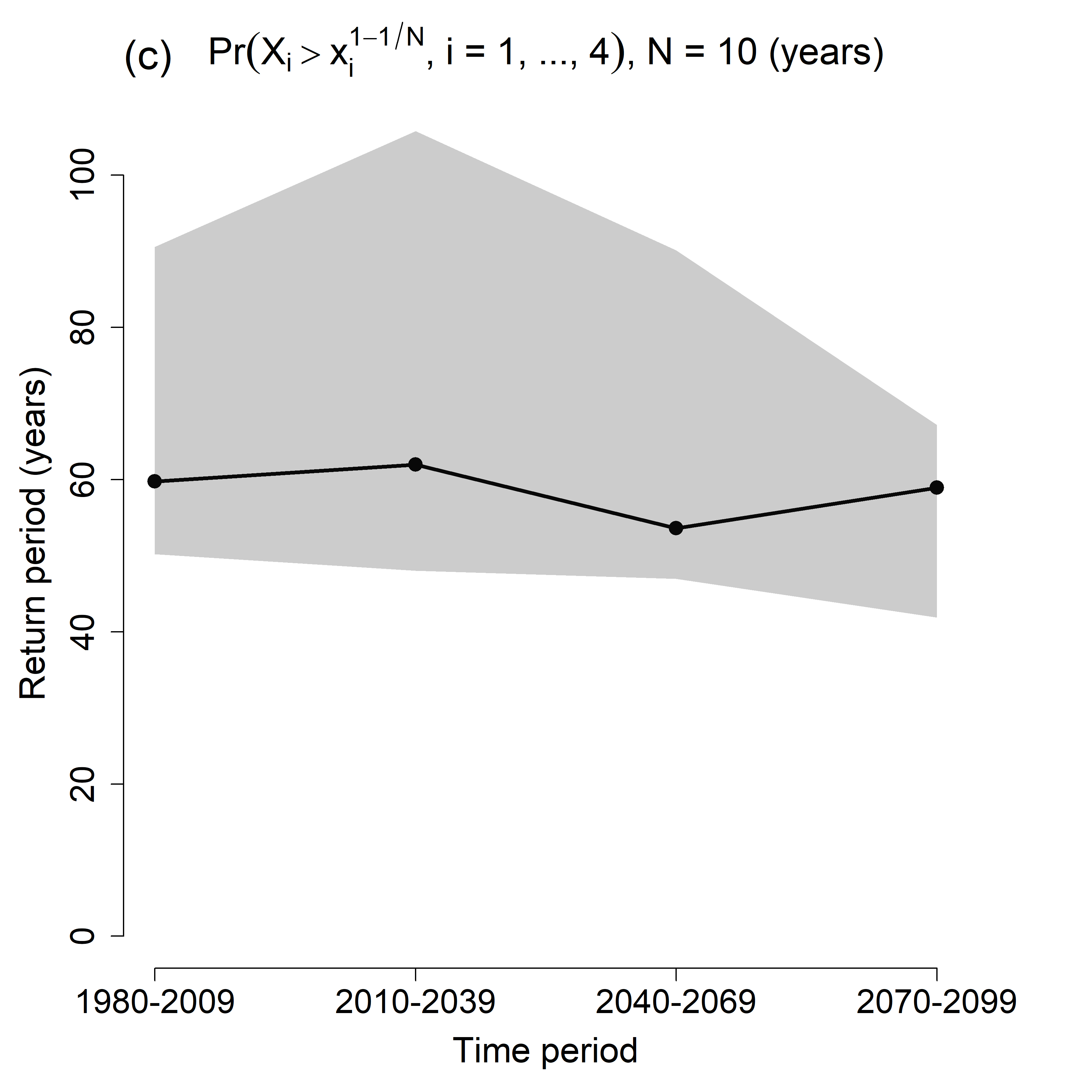}
    \end{minipage}%
     \hfill
       \begin{minipage}{0.49\textwidth} 
        \centering
        \includegraphics[width=\textwidth]{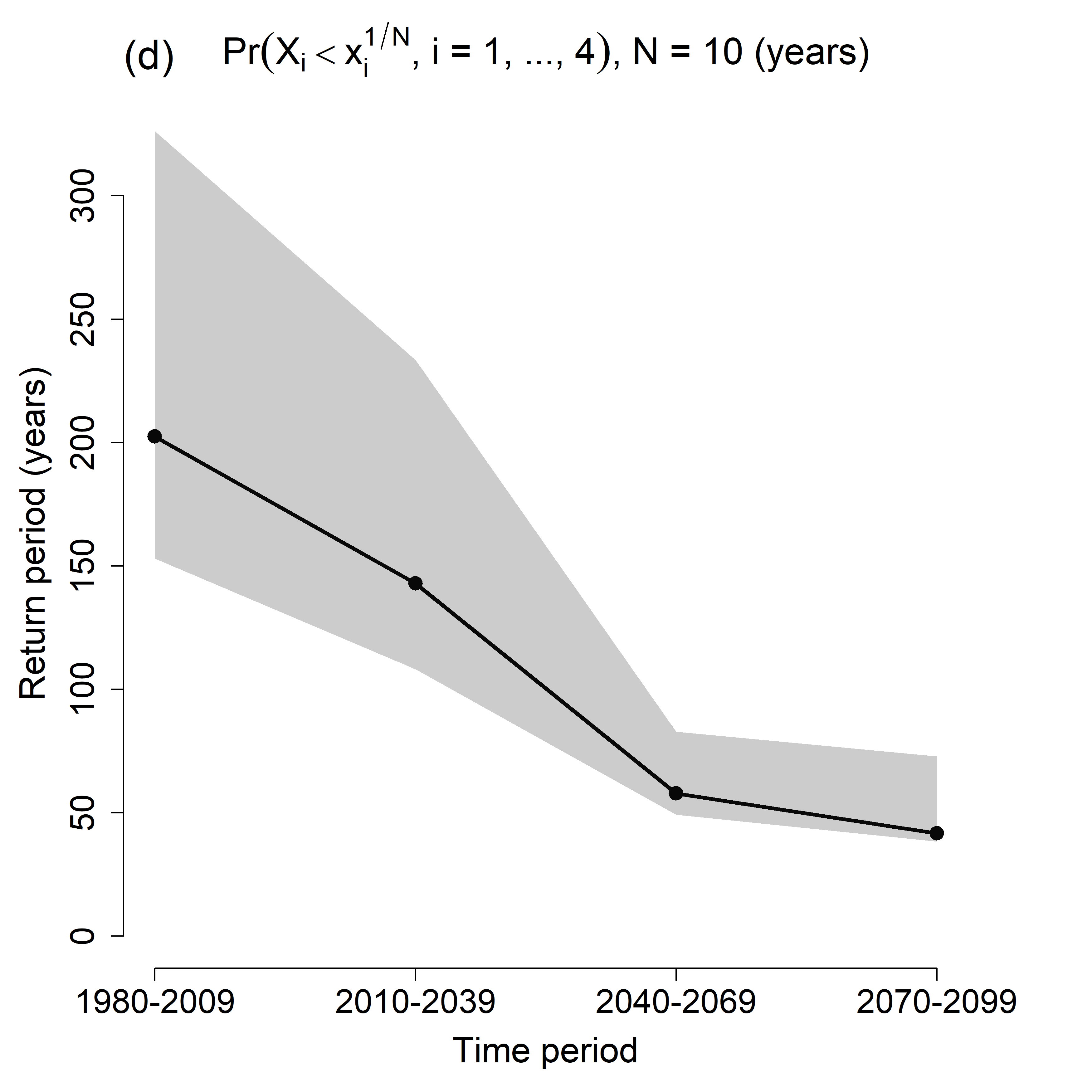}
    \end{minipage}%
    \caption{Estimated tail probability return periods over each time period with 95\% confidence intervals in grey. Panels (a) and (b) correspond to upper and lower tail probabilities of the sum $S$, respectively. Panels (c) and (d) illustrate probabilities of all sites experiencing simultaneously high or low weekly maxima values, respectively.}
    \label{fig:probs_cis}
\end{figure}

Estimated return periods for the different considered tail probabilities show strong temporal trends (Figure~\ref{fig:probs_cis}). In particular, the probability of extremes in both the upper and lower tails of $S$ appears to increase considerably over time, with the return periods in the latter time intervals often orders of magnitude smaller than the corresponding return periods for the earlier time intervals (Figures~\ref{fig:probs_cis}a-b). This is especially the case for the lower tail of $S$, where the best estimate for the return periods decreases from $\approx 718$ years to $\approx 1.6$ years.  Moreover, a similar trend appears to be present when one considers the joint lower tail of the distribution function, that is, the probability of all locations exhibiting extremely low discharge simultaneously is increasing substantially over time (Figure~\ref{fig:probs_cis}d). However, this trend does not appear present when one considers the joint upper tail of the survivor function, for which there is very little difference between the estimated return periods (Figure~\ref{fig:probs_cis}c). While the marginal return level magnitudes are increasing over time, the probability of all sites being extremely large simultaneously does not appear to change much over the considered simulation period, 1980--2099. 
These findings illustrate the nuanced and complex nature in which the tail behaviour of Alpine discharges can evolve over time. In particular, they highlight that changes in the extremal dependence structure over time are highly relevant for risk assessments and decision making. Furthermore, the advantages from modelling joint tail behaviour in a highly general and flexible manner, as is enabled via the SPAR method, are immediately apparent.  

As an additional check, we test the robustness of the tail probability estimates against smaller sample sizes. These results are given in the Supporting Information. In general, the tail probabilities from the full sample size are well captured within the confidence intervals obtained using the lower sample sizes. Naturally, the uncertainty associated with the estimates is much higher as the sample size is decreased. We note that there are some cases with a seemingly clear difference in the tail probability estimates, which is related to both the sensitivity of such statistics to the fitted GPD model alongside the fact that subsamples do not always represent the stochastic variability across all ensemble members.

\section{Discussion} \label{sec:discuss}

Our application of the deep SPAR framework to simulated discharge from the Alpine Foreland of the Upper Danube shows that contemporary statistical deep learning techniques can provide reliable and interpretable inference for both marginal and joint discharge extremes. The SPAR model accurately reproduces marginal tail behaviours and pairwise dependence relationships while simultaneously respecting physical properties. Representing the GPD threshold, scale, and shape functions via MLPs yields a parsimonious and flexible non-parametric approximation that avoids restrictive parametric assumptions and permits inference on marginal and joint tail regions on the data scale, reducing propagated marginal modelling uncertainty that occurs with most multivariate extreme value techniques. Computational cost for our framework is modest: fits for a single time window complete within an hour on typical desktop hardware, making applications of this framework feasible for practitioners. Our work paves the way for further applications of the deep SPAR framework in modelling compound extremes. 

The hydrological interpretation of the fitted SPAR models highlights notable changes in joint extreme behaviour over time. Tail probability estimates indicate substantially increased probabilities of certain concurrent extremes in later time windows, implying greater likelihood of compound flood and drought-like conditions in the simulated future climate. The greater likelihood of compound floods might be driven by the increasing intensity \cite{poschlod2021internal} and increasing spatial extent \cite{matte2022spatial} of heavy rainfall events in a warmer climate over Central Europe. This is somewhat consistent with \citeA{Fang2025}, who find that very large floods are projected to increase across Europe with climate change and pose challenges to hydrological risk management. The estimated probabilities that all four catchments exhibit drought conditions simultaneously also strongly increase under climate change (Figure~\ref{fig:probs_cis}d). \citeA{poschlod2026} find, for the Ammer river, that the river drought type shifts from dominating winter low-flow until the 2010s, to dominating summer low-flow in the future. Winter low-flows in the Alpine Foreland are governed by snow dynamics, which depend on catchment elevation profiles and thus exhibit a limited degree of synchrony across basins. Summer low-flows, by contrast, are primarily driven by large-scale hot-and-dry conditions, where the dependence between heat and drought is projected to increase in the driving climate of the CRCM5-LE \cite{bohnisch2025, felsche2024}. \citeA{poschlod2026} further project a strong increase in summer low-flow event durations. These processes underpin the observed increases in probabilities associated with spatially coherent river droughts. At the same time, the estimated probabilities of all sites being simultaneously high exhibit little systematic change over time, despite increasing marginal return levels; this suggests a complex evolving dependence structure rather than uniform magnification of synchronous extremes. Such nuanced behaviour underscores the value of the SPAR model; one can assess a wide variety of extreme scenarios, including where only subsets of catchments are extreme. Such cases are of high practical importance for emergency response and infrastructure resilience.

While deviations between the SPAR and the hydro-SMILE are to a limited extent, we observe some discrepancies for extreme lower tail quantiles. As noted in Section~\ref{subsec:data_set}, the de-clustered time series in such cases still exhibit some residual temporal dependence due to long intervals of consecutive low discharge \cite{poschlod2026}. This invalidates our modelling assumptions and helps to explain why the GPD fits in such cases exhibit slight deviations. Another possible reason for this observation is the GPD loss surface; as noted in the Supporting Information, this loss surface is highly irregular and can be difficult to optimise, especially for bounded variables (i.e., the lower tails of discharge data).

While the weekly maxima series (Section~\ref{subsec:data_set}) is best suited for characterising compound flooding events, our analysis also provides insight into the evolving behaviour of concurrent drought events which are broadly consistent with other findings in the literature. Our work demonstrates that the SPAR model provides a highly accurate and flexible framework for characterising compound extremes, capturing marginal and joint behaviour in both upper and lower tails; therefore, this framework could also be used to model other datasets that better represent drought events (e.g., monthly or seasonal minima series). 

Reduced sample sizes (subsamples of the ensemble) inflate uncertainty as expected, but the SPAR model diagnostics remain broadly comparable with results obtained for the full sample, indicating practical utility for the proposed framework even in lower data settings. Nevertheless, all of our results are conditional on the hydrological and climate models used for forcing, and structural or forcing biases in those simulations will propagate. Focusing on subsets of ensembles can significantly alter tail behaviour, implying model fits for smaller subsamples may not represent the full stochastic behaviour across all ensembles. Expert judgement is therefore advisable when using, for instance, extrapolated probabilities for high‑stakes decision making. Future work could further validate our modelling approach on observational data, and compare those model fits with the simulated discharge. 

Finally, we note that while the deep SPAR framework works well here in a four-dimensional setting, higher dimensional applications of the model will require more complex formulations and model fitting procedures. This is especially true for the angular density, for which empirical estimation is not viable in high dimensional settings \cite{Wessel2025}. This represents an open area of research, and recent work by \citeA{Kakampakou2025} and \citeA{Papastathopoulos2026} provide the first attempts at extending the angular-radial modelling paradigm to high-dimensional spatial and graphical settings, respectively.

\section{Conclusions} \label{sec:conclusion}

The deep SPAR framework offers a flexible, accurate, and computationally feasible approach to joint extreme inference on river networks. Applied to simulated discharge in the Alpine Foreland from a 50-member hydro-SMILE, the method reproduces marginal and multivariate tail behaviour and reveals meaningful temporal changes: aggregate extreme discharge (both high and low) becomes substantially more probable in later time windows, while concurrent floods change less, reflecting complex evolving dependence. This unified marginal and multivariate tail inference, together with quantifiable uncertainty and practical scalability, makes deep SPAR a promising tool for compound flood and drought risk assessment. Future work should prioritise validation against observations and multi‑model forcings, adaptation of the deep SPAR framework for higher dimensions, and systematic propagation of forcing and structural uncertainties into downstream impact assessments.

%
%

\section*{Open Research Section}
The simulated discharge data series used for modelling in this study are available at \url{https://github.com/callumbarltrop/DeepSPAR-hydroSMILE/data}. Furthermore, code for fitting our modelling approach and reproducing figures from the article is available at \url{https://github.com/callumbarltrop/DeepSPAR-hydroSMILE}.

\section*{Conflict of Interest declaration}
The authors declare there are no conflicts of interest for this manuscript.

\acknowledgments
The authors acknowledge the financial support by the Federal Ministry of Research, Technology and Space of Germany and by Sächsische Staatsministerium für Wissenschaft, Kultur und Tourismus in the programme Center of Excellence for AI-research "Center for Scalable Data Analytics and Artificial Intelligence Dresden/Leipzig", project identification number: ScaDS.AI.  This work has made use of the resources provided by the Edinburgh Compute and Data Facility (\url{https://www.ecdf.ed.ac.uk/}).
The CRCM5-LE and WaSiM-LE were developed within the ClimEx Project, funded by the Bavarian Ministry for the Environment and Consumer Protection. The authors thank the Leibniz Supercomputing Centre (Bavarian Academy of Sciences and Humanities) for providing HPC infrastructure and computation time for the WaSiM-LE and CRCM5-LE. We further acknowledge Environment and Climate Change Canada for providing the CanESM2-LE driving data.

%
%

\appendix

\renewcommand{\thefigure}{\AlphAlph{\value{figure}}}
\setcounter{figure}{0} 
\renewcommand{\thetable}{\Alph{table}}
\setcounter{table}{0} 
\renewcommand{\thesubsection}{A\arabic{subsection}}
\renewcommand{\theequation}{A\arabic{equation}}

\section*{Appendix} \label{appen}

\subsection{Marginal pre-processing}\label{appen:marg_stand}

Let $\boldsymbol{X}^o:=(X_1^o,\dots,X_d^o)$ denote the hydro-SMILE discharge data described in Section~\ref{subsec:data_set} with $d=4$. As negative discharge values are infeasible, we have $X_i^o \geq 0$ for all $i \in \{1,\hdots,d\}$. The SPAR model requires that data be centred around the origin $\boldsymbol{0}$, which is clearly not the case for non-negative data. Owing to these features, pre-processing of the data is required prior to using the SPAR framework.  

The aim of pre-processing is to define an invertible map for each marginal variable that `centres' the data at $\boldsymbol{0}$ while preserving non-negativity. While fitting on the observed scale of data is possible, this would run the risk of generating physically-implausible negative observations from the fitted SPAR model. We stress here that this step is different from statistical modelling of the margins since our pre-processing procedure is deterministic and does not require one to specify a probabilistic model for the marginal tail behaviour. 

To begin, we set $X_i^* = \log( \exp(X_i^o/\nu_i) - 1),i=1,\dots,d,$ where $\nu_i>0$ denotes the standard deviation of $X_i^o$. This step serves three purposes. First, division by the standard deviation standardises the scale across margins, i.e., $\text{Var}(X_i^o/\nu_i) = 1$ for all $i=1,\dots,d$. Next, evaluating $\exp(X_i^o/\nu_i)$ rather than $\exp(X_i^o)$ prevents numerical overflow, which is a problem that could otherwise arise for observations with large magnitudes. Finally, the transformation $\phi(x) := \log( \exp(x) - 1)$ provides an invertible mapping from $\RR^+$ to $\RR$, which does not alter the marginal upper tail behaviour of $\mathbf{X}^o$,  as $\phi(x) \sim x$ as $x \to \infty$ \cite{Cooley2019a}. However, it does make the marginal lower tails heavier, as $\phi(x) \sim \log(x)$ as $x \to 0$. This latter feature is useful when the data have a natural lower bound (as in the case for our discharge data), as deep GPD models can be numerically difficult to estimate in this case \cite{Richards2022a}.

To centre the data at $\boldsymbol{0}$, we must specify a star-centre for $\boldsymbol{X}^*:=(X_1^*,\hdots,X_d^*)$ and subtract this from the data. This step is non-trivial, especially as the dimension $d$ increases, and selecting an appropriate star-centre is vital to ensure the accuracy of the SPAR approach. Following \citeA{Murphy-Barltrop2024} and \citeA{Mackay2025b}, we tested vectors of componentwise means and medians of $\boldsymbol{X}^*$, but found the resulting model fits to be unreliable, with few observations in portions of the hypersphere $\mathbb{S}^{d-1}$. For a robust model fit, we desire that mass be placed on all orthants of the hypersphere to ensure accurate extrapolation in all directions. 

We thus propose to use the \textit{geometric median}, or Fermat--Weber point (see, e.g., \citeA{Minsker2015}), as the star-centre for $\boldsymbol{X}^*$. Given observations $\boldsymbol{x}^*_1,\dots,\boldsymbol{x}^*_n \in \mathbb{R}^d$, the geometric median  $\boldsymbol{m}^*$ is defined as
\begin{equation*}
\boldsymbol{m}^* = \arg\min_{\boldsymbol{y} \in \mathbb{R}^d} \sum_{t=1}^n \|\boldsymbol{y} - \boldsymbol{x}^*_t\|,
\end{equation*}
which is the minimiser of the sum of pairwise Euclidean distances across all sample points. This ensures that $\boldsymbol{m}^*$ lies in the region of maximal overall attraction of the data and therefore tends to fall in the interior of the point cloud, thus providing adequate angular observations over the hypersphere.  A popular method for computing $\boldsymbol{m}^*$ is Weiszfeld’s iterative algorithm
\cite{Weiszfeld2009}, which guarantees convergence under mild conditions. Using this algorithm, we estimate the geometric median and use this as our star-centre, setting $\boldsymbol{X} = \boldsymbol{X}^* - \boldsymbol{m}^*$, i.e., centring observations around $\boldsymbol{0}$. We proceed to fit the deep SPAR model to $\boldsymbol{X}$. 

\subsection{Simulation and probability estimation via the SPAR model} \label{appen:sim_prob}

For estimating marginal tail and multivariate probabilities using the SPAR model, we require a scheme for simulating an observation $\boldsymbol{x}^s$ in the joint tail of $\boldsymbol{X}$, i.e., $ \boldsymbol{X}$ given $\|\boldsymbol{X}\| > u(\boldsymbol{X}/\|\boldsymbol{X}\|)$. This is achieved using the following procedure:
\begin{enumerate}
    \item Draw an angle $\boldsymbol{w}$ from the density $f_{\boldsymbol{W}}$ of the angular variable $\boldsymbol{W}$. In our case, where we use the empirical density, we randomly sample from the observed angles in the joint tail, i.e., $\boldsymbol{W} \mid R > u (\boldsymbol{W})$. 
    \item Draw an observation $y$ from a GPD with parameters $\sigma(\boldsymbol{w}),\xi(\boldsymbol{w})$, and add $u(\boldsymbol{w})$ to obtain a sample $r := y + u(\boldsymbol{w})$ from the tail of $R \mid \boldsymbol{W} = \boldsymbol{w}$. 
    \item Set $\boldsymbol{x}^s := r\boldsymbol{w}$. 
\end{enumerate}
The last step transforms the sample from the angular-radial scale back to Cartesian coordinates. This results in a sample from the joint tail region $\mathbb{Q}^c_{u}$. We refer to \citeA{Murphy-Barltrop2024} for further details about this sampling scheme. 

Suppose now we have a probability region $\mathcal{A}$ that is of practical interest (e.g., a failure region) for which we wish to evaluate the probability $\Pr(\boldsymbol{X} \in \mathcal{A})$. Applying the law of total probability, we have
\begin{equation*}
    \Pr(\boldsymbol{X} \in \mathcal{A}) = \Pr(\boldsymbol{X} \in \mathcal{A} \mid \boldsymbol{X} \in \mathbb{Q}_{u})\Pr(\boldsymbol{X} \in \mathbb{Q}_{u}) + \Pr(\boldsymbol{X} \in \mathcal{A} \mid \boldsymbol{X} \in \mathbb{Q}^c_{u})\Pr(\boldsymbol{X} \in \mathbb{Q}^c_{u}).
\end{equation*}
Furthermore, one can show that ${\Pr(\boldsymbol{X} \in\mathbb{Q}_{u}) = 1 - \alpha} \Rightarrow {\Pr(\boldsymbol{X} \in \mathbb{Q}^c_{u}) = \alpha}$ \cite{Papastathopoulos2025}, hence 
\begin{equation}\label{eqn:prob_est}
    \Pr(\boldsymbol{X} \in \mathcal{A}) = \Pr(\boldsymbol{X} \in \mathcal{A} \mid \boldsymbol{X} \in \mathbb{Q}_{u})(1 - \alpha) + \Pr(\boldsymbol{X} \in \mathcal{A} \mid \boldsymbol{X} \in \mathbb{Q}^c_{u})\alpha.
\end{equation}
Both probabilities on the RHS of equation~\eqref{eqn:prob_est} can be approximated using Monte-Carlo techniques. For the first probability $\Pr(\boldsymbol{X} \in \mathcal{A} \mid \boldsymbol{X} \in \mathbb{Q}_{u})$, we require a sample from $\mathbb{Q}_{u}$. By construction, $\mathbb{Q}_{u}$ represents the joint body of $\boldsymbol{X}$ (see Figure~\ref{fig:SPAR_illustrative}c), and thus it is reasonable, given a sufficiently large sample size $n$, to empirically approximate probabilities in this region. Suppose $\{\boldsymbol{x}_t\}_{t=1}^n$ denotes a sample of size $n$ from $\boldsymbol{X}$, where $n$ is large. Setting $\mathcal{I} := \{t \in \{1,\hdots,n \} \mid \|\boldsymbol{x}_t \| \leq u(\boldsymbol{x}_t/\|\boldsymbol{x}_t \|) \}$, i.e., the indices of non-threshold exceeding observations, we have that $$\Pr(\boldsymbol{X} \in \mathcal{A} \mid \boldsymbol{X} \in \mathbb{Q}_{u}) \approx (1/| \mathcal{I}|) \sum_{t\in\mathcal{I}}\mathbbm{1}(\boldsymbol{x}_t\in \mathcal{A}),$$ where $|\mathcal{I}|$ denotes the cardinality of $\mathcal{I}$. 

For the latter probability $\Pr(\boldsymbol{X} \in \mathcal{A} \mid \boldsymbol{X} \in \mathbb{Q}^c_{u})$, sample $m$ realisations, denoted $\{\boldsymbol{x}^{*}_j\}_{j=1}^m$, from the fitted SPAR model. For large $m$, we have, under our modelling assumptions, $\Pr(\boldsymbol{X} \in \mathcal{A} \mid \boldsymbol{X} \in \mathbb{Q}^c_{u}) \approx (1/m)\sum_{j\leq m}\mathbbm{1}(\boldsymbol{x}^s_j \in \mathcal{A})$. Thus, the resulting estimator of $\Pr(\boldsymbol{X} \in \mathcal{A})$ is given by
\begin{equation*}
    \widehat{\Pr(\boldsymbol{X} \in \mathcal{A})} = \frac{1-\alpha}{| \mathcal{I}|} \sum_{t\in\mathcal{I}}\mathbbm{1}(\boldsymbol{x}_t \in \mathcal{A}) + \frac{\alpha}{m}\sum_{j\leq m}\mathbbm{1}(\boldsymbol{x}^s_j \in \mathcal{A}).
\end{equation*}

In many cases, we are likely to have $\mathcal{A} \cap \mathbb{Q}_{u} = \emptyset$, i.e., the tail region does not intersect with the joint body of the data. On such occasions, probability estimates are computed solely from the fitted SPAR model (i.e., no empirical sampling from $\mathbb{Q}_{u}$ is required). For example, setting $q_i^M = \max_{\boldsymbol{w}\in\hypsphere} w_iu(\boldsymbol{w})$ and $q_i^m$ $=$ $\min_{\boldsymbol{w}\in\hypsphere} w_iu(\boldsymbol{w})$ for each $i = 1,\hdots,d$, i.e., the coordinate-wise maxima and minima of the quantile set $\mathbb{Q}_{u}$, one can consider the regions given by $\mathcal{A}^h_i := \RR \times \cdots \times \RR \times  (h_i ,\infty) \times \RR \times \cdots \times \RR $ and $\mathcal{A}^l_i := \RR \times \cdots \times \RR \times  (-\infty,l_i) \times \RR \times \cdots \times \RR $ for any $h_i > q_i^M$ and $l_i < q_i^m$. By construction, we have that  $\mathcal{A}_i^h \cap \mathbb{Q}_{u} = \mathcal{A}_i^l \cap \mathbb{Q}_{u} = \emptyset$, implying $\Pr(\boldsymbol{X} \in \mathcal{A}_i^h) = \Pr(\boldsymbol{X} \in \mathcal{A}_i^h \mid \boldsymbol{X} \in \mathbb{Q}^c_{u})\alpha$ and $\Pr(\boldsymbol{X} \in \mathcal{A}_i^l) = \Pr(\boldsymbol{X} \in \mathcal{A}_i^l \mid \boldsymbol{X} \in \mathbb{Q}^c_{u})\alpha$. Moreover, we have $\Pr(\boldsymbol{X} \in \mathcal{A}_i^h) = \Pr(X_i > h_i)$ and $\Pr(\boldsymbol{X} \in \mathcal{A}_i^l) = \Pr(X_i < l_i)$, thus demonstrating how the SPAR framework can be used to construct a model for marginal tail probabilities. Three examples of failure sets $\mathcal{A}$ satisfying $\mathcal{A} \cap \mathbb{Q}_{u} = \emptyset$ are illustrated in Figure~\ref{fig:SPAR_illustrative}d.

\begin{figure}[h]
   \centering
   \includegraphics[width=\linewidth]{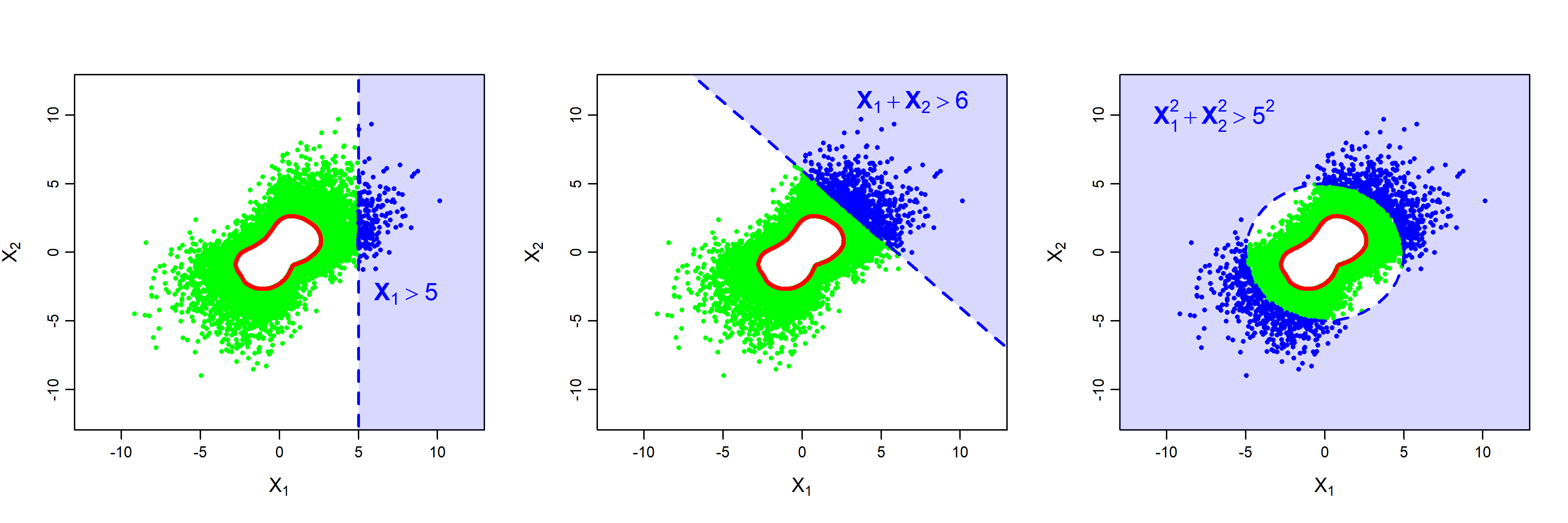}
   \caption{Example event sets in $\RR^2$ that one could generate from the SPAR model. See also Figure~\ref{fig:SPAR_illustrative}. }
   \label{fig:SPAR_event_sets}
\end{figure}

We remark that unlike some competing approaches to modelling multivariate extremes (e.g., \citeA{Heffernan2004,Wadsworth2013}), we can define the region $\mathcal{A}$ in full generality, thus increasing the scope of possible applications. We also note that the ability to simulate from the model is not just relevant from a probability estimation perspective; many environmental applications require large stochastic extreme events sets for performing risk assessments \cite{Keef2013a,Gouldby2017,Quinn2019,Jane2020}. Unlike competing approaches, our proposed framework can generate very general events sets encompassing both upper and lower tail behaviour, and does not require specification of a conditioning site. To demonstrate this, Figure~\ref{fig:SPAR_event_sets} illustrates a range of example event sets that can be generated from a fitted SPAR model. Such sets are applicable across a broad range of applications.

\clearpage

\bibliography{LatexFiles/library}

\clearpage 

\renewcommand{\thefigure}{S\arabic{figure}}
\renewcommand{\thesection}{S\arabic{section}}
\setcounter{figure}{0} 
\renewcommand{\thesubsection}{S\arabic{subsection}}
\renewcommand{\theequation}{S\arabic{equation}}

\section*{Supplementary Information} \label{SI}

This Supplementary Information contains additional supporting figures and details about the methods used for neural network parameter estimation. Section~\ref{appen:nn_fitting} describes the optimisation techniques used to estimate both the threshold and GPD parameter functions, the choices of optimisation tuning parameters, and the numerical problems that can arise when optimising the log-likelihood associated with the GPD. Section~\ref{appen:add_diagnostics} illustrates the remaining diagnostic plots for the full sample size model fits (i.e., those not in the main article). Section~\ref{appen:diagnostics_smaller_ss} illustrates the remaining diagnostic plots for smaller subsample sizes, and the comparisons of return level estimates across sample sizes. Finally, Section~\ref{appen:probs_smaller_ss} illustrates tail probability estimates across sample sizes and time windows. 

\section{Optimisation and parameter estimation for neural networks} \label{appen:nn_fitting}

Let $\mathcal{W}_u$ and $\mathcal{W}_{(\sigma,\xi)}$ denote the parameter sets for the MLPs associated with the threshold function $u(\boldsymbol{w})$ and GPD parameter functions $(\sigma(\boldsymbol{w}),\xi(\boldsymbol{w}))$, respectively. These parameter sets are estimated through empirical minimisation of appropriately chosen loss functions. 

Let $\{(r_t,\boldsymbol{w}_t); t=1,...,n\}$ denote radial and angular observations, with sample size $n$. By definition, $u(\boldsymbol{w})$ is the quantile of $R|(\boldsymbol{W}=\boldsymbol{w})$ at exceedance probability $\alpha$ close to zero. Thus, we can estimate the parameter set $\mathcal{W}_u$ associated with $u(\boldsymbol{w})$ using quantile regression techniques \cite{Koenker2017}. In this case, the appropriate loss function for $u(\boldsymbol{w})$ is the tilted loss function
\begin{equation}
\label{eqn:titled_loss}
    \mathcal{L}_u(\mathcal{W}_u) := \sum_{t=1}^n \rho_{1-\alpha}\left\{r_t-u\left(\boldsymbol{w}_t\right)\right\},
\end{equation}
where $\rho_{1-\alpha}(t):=t(1-\alpha-\mathbbm{1}\{t<0\})$ and $\mathbbm{1}$ is the indicator function. For ease of notation, dependency of $u(\boldsymbol{w})$ on $\mathcal{W}_u$ has been suppressed. 

Given an estimate $\hat{u}(\boldsymbol{w})$ of $u(\boldsymbol{w})$, we let
$I_u := \{t\in\{1,...,n\} : r_t > \hat{u} (\boldsymbol{w}_t) \}$, i.e., the set of indices for which the radial observation exceeds the estimated threshold function. To estimate the MLP corresponding to the GPD parameter functions, we minimise the negative log-likelihood function associated with the GPD. This corresponds to a `conditional density estimation network', for which a wide range of literature is available -- see the review by \citeA{Rothfuss2019}. In this case, the loss function is given by 
\begin{equation} \label{eqn:gp_ll}
   \mathcal{L}_{\rm GP}(\mathcal{W}_{(\sigma,\xi)}):= - \sum_{t \in I_u} \log \left[ h\left(r_t - \hat{u}(\boldsymbol{w}_t); \sigma(\boldsymbol{w}_t), \xi(\boldsymbol{w}_t)\right) \right].
\end{equation}

Hard-parameter sharing is employed, whereby all hidden layers are shared between the parameter functions. This reduces the risk of over-fitting while maintaining task-specific heads for each parameter \cite{Ruder2017}. 

Both loss functions, equations~\eqref{eqn:titled_loss} and \eqref{eqn:gp_ll}, are optimised via stochastic gradient descent using the Adam optimisation algorithm \cite{Kingma2017}, which is widely adopted, well-studied, and has demonstrated strong empirical performance and convergence in non-convex optimization problems. Further, to mitigate overfitting, data are split into training (90\%) and validation (10\%) sets, with the validation set used to monitor out-of-sample performance and assess convergence of Adam during training.

Adam requires specification of a \textit{learning rate} which controls the step size of the parameter updates during optimisation and balances stable convergence against computational efficiency. In addition, training is performed over multiple epochs, where each epoch corresponds to a full pass through the training data, using mini-batches of fixed size to compute stochastic gradient updates. Selection of these tuning parameters is non-trivial and typically involves some degree of trial and error. Training is initiated with a learning rate of $10^{-3}$, which is a standard choice in the deep learning literature, and early stopping with a patience of five epochs is employed to identify convergence of the optimisation algorithm. The learning rate is then progressively reduced in small increments down to $5 \times 10^{-5}$ to allow for potential further reductions in the empirical loss. Regarding the number of epochs and batch size, we use $500$ and $1024$ for the threshold MLP $u(\cdot)$, respectively, and $750$ and $| \mathcal{I}_u |$ for the GPD parameter MLP; such values appeared sufficient for ensuring convergence of the Adam algorithm in our application.  

Equation~\eqref{eqn:gp_ll} is sensitive to the value of the shape parameter, as the GPD support is dependent on this parameter: for any $\boldsymbol{w}$ satisfying $\xi(\boldsymbol{w}) <  0$, the tail of $R|(\boldsymbol{W}=\boldsymbol{w})$ will have the finite upper endpoint $u(\boldsymbol{w})-\sigma(\boldsymbol{w})/\xi(\boldsymbol{w})$. This results in a highly irregular loss surface, and training schemes that permit negative shape parameters can therefore be computationally troublesome \cite{Richards2024}. To circumvent this issue, we follow \citeA{Mackay2025b} and use a bespoke training scheme which requires a judicious choice of initial values and an iterative decay of the learning rate. We initialise the MLP for $(\sigma(\boldsymbol{w}),\xi(\boldsymbol{w}))$ to ensure that the shape parameter is non-negative for all angles. This ensures at the outset of the training procedure that the loss function is guaranteed to be finite for all threshold exceeding observations. If the gradient descent algorithm becomes unstable (i.e., we observe an infinite value for the loss in equation~\eqref{eqn:gp_ll}), we restart training at the previous finite loss value and reduce the learning rate. For our application, this allowed for negative values of the shape parameter function on certain parts of the hypersphere without optimisation issues arising. We stress here that this issue occurs for all GPD regression-based procedures, and is not a problem specific to our approach.

\section{Additional diagnostic plots} \label{appen:add_diagnostics}

\begin{figure}[ht] 
    \centering
    \begin{minipage}{0.32\textwidth} 
        \centering
        \includegraphics[width=\textwidth]{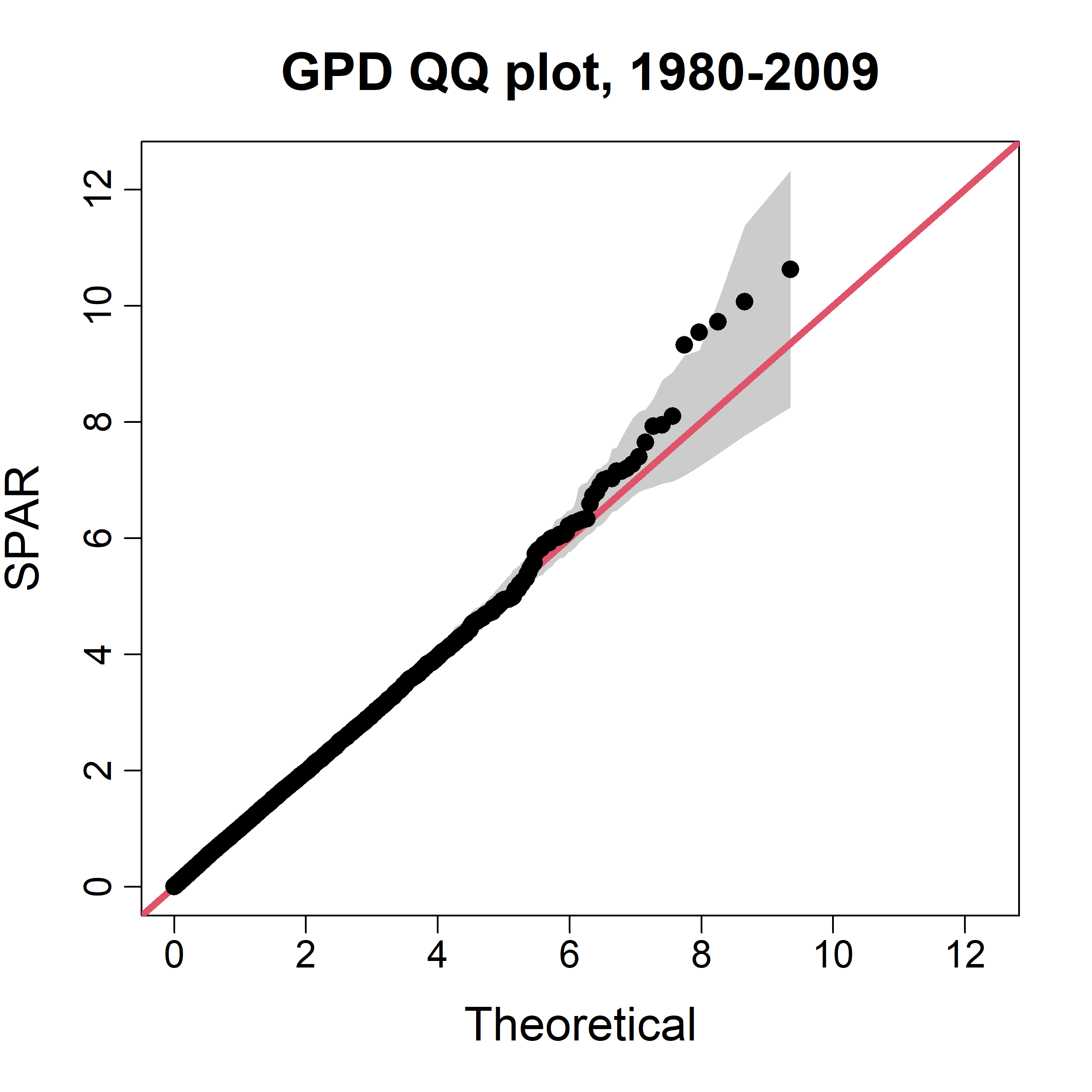}
    \end{minipage}
    \hfill 
    \begin{minipage}{0.32\textwidth} 
        \centering
        \includegraphics[width=\textwidth]{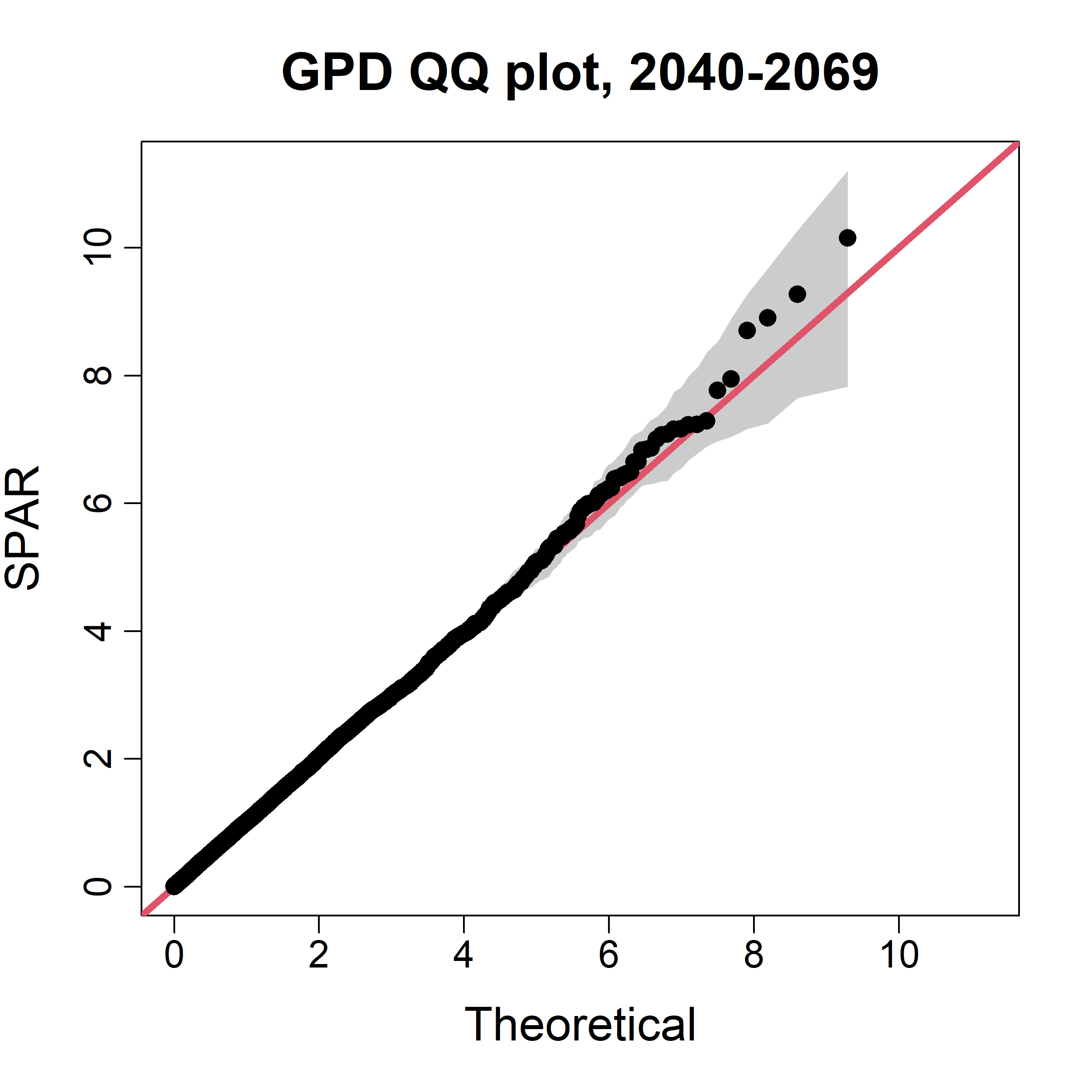}
    \end{minipage}
    \hfill
    \begin{minipage}{0.32\textwidth} 
        \centering
        \includegraphics[width=\textwidth]{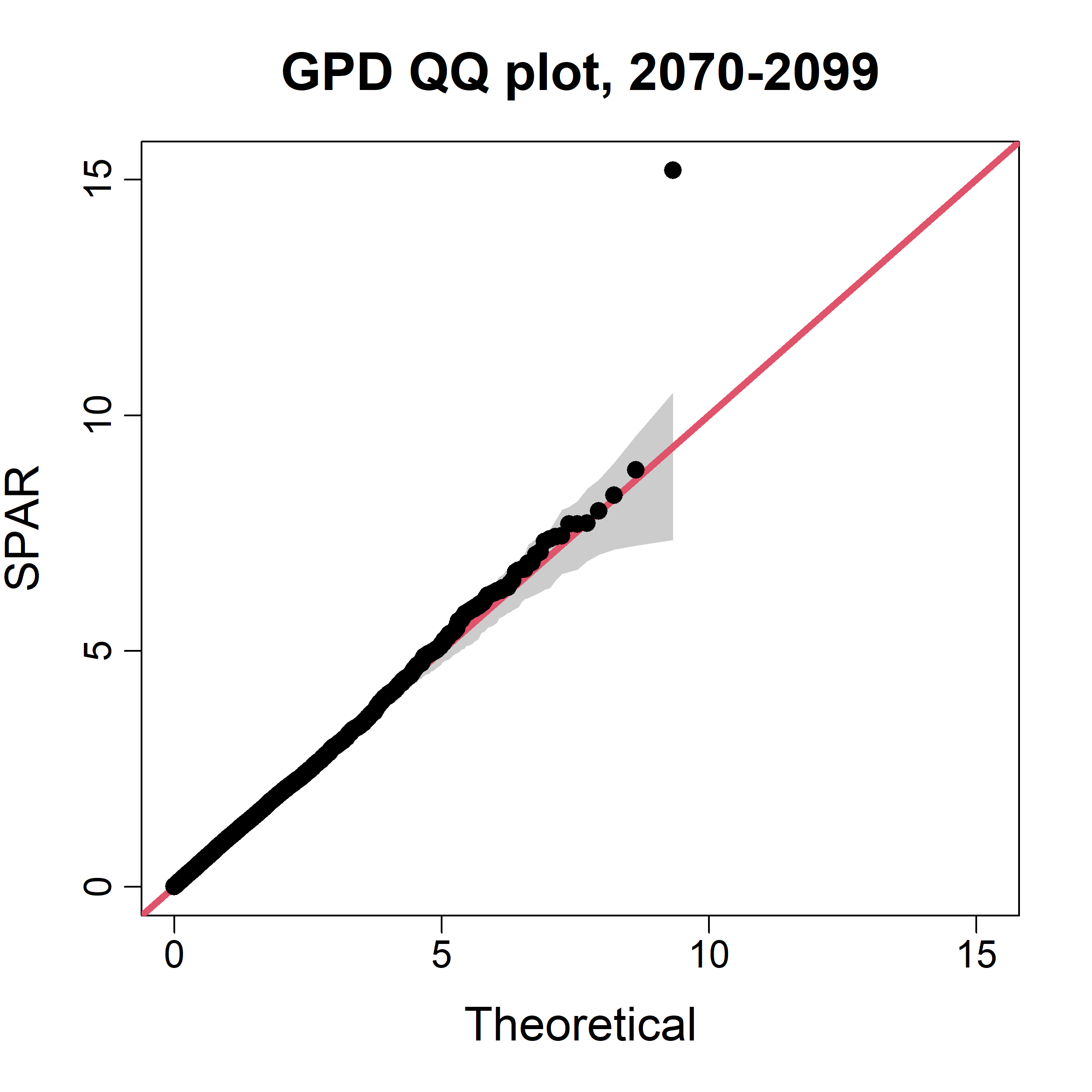}
    \end{minipage}

    \caption{GPD QQ plots for sequential time windows, ordered chronologically from left to right. Grey shading represents 95\% confidence intervals computed via bootstrapping.}
    \label{fig:gpd_qq_plots}
\end{figure}

\begin{figure}[h] 
    \centering
    \includegraphics[width=\textwidth]{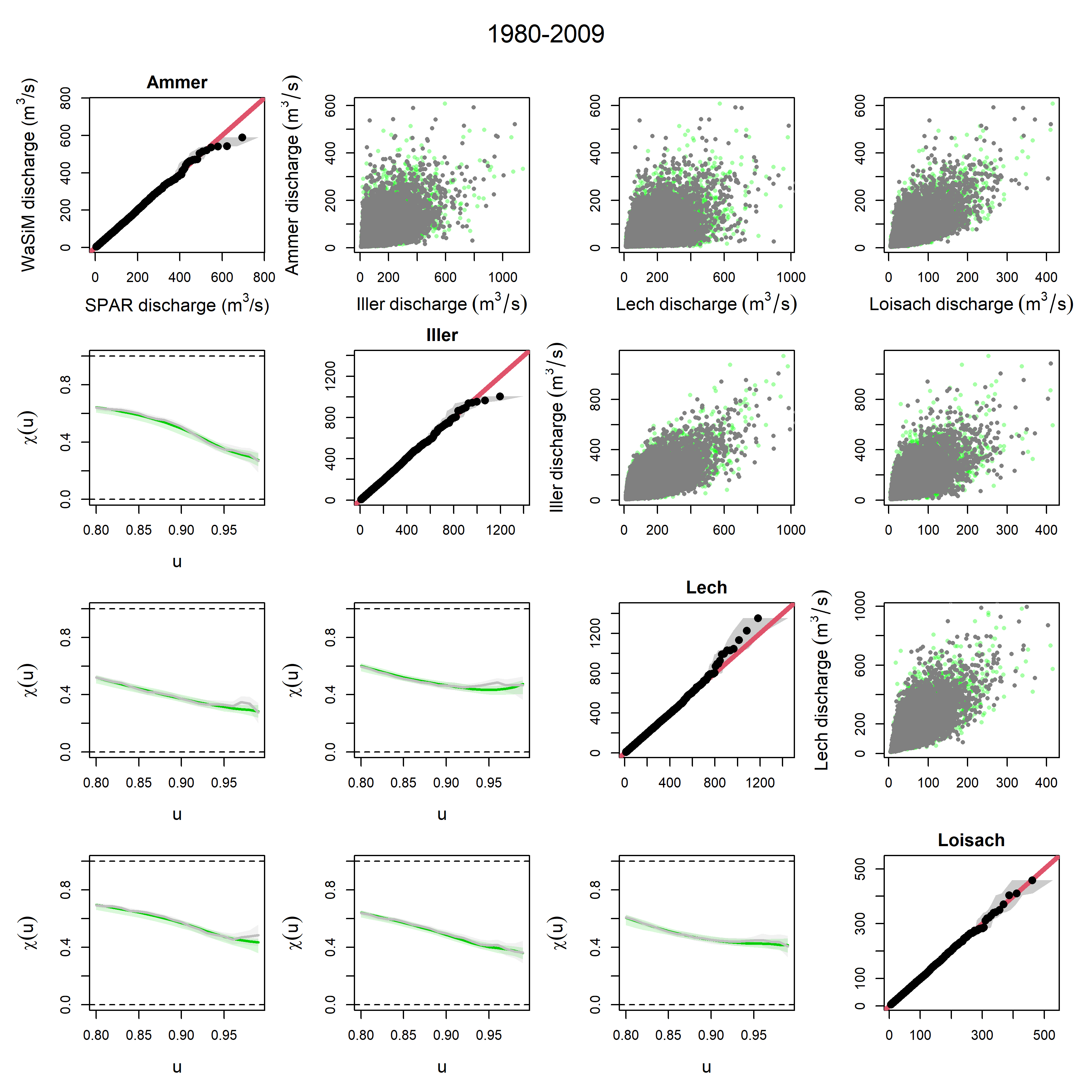}
    \caption{Model performance in the joint tail region $\mathbb{Q}^c_{u}$ for the 1980--2009 time period.  See Figure~5 in the main manuscript for details on the interpretation of individual panels.
    }
    \label{fig:scatter_1}
\end{figure}

\begin{figure}[h] 
    \centering
    \includegraphics[width=\textwidth]{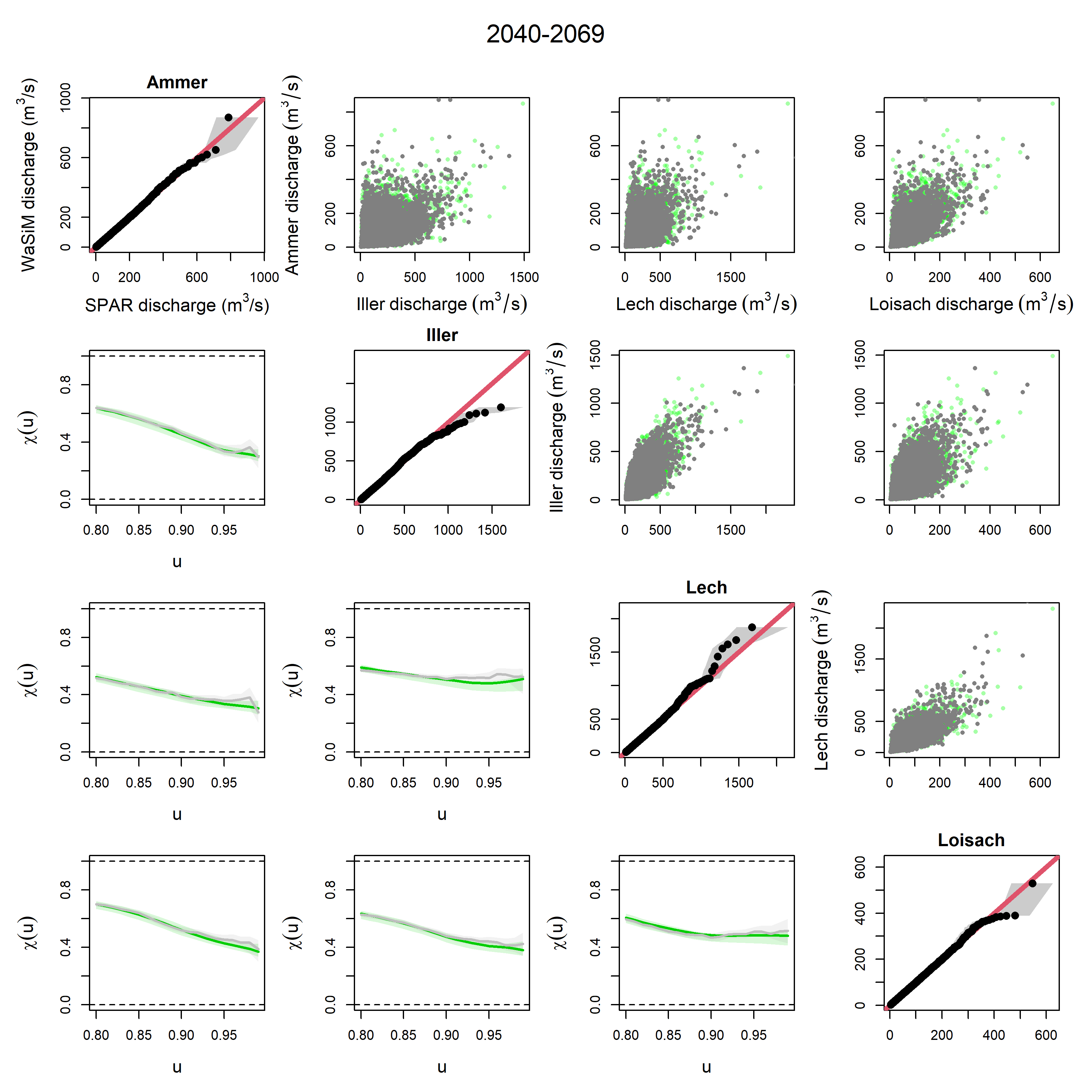}
    \caption{Model performance in the joint tail region $\mathbb{Q}^c_{u}$ for the 2040--2069 time period. See Figure~5 in the main manuscript for details on the interpretation of individual panels.
    }
    \label{fig:scatter_3}
\end{figure}

\begin{figure}[h] 
    \centering
    \includegraphics[width=\textwidth]{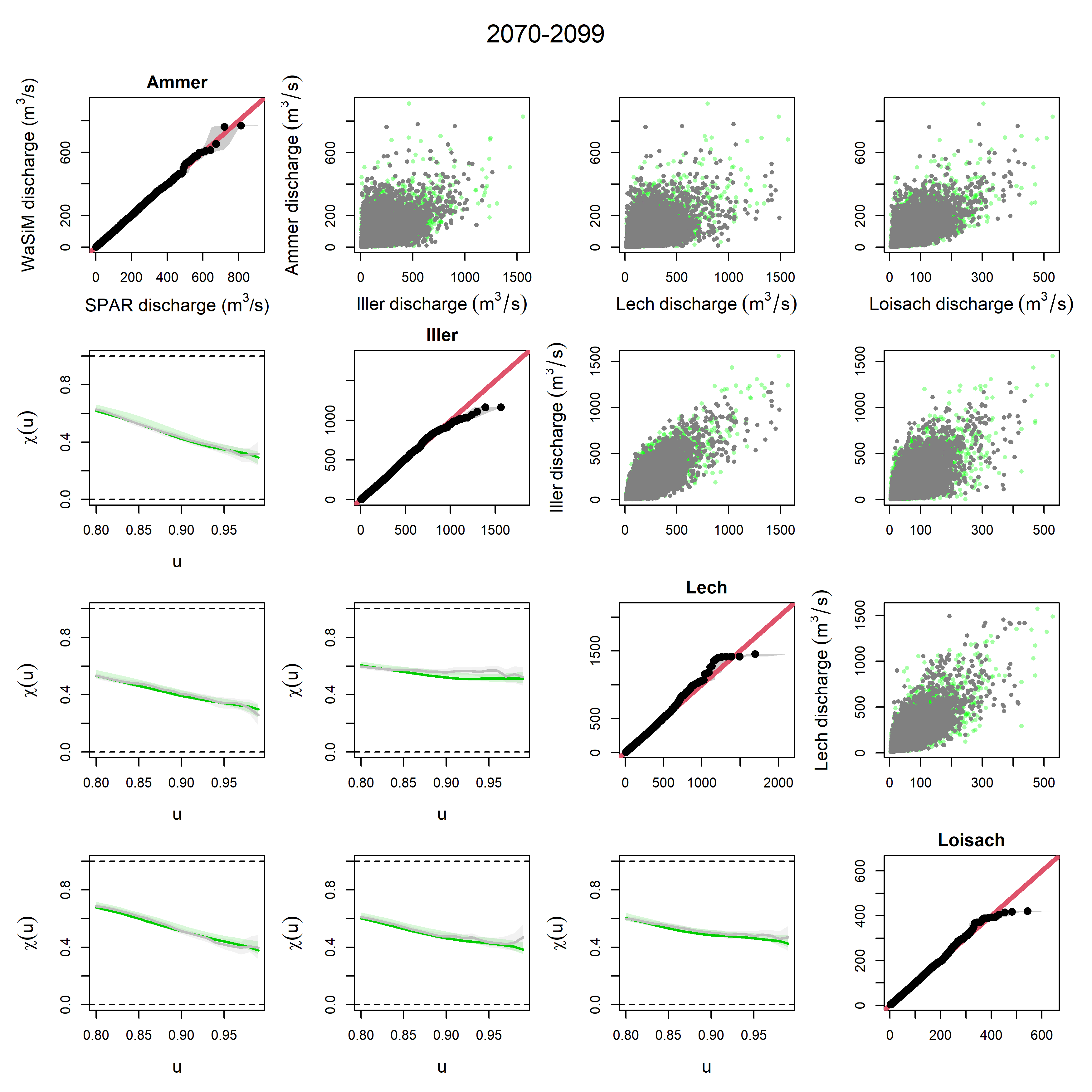}
    \caption{Model performance in the joint tail region $\mathbb{Q}^c_{u}$ for the 2070--2099 time period. See Figure~5 in the main manuscript for details on the interpretation of individual panels.
    }
    \label{fig:scatter_4}
\end{figure}

\begin{figure}[h] 
    \centering
    \begin{minipage}{0.48\textwidth} 
        \centering
        \includegraphics[width=\textwidth]{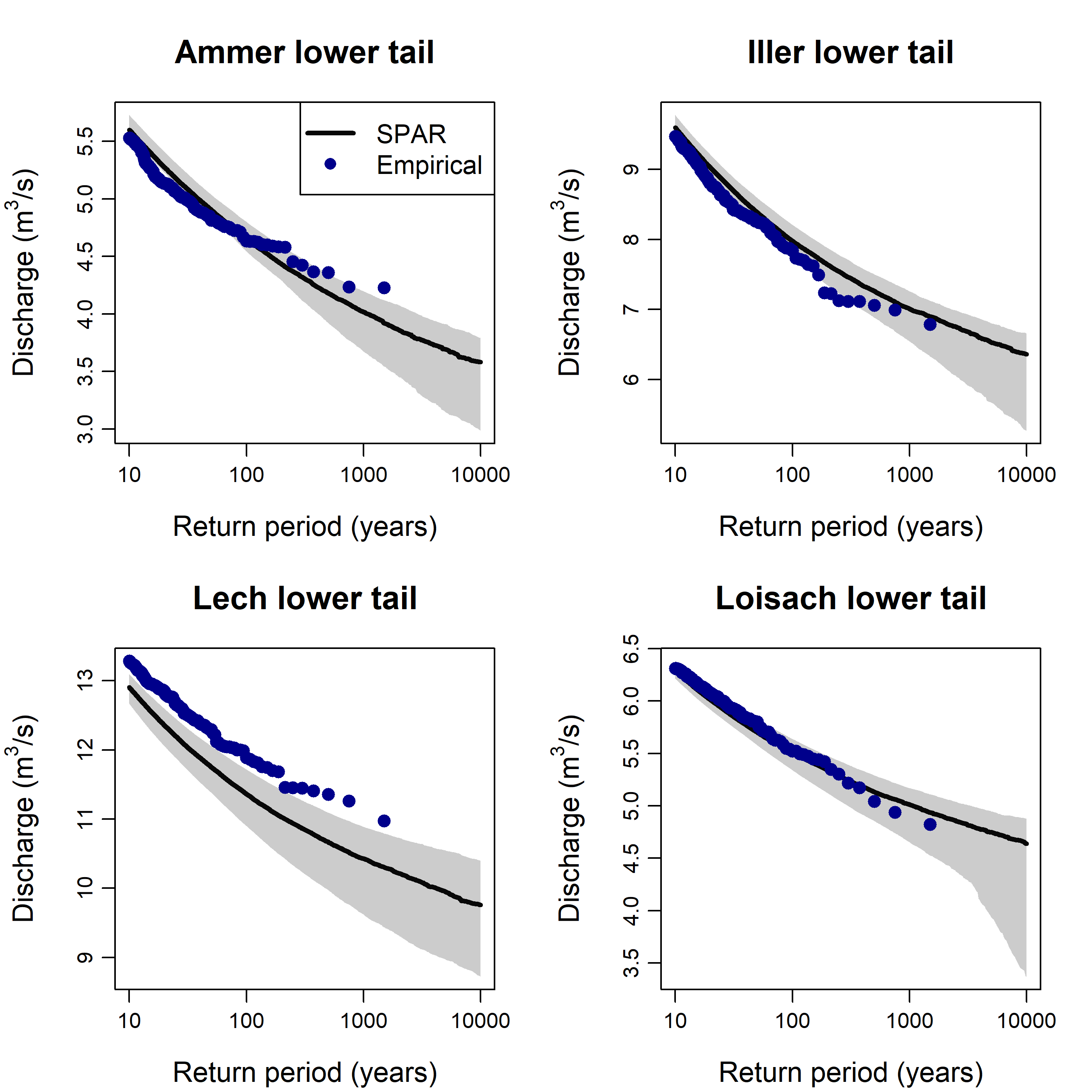}
    \end{minipage}%
    \hfill
   \begin{minipage}{0.48\textwidth} 
        \centering
        \includegraphics[width=\textwidth]{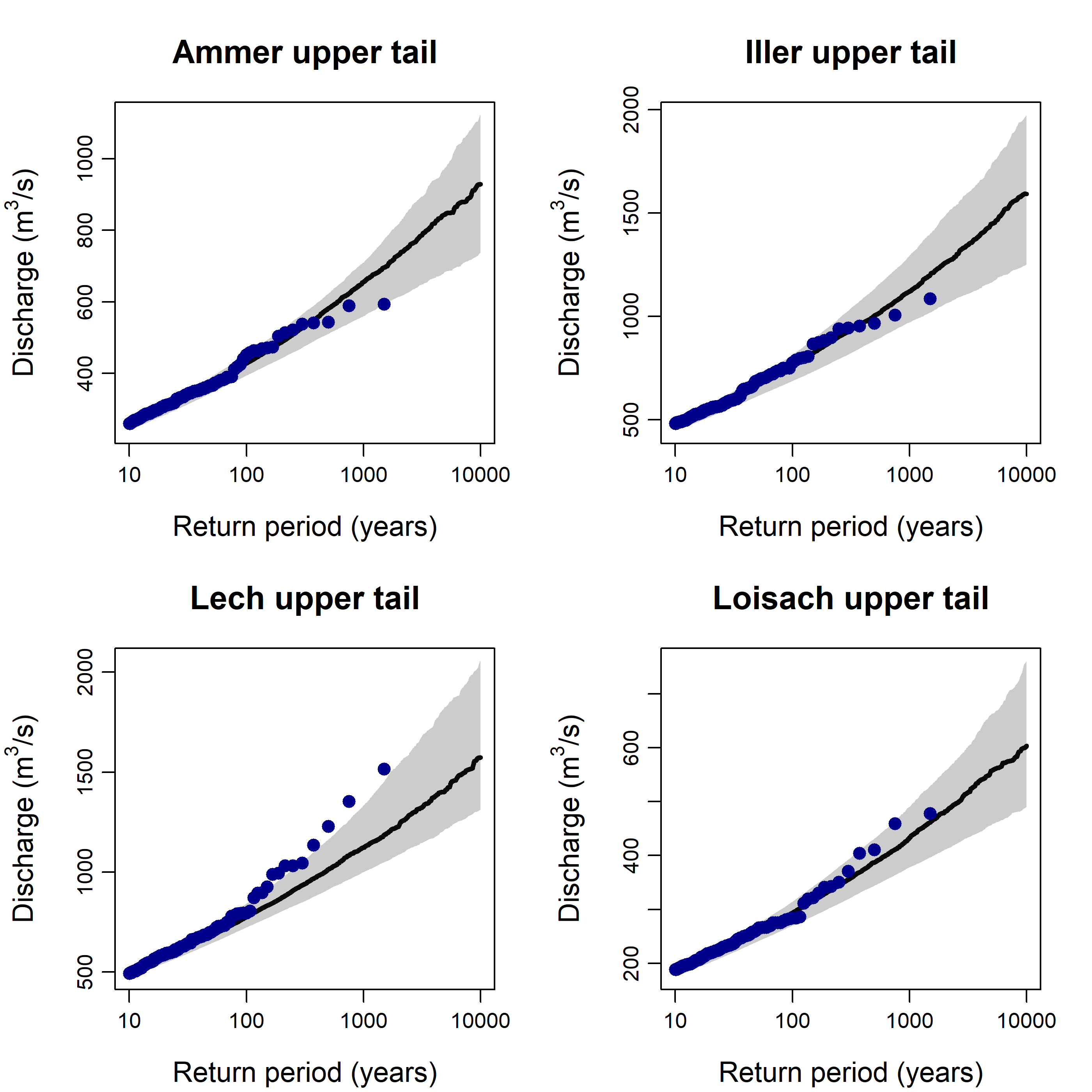}
    \end{minipage}%
    \caption{Return level plots in the lower and upper tails of each marginal variable for the 1980--2009 time period. See Figure~6 in the main manuscript for details on the interpretation of individual panels.  }
    \label{fig:ret_level_1}
\end{figure}

\begin{figure}[h] 
    \centering
    \begin{minipage}{0.48\textwidth} 
        \centering
        \includegraphics[width=\textwidth]{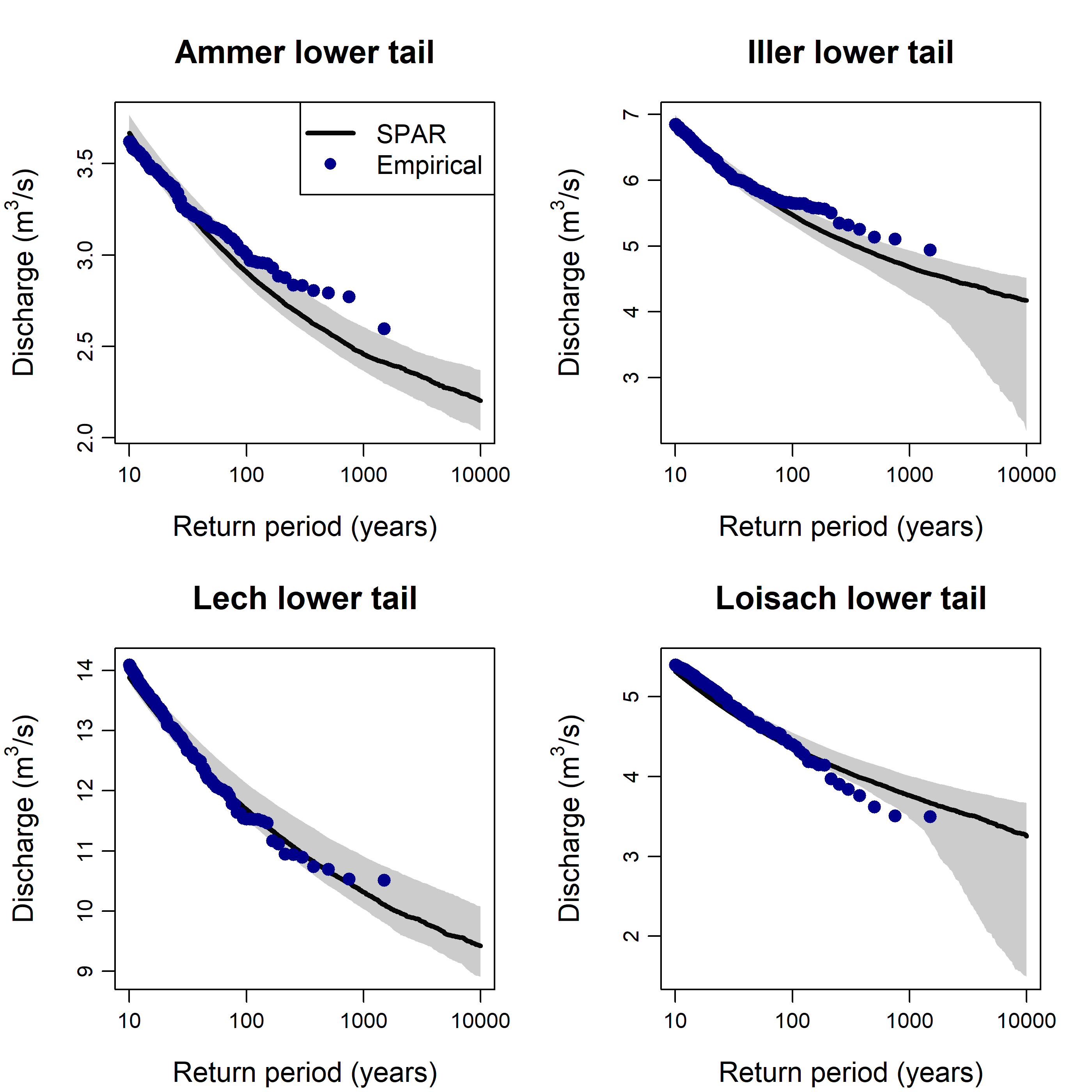}
    \end{minipage}%
    \hfill
   \begin{minipage}{0.48\textwidth} 
        \centering
        \includegraphics[width=\textwidth]{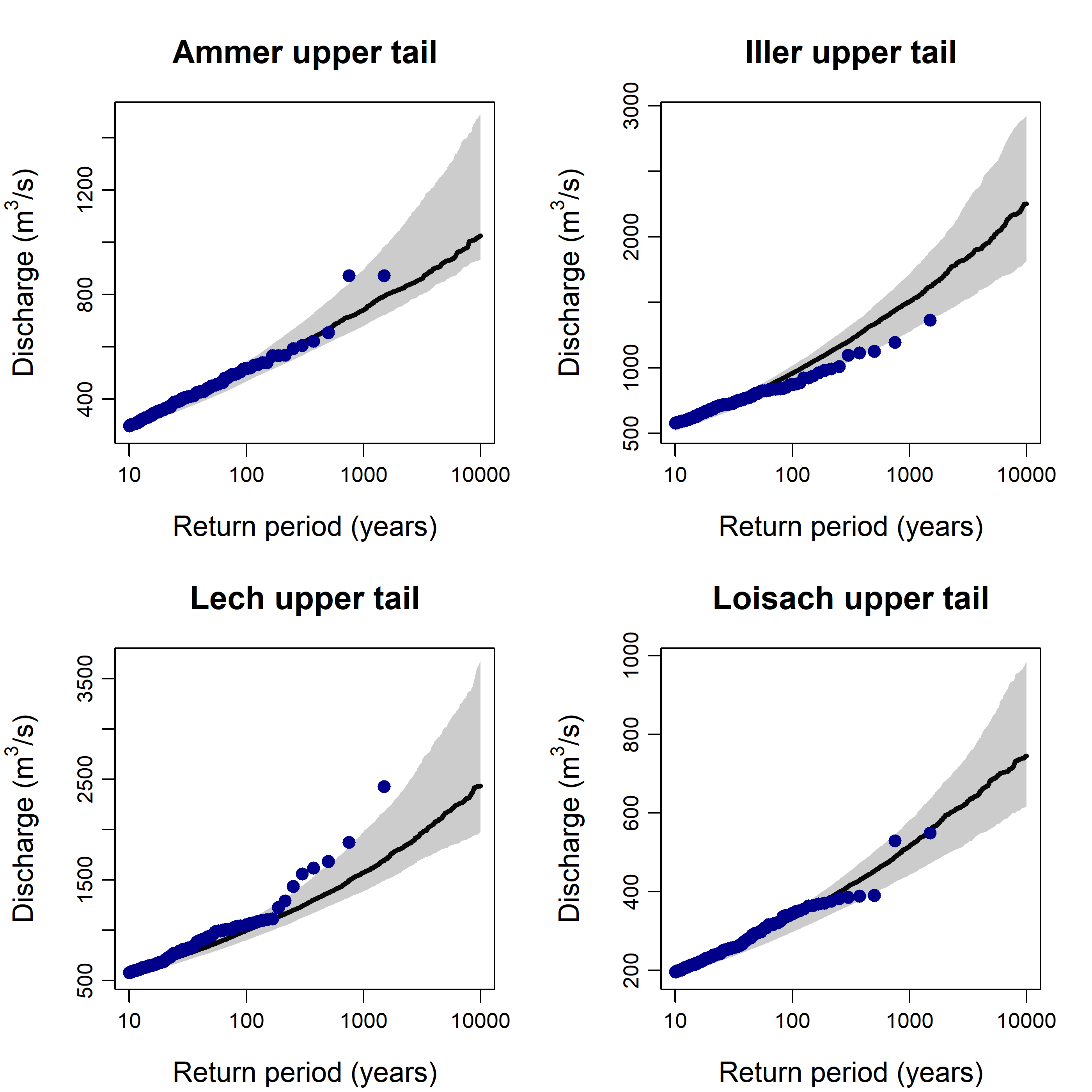}
    \end{minipage}%
    \caption{Return level plots in the lower and upper tails of each marginal variable for the 2040--2069 time period. See Figure~6 in the main manuscript for details on the interpretation of individual panels. }
    \label{fig:ret_level_3}
\end{figure}

\begin{figure}[h] 
    \centering
    \begin{minipage}{0.48\textwidth} 
        \centering
        \includegraphics[width=\textwidth]{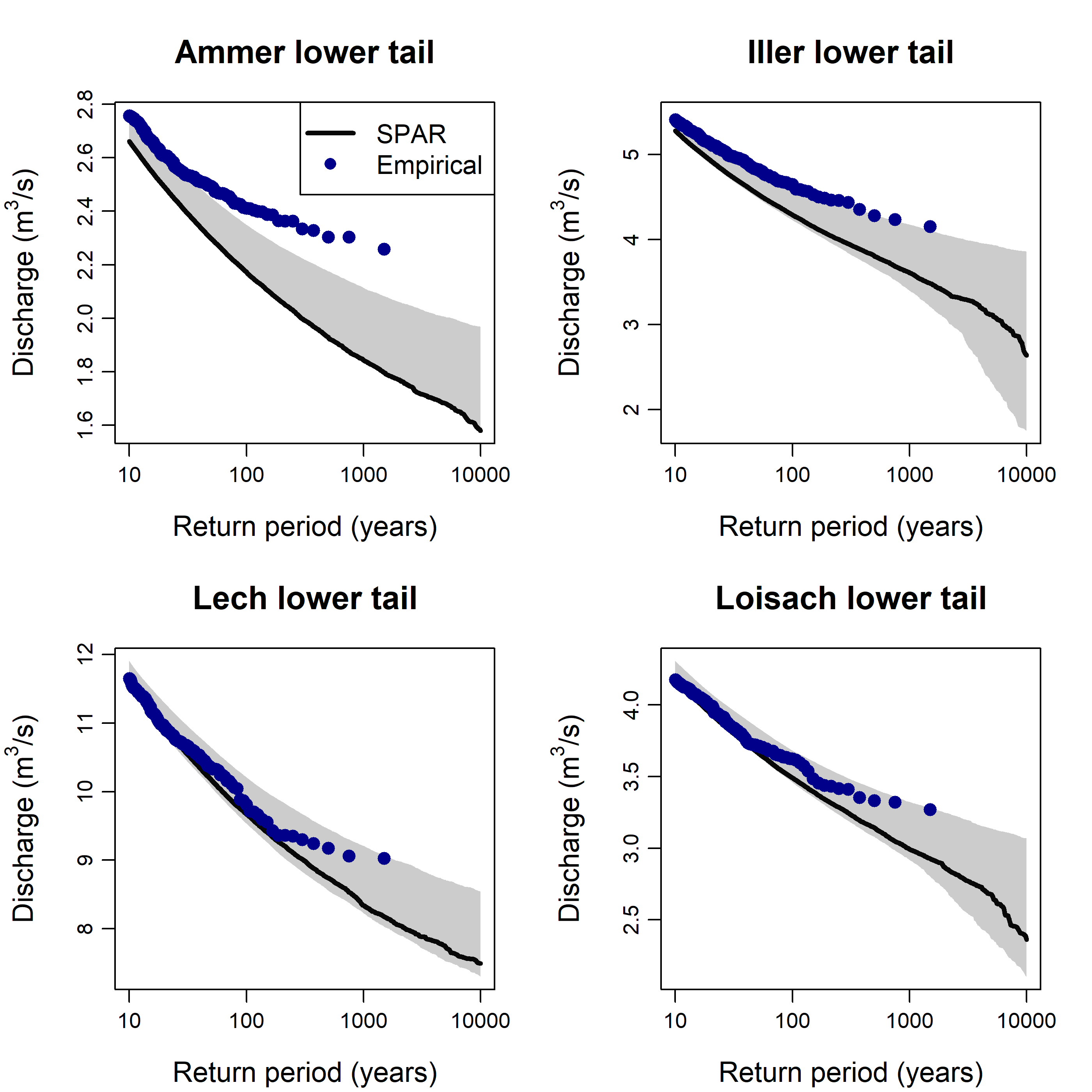}
    \end{minipage}%
    \hfill
   \begin{minipage}{0.48\textwidth} 
        \centering
        \includegraphics[width=\textwidth]{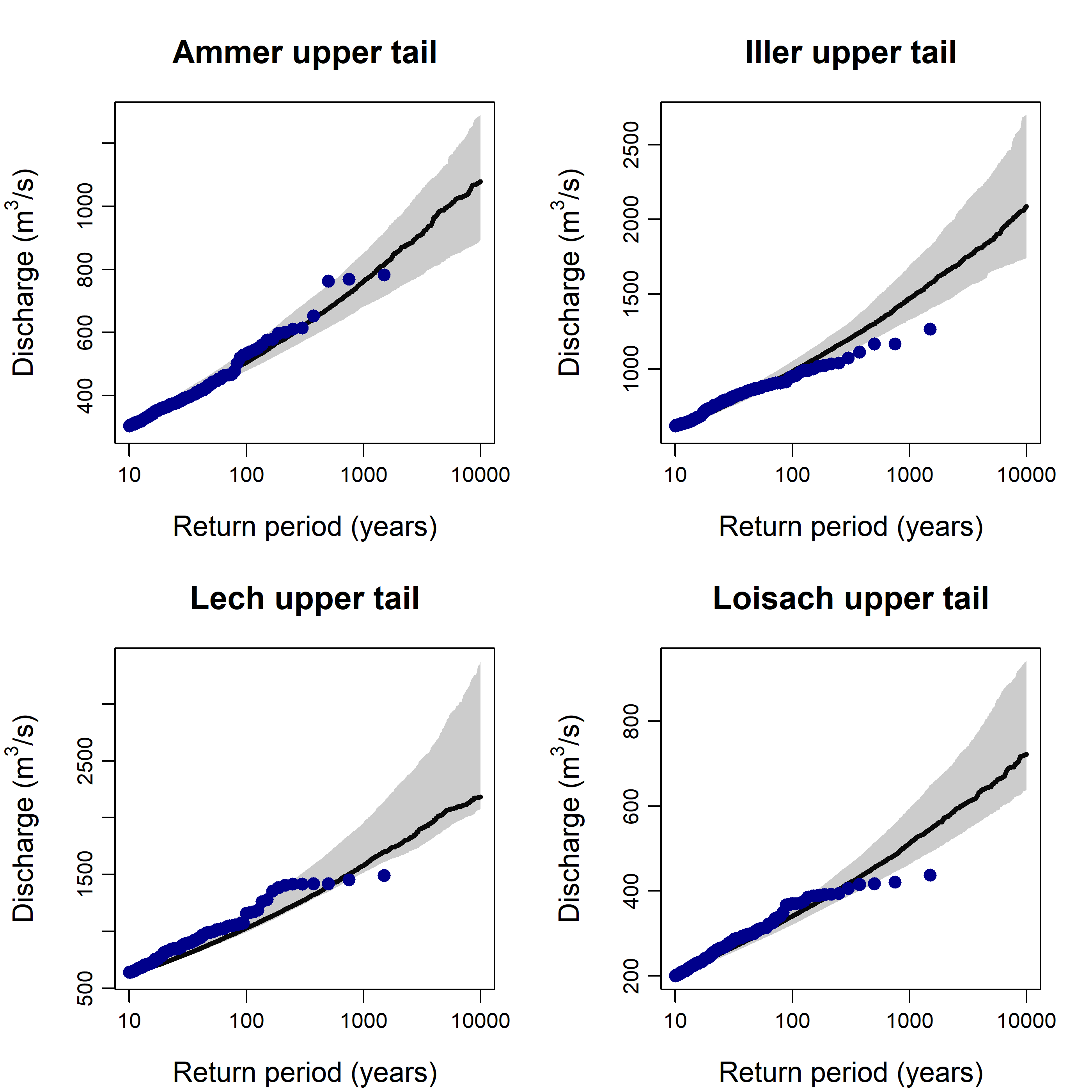}
    \end{minipage}%
    \caption{Return level plots in the lower and upper tails of each marginal variable for the 2070--2099 time period. See Figure~6 in the main manuscript for details on the interpretation of individual panels. }
    \label{fig:ret_level_4}
\end{figure}

Figures~\ref{fig:gpd_qq_plots}-\ref{fig:ret_level_4} illustrate the remaining diagnostic plots for the full sample size model fits. 

\clearpage

\section{Diagnostic plots for smaller sample sizes} \label{appen:diagnostics_smaller_ss}

\begin{figure}[h] 
    \centering
    \begin{minipage}{0.24\textwidth} 
        \centering
        \includegraphics[width=\textwidth]{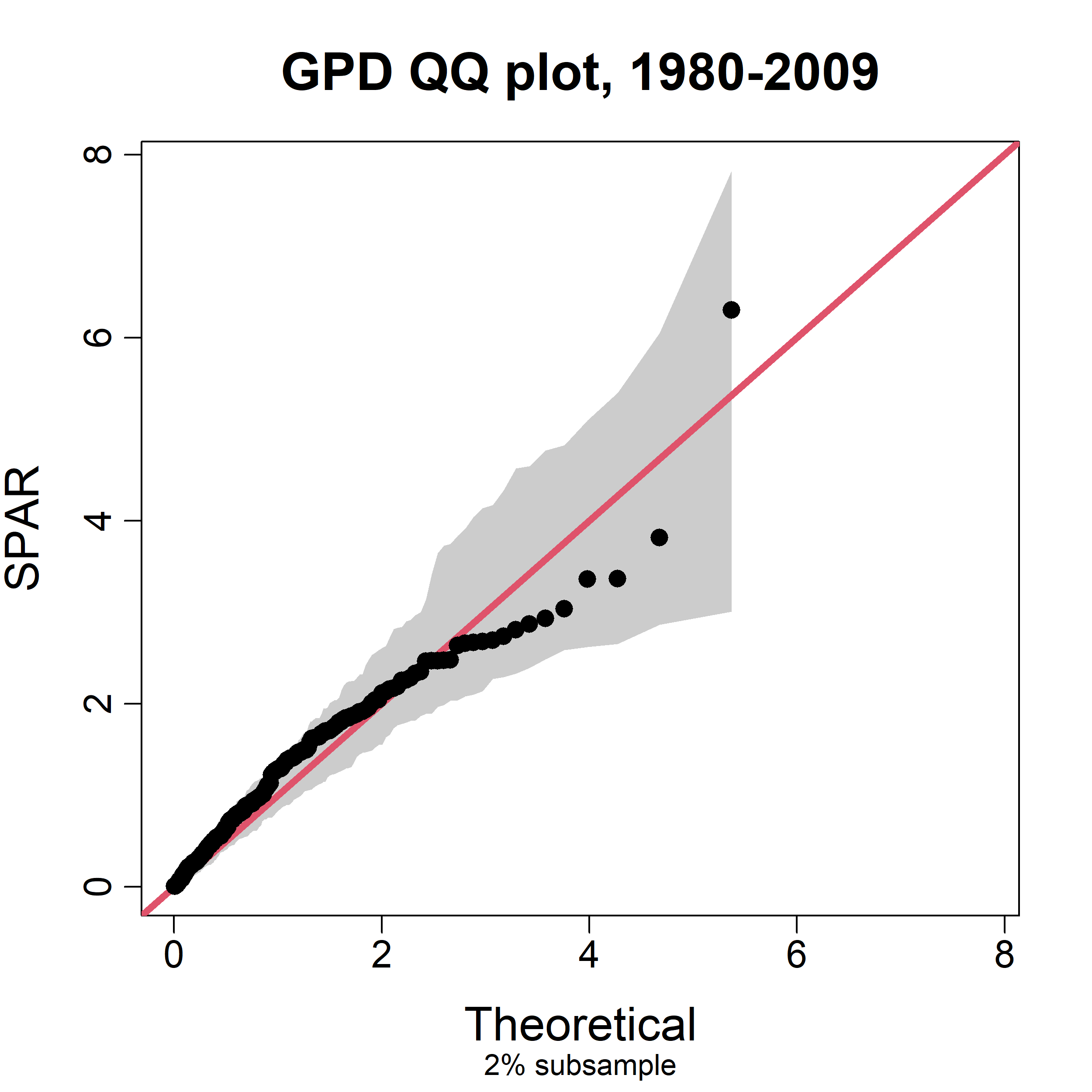}
    \end{minipage}%
    \hfill
   \begin{minipage}{0.24\textwidth} 
        \centering
        \includegraphics[width=\textwidth]{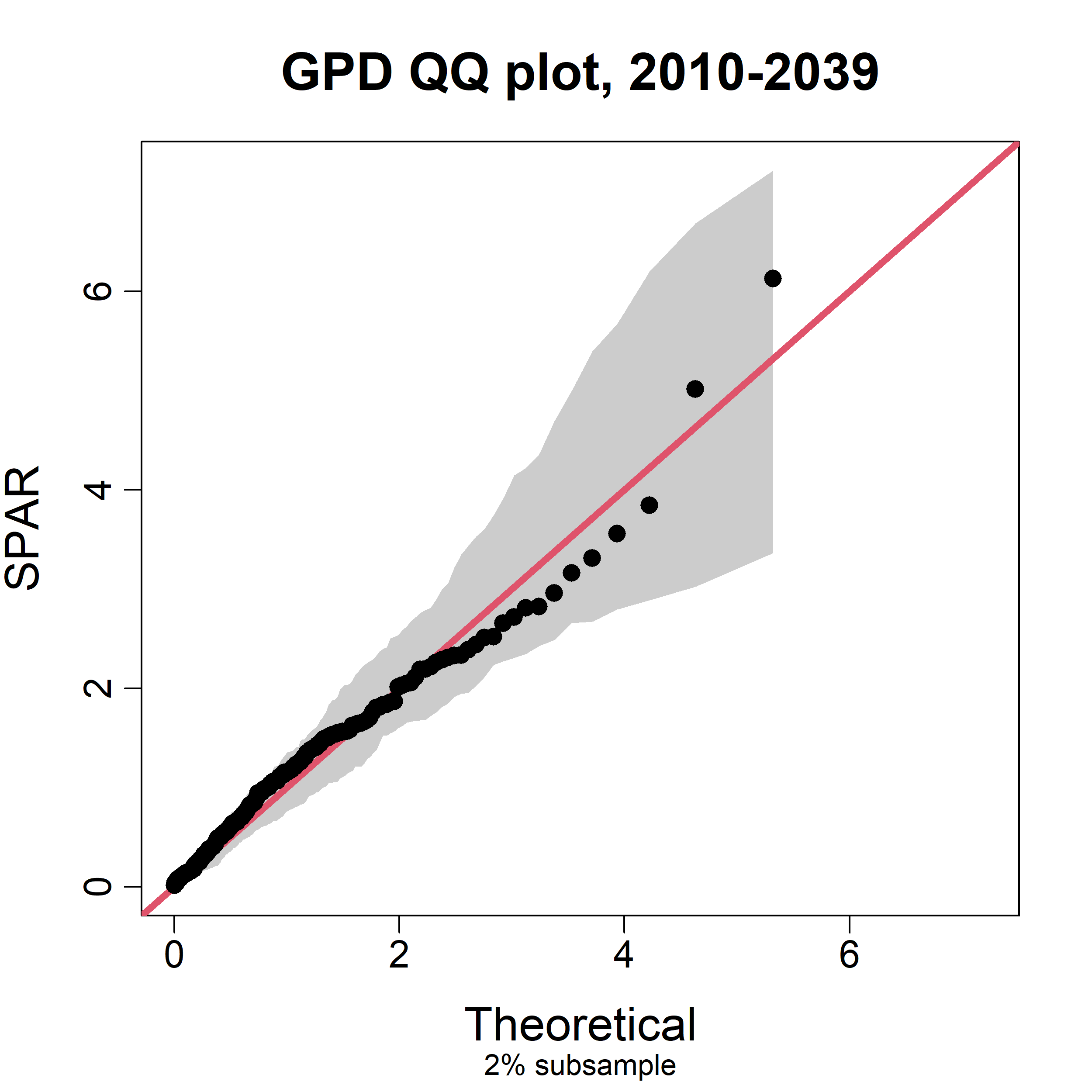}
    \end{minipage}%
    \hfill
    \begin{minipage}{0.24\textwidth} 
        \centering
        \includegraphics[width=\textwidth]{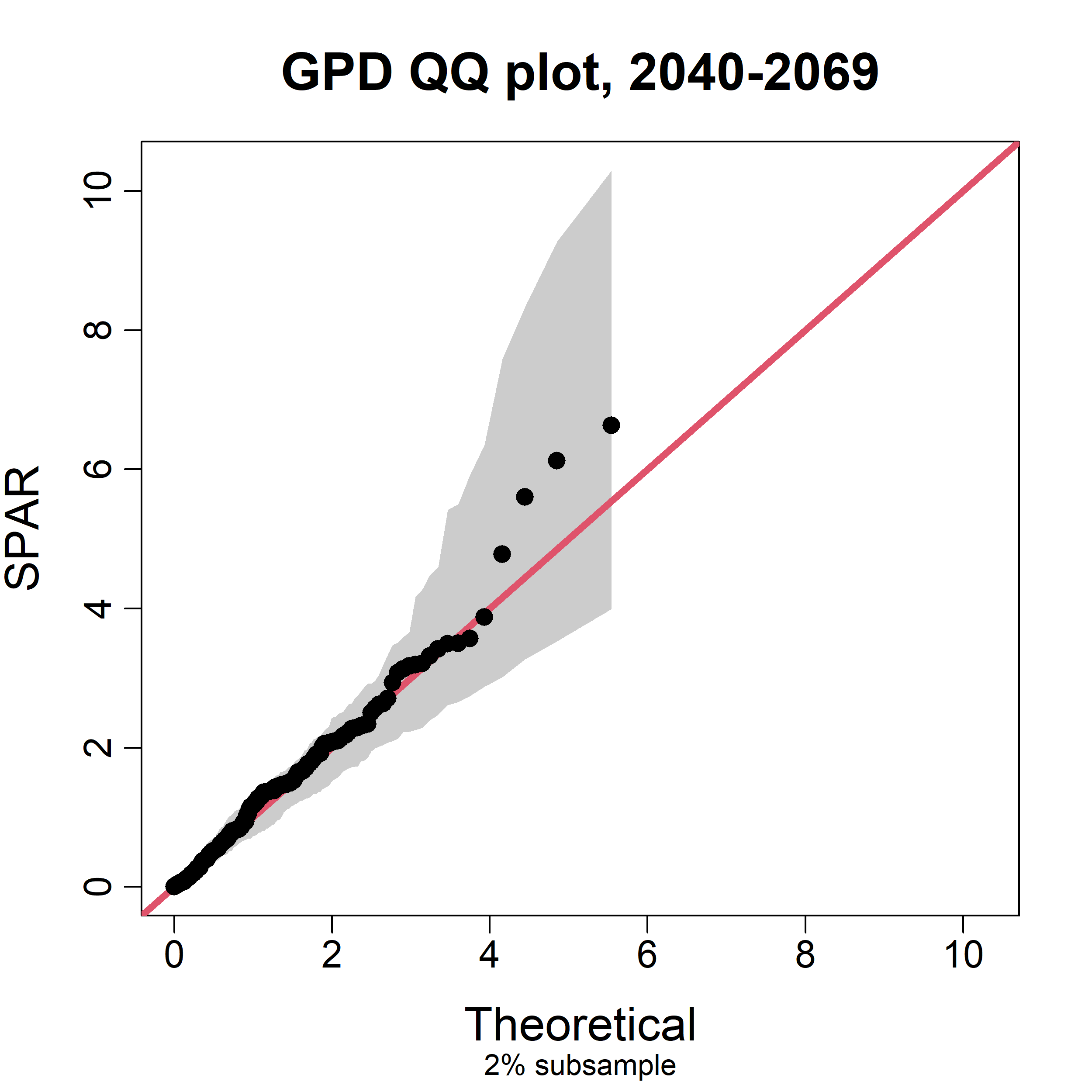}
    \end{minipage}%
    \hfill
    \begin{minipage}{0.24\textwidth} 
        \centering
        \includegraphics[width=\textwidth]{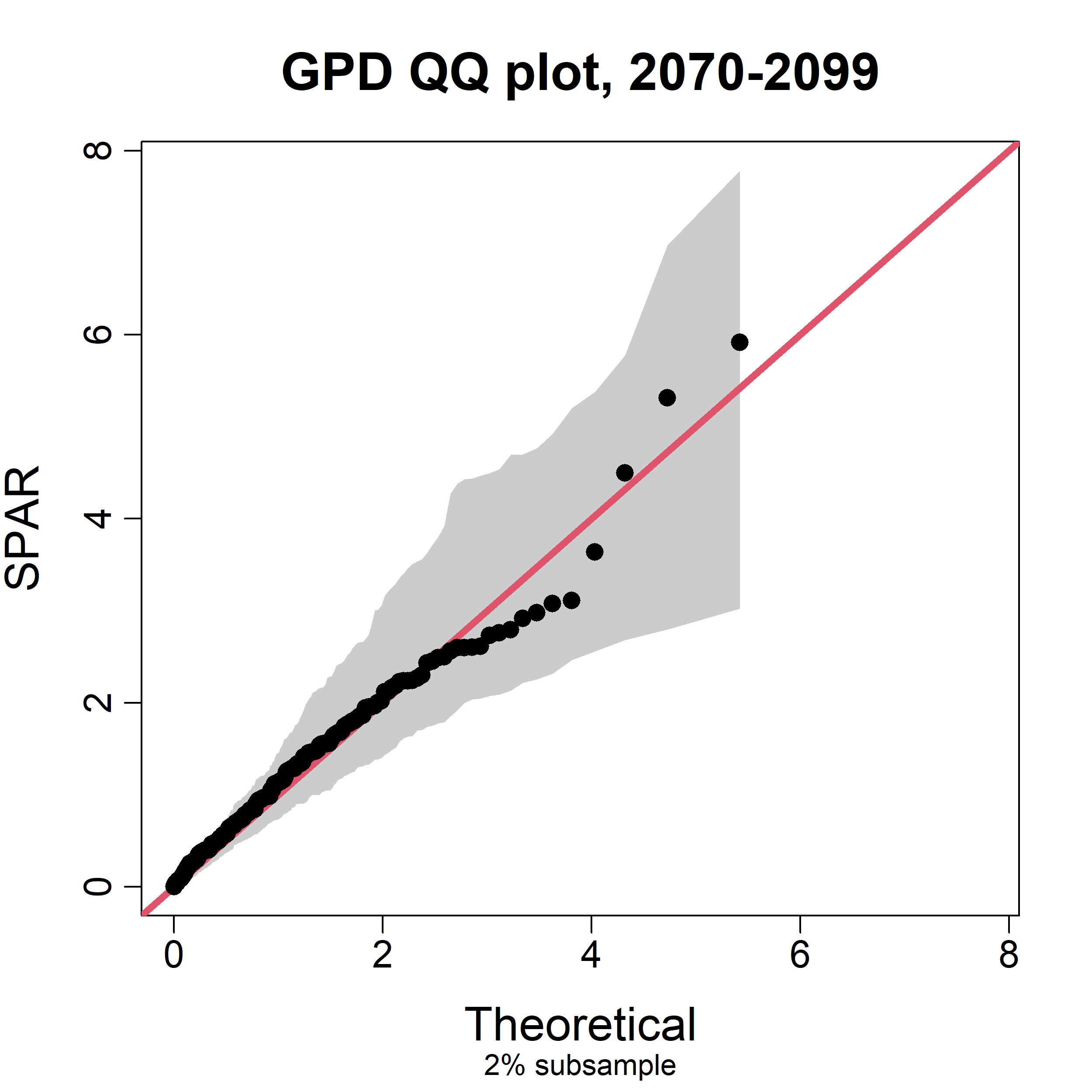}
    \end{minipage}%
    \caption{GPD QQ plots for sequential time windows, ordered chronologically from left to right, for the subsample containing $2\%$ of the full data.}
    \label{fig:gpd_qq_plots_2}
\end{figure}

\begin{figure}[h] 
    \centering
    \begin{minipage}{0.24\textwidth} 
        \centering
        \includegraphics[width=\textwidth]{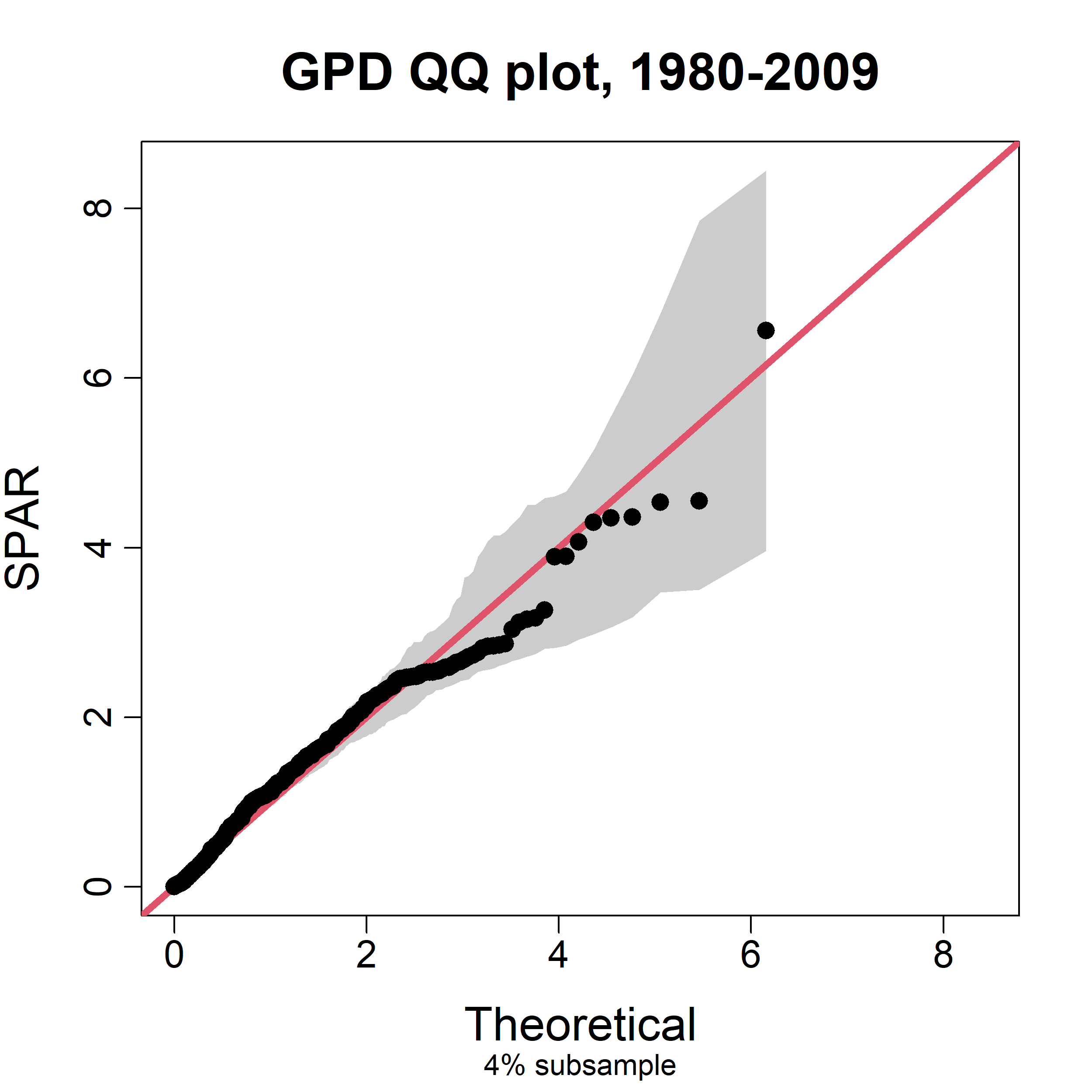}
    \end{minipage}%
    \hfill
   \begin{minipage}{0.24\textwidth} 
        \centering
        \includegraphics[width=\textwidth]{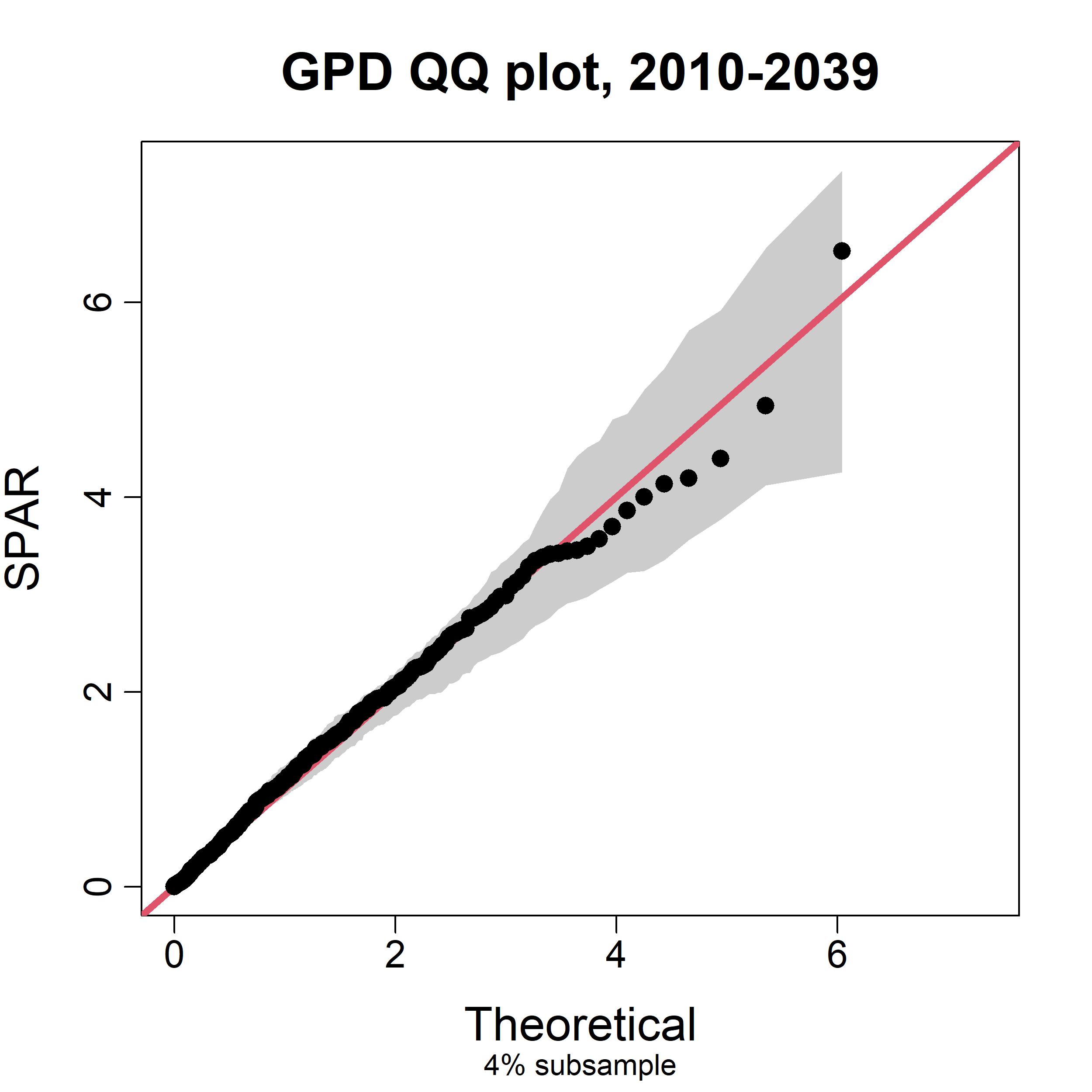}
    \end{minipage}%
    \hfill
    \begin{minipage}{0.24\textwidth} 
        \centering
        \includegraphics[width=\textwidth]{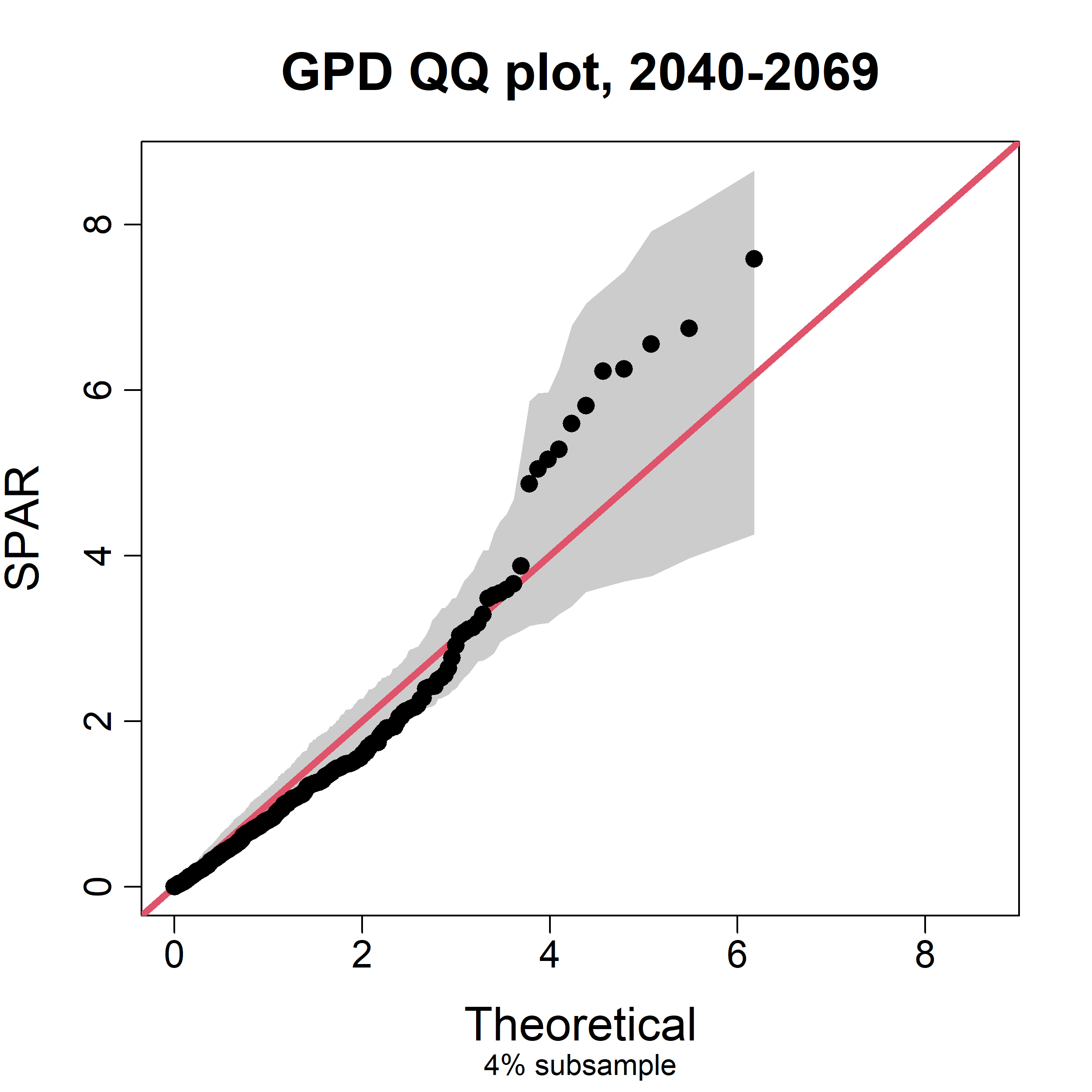}
    \end{minipage}%
    \hfill
    \begin{minipage}{0.24\textwidth} 
        \centering
        \includegraphics[width=\textwidth]{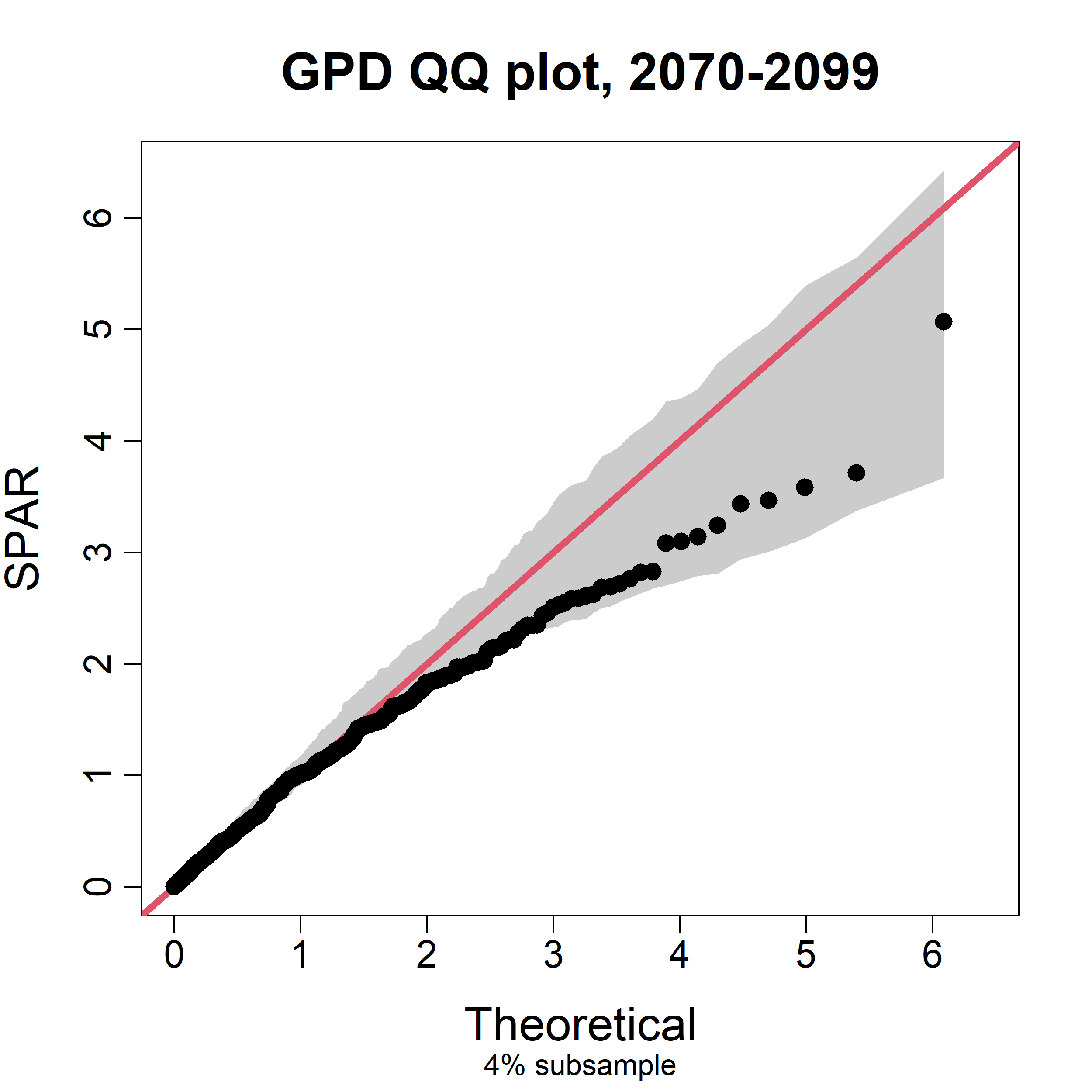}
    \end{minipage}%
    \caption{GPD QQ plots for sequential time windows, ordered chronologically from left to right, for the subsample containing $4\%$ of the full data.}
    \label{fig:gpd_qq_plots_4}
\end{figure}

\begin{figure}[h] 
    \centering
    \begin{minipage}{0.24\textwidth} 
        \centering
        \includegraphics[width=\textwidth]{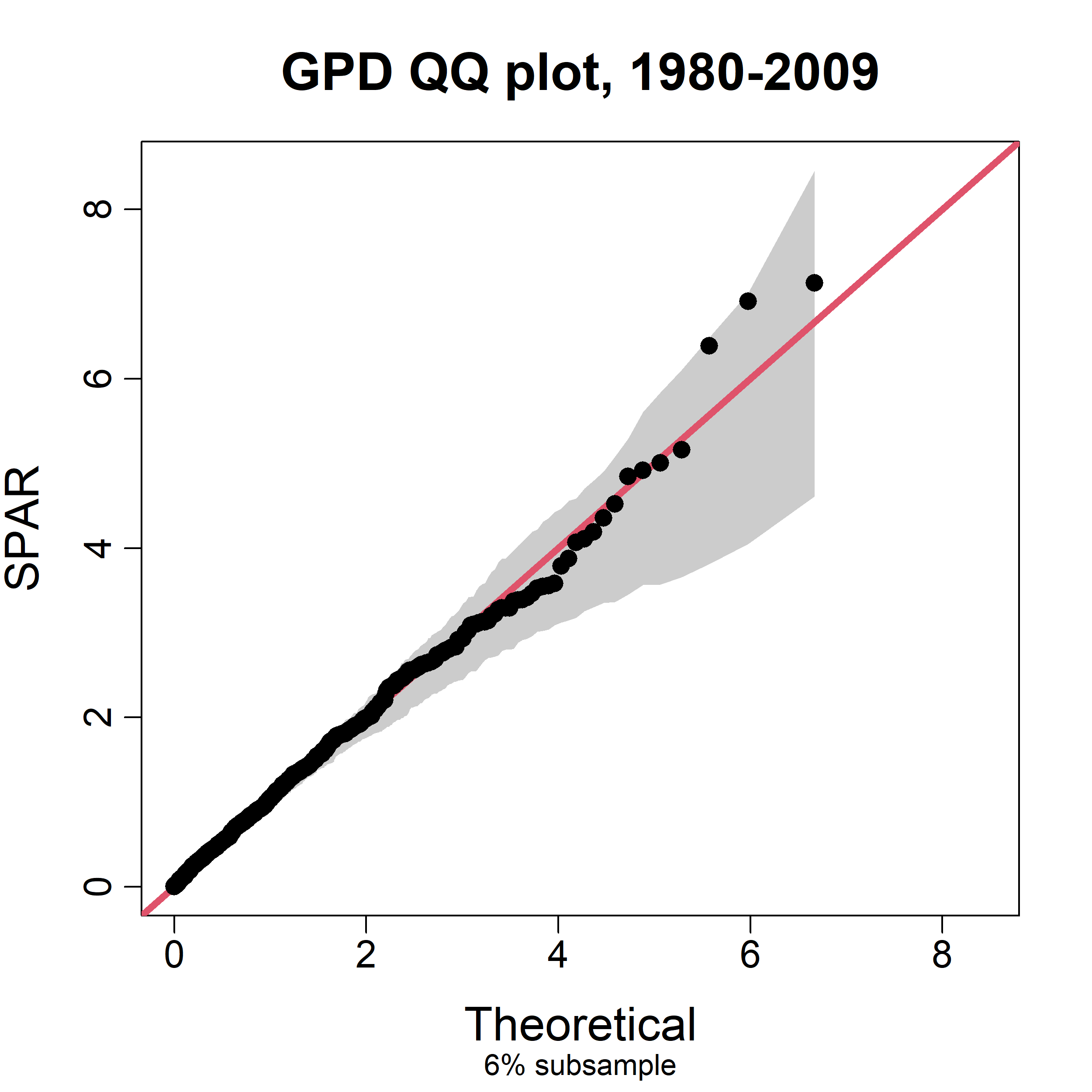}
    \end{minipage}%
    \hfill
   \begin{minipage}{0.24\textwidth} 
        \centering
        \includegraphics[width=\textwidth]{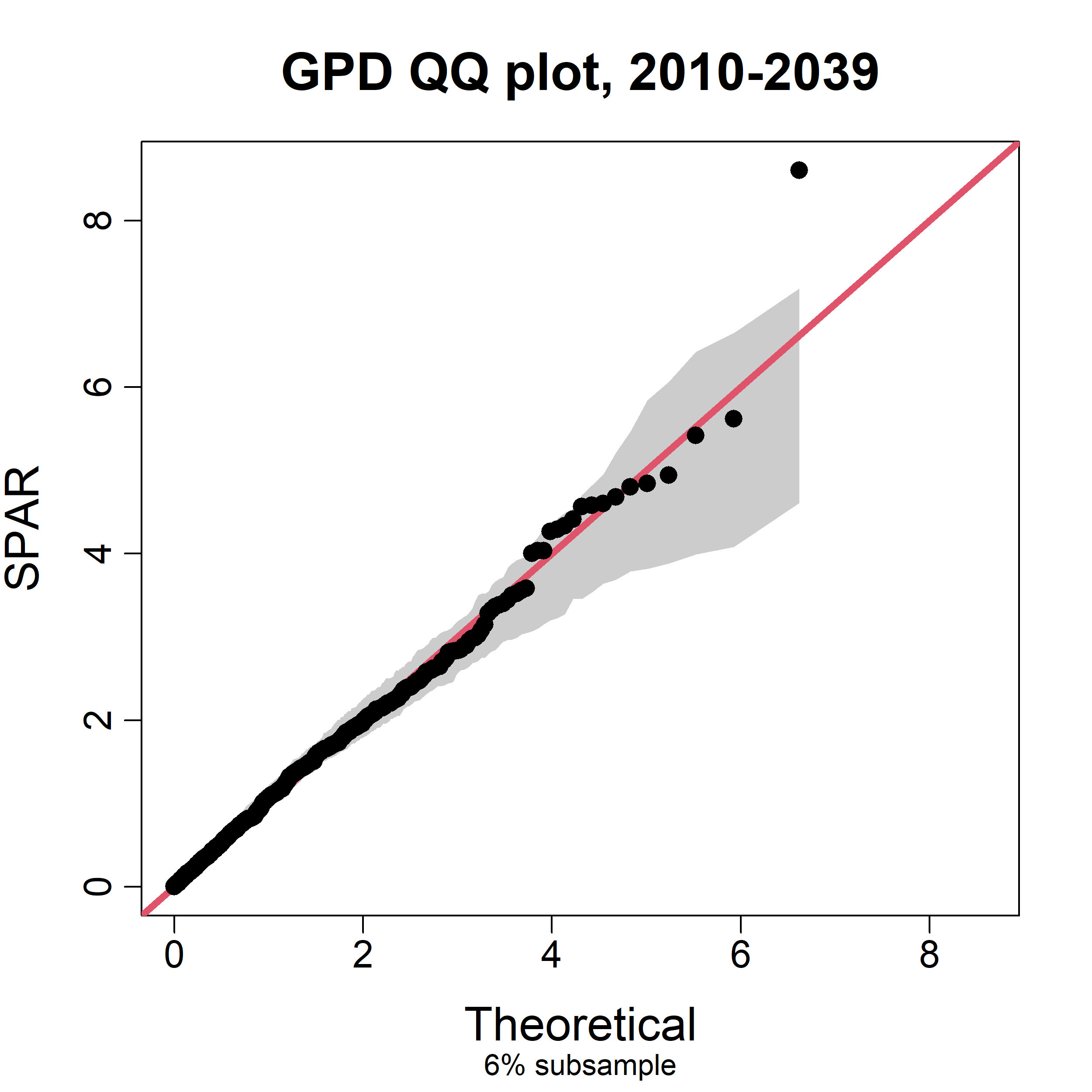}
    \end{minipage}%
    \hfill
    \begin{minipage}{0.24\textwidth} 
        \centering
        \includegraphics[width=\textwidth]{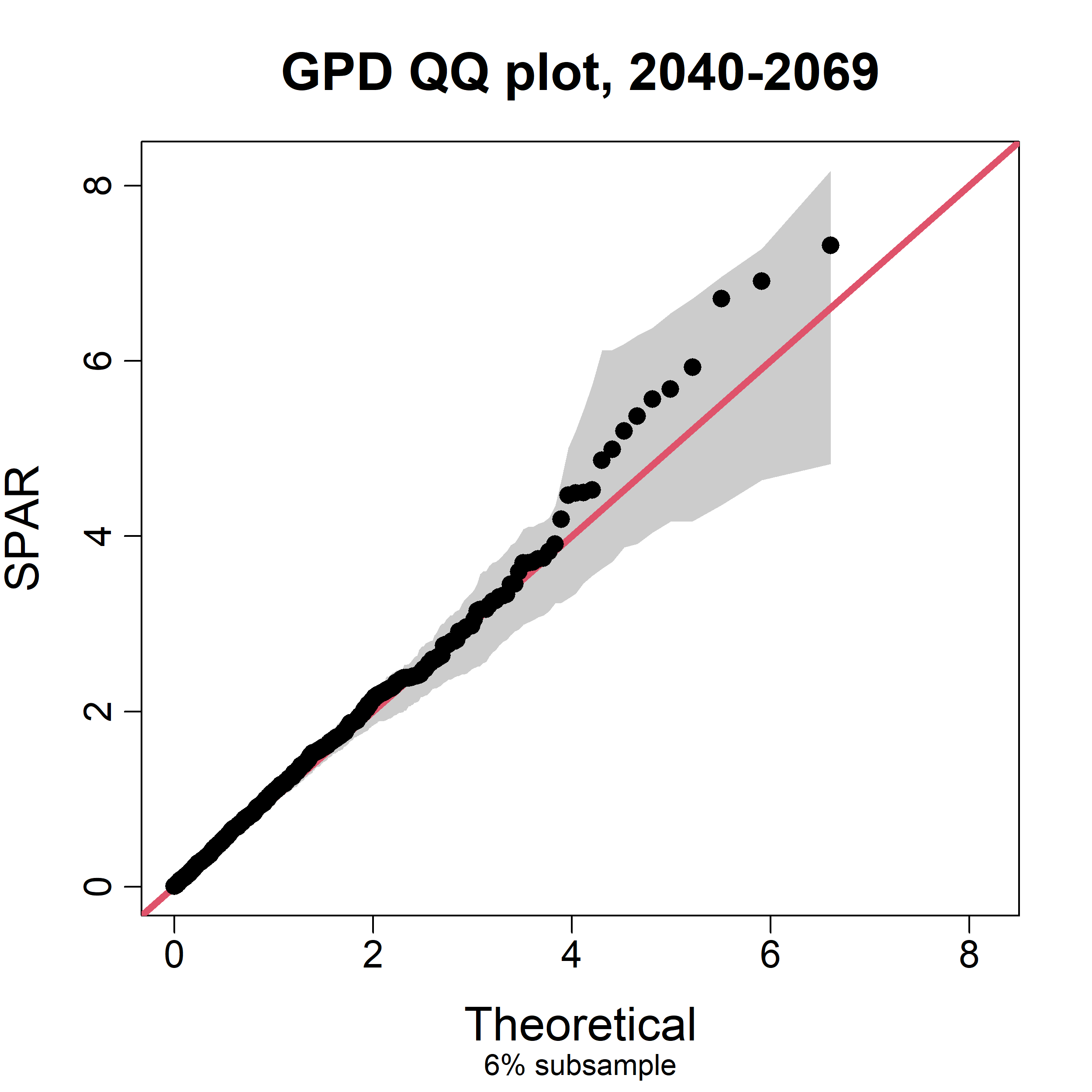}
    \end{minipage}%
    \hfill
    \begin{minipage}{0.24\textwidth} 
        \centering
        \includegraphics[width=\textwidth]{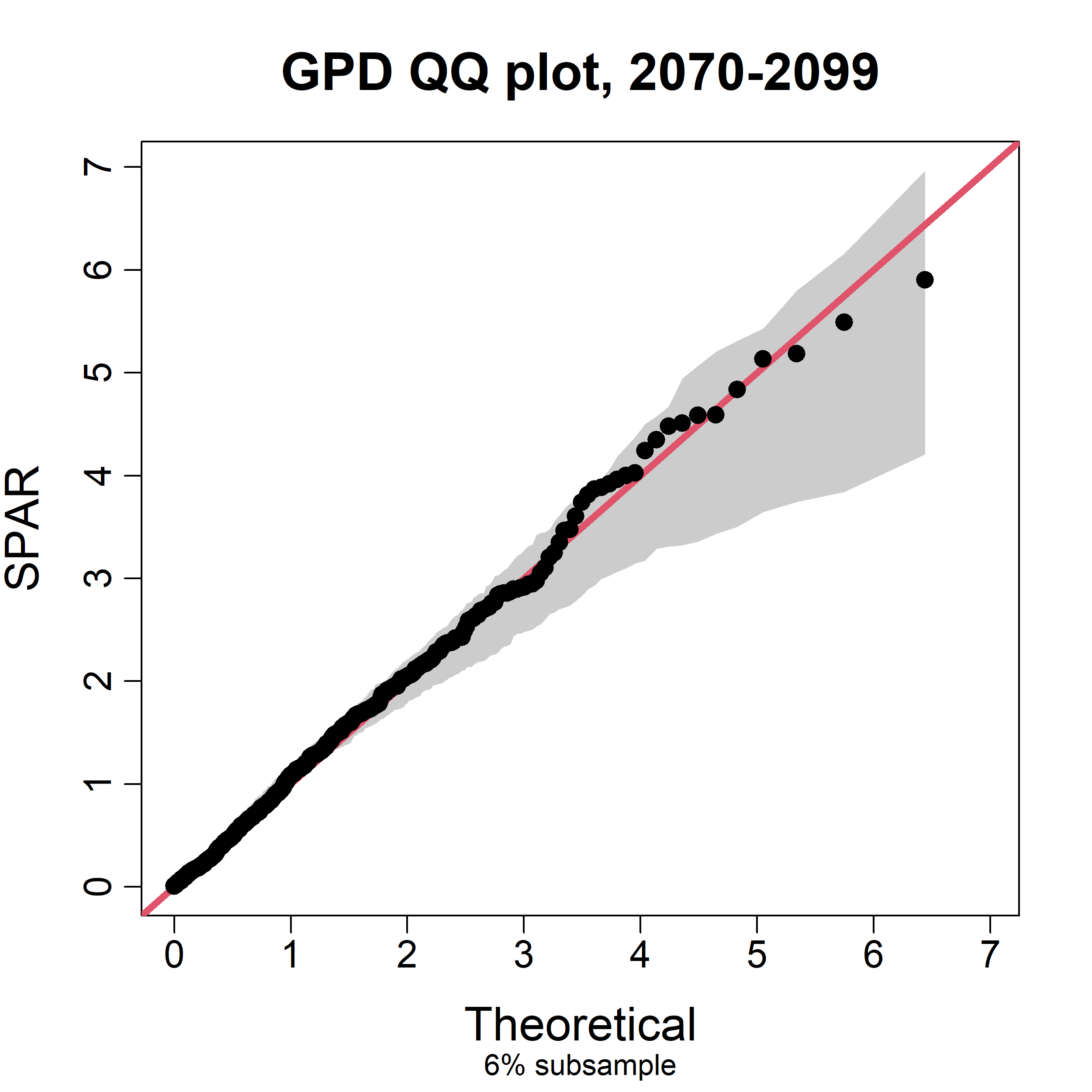}
    \end{minipage}%
    \caption{GPD QQ plots for sequential time windows, ordered chronologically from left to right, for the subsample containing $6\%$ of the full data.}
    \label{fig:gpd_qq_plots_6}
\end{figure}

\begin{figure}[h] 
    \centering
    \begin{minipage}{0.24\textwidth} 
        \centering
        \includegraphics[width=\textwidth]{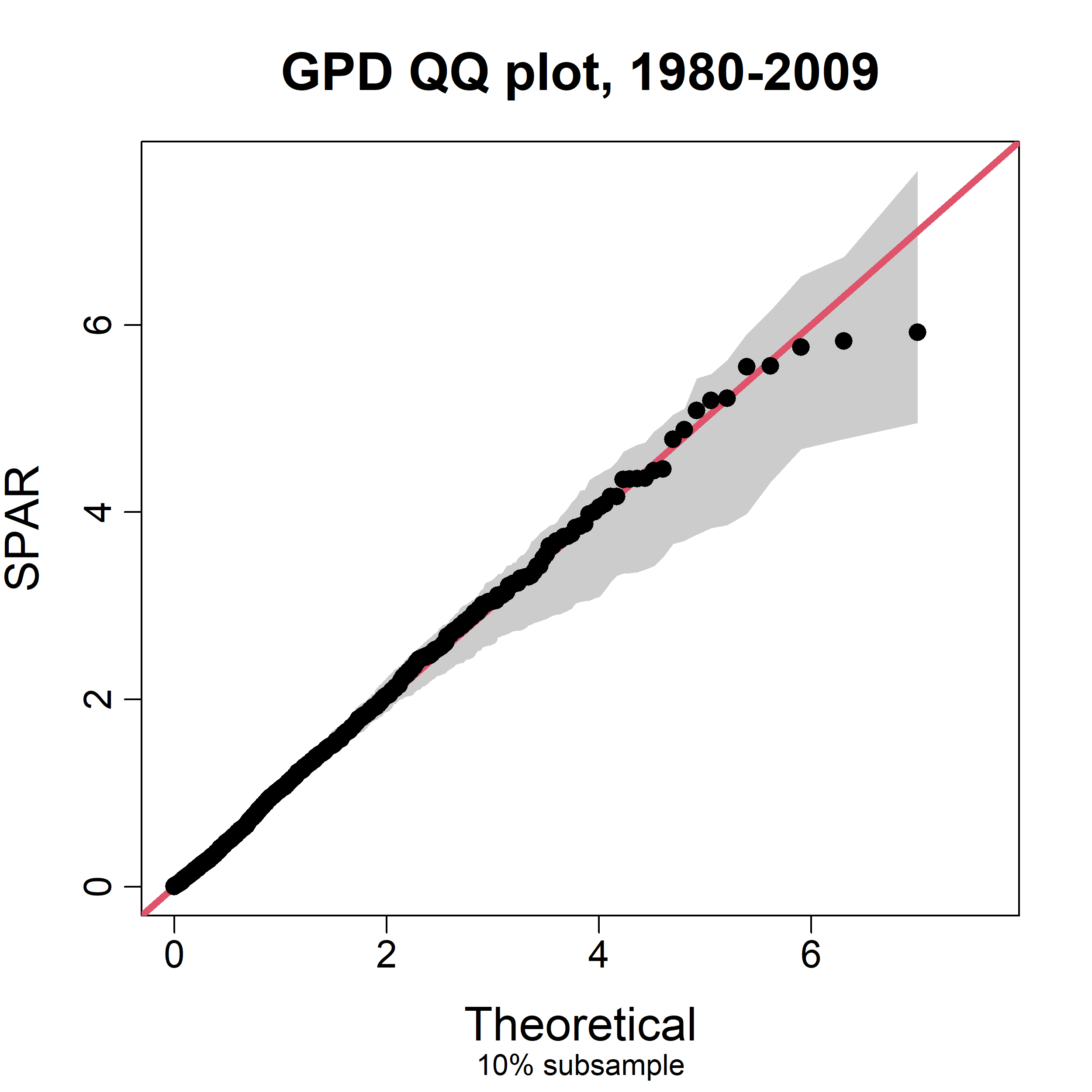}
    \end{minipage}%
    \hfill
   \begin{minipage}{0.24\textwidth} 
        \centering
        \includegraphics[width=\textwidth]{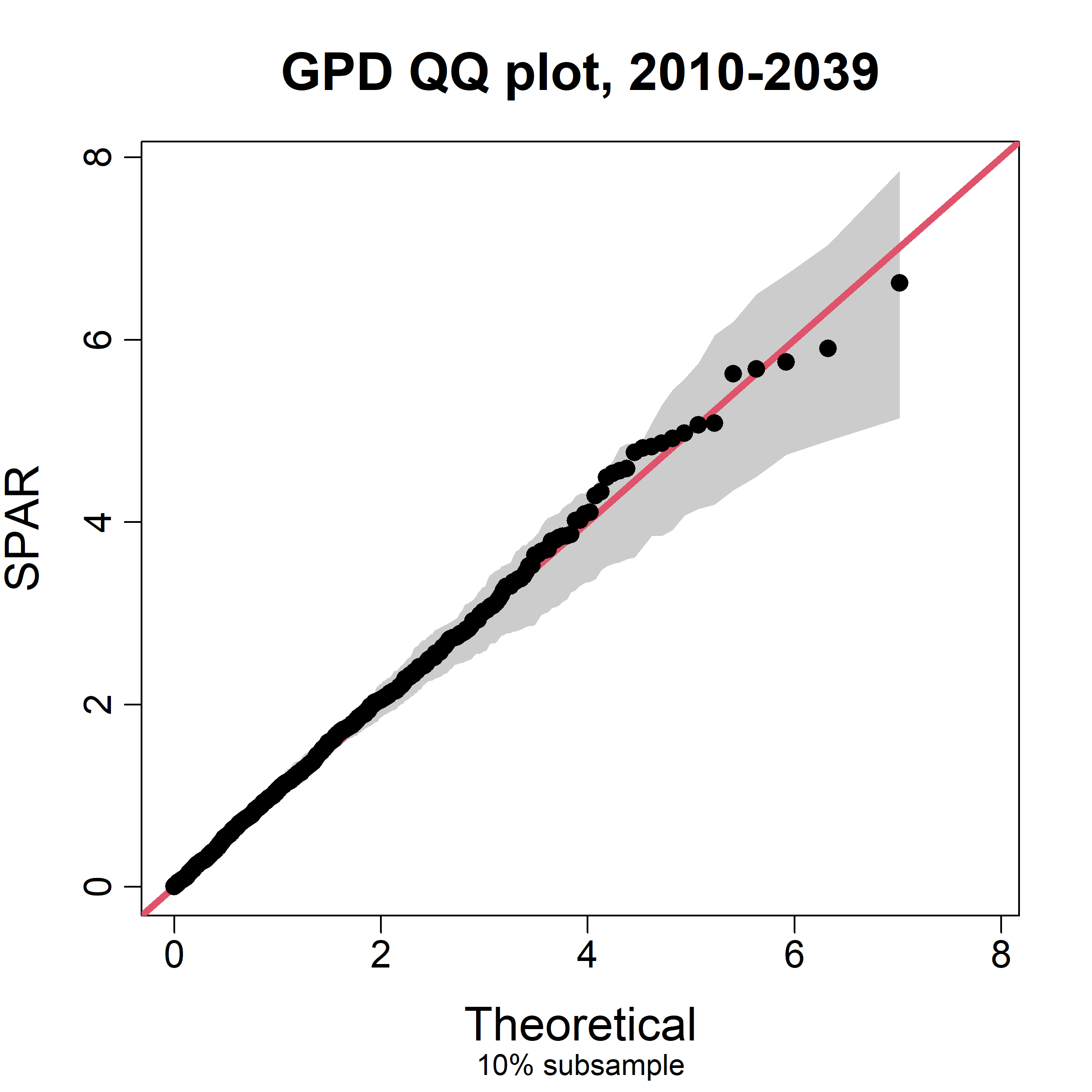}
    \end{minipage}%
    \hfill
    \begin{minipage}{0.24\textwidth} 
        \centering
        \includegraphics[width=\textwidth]{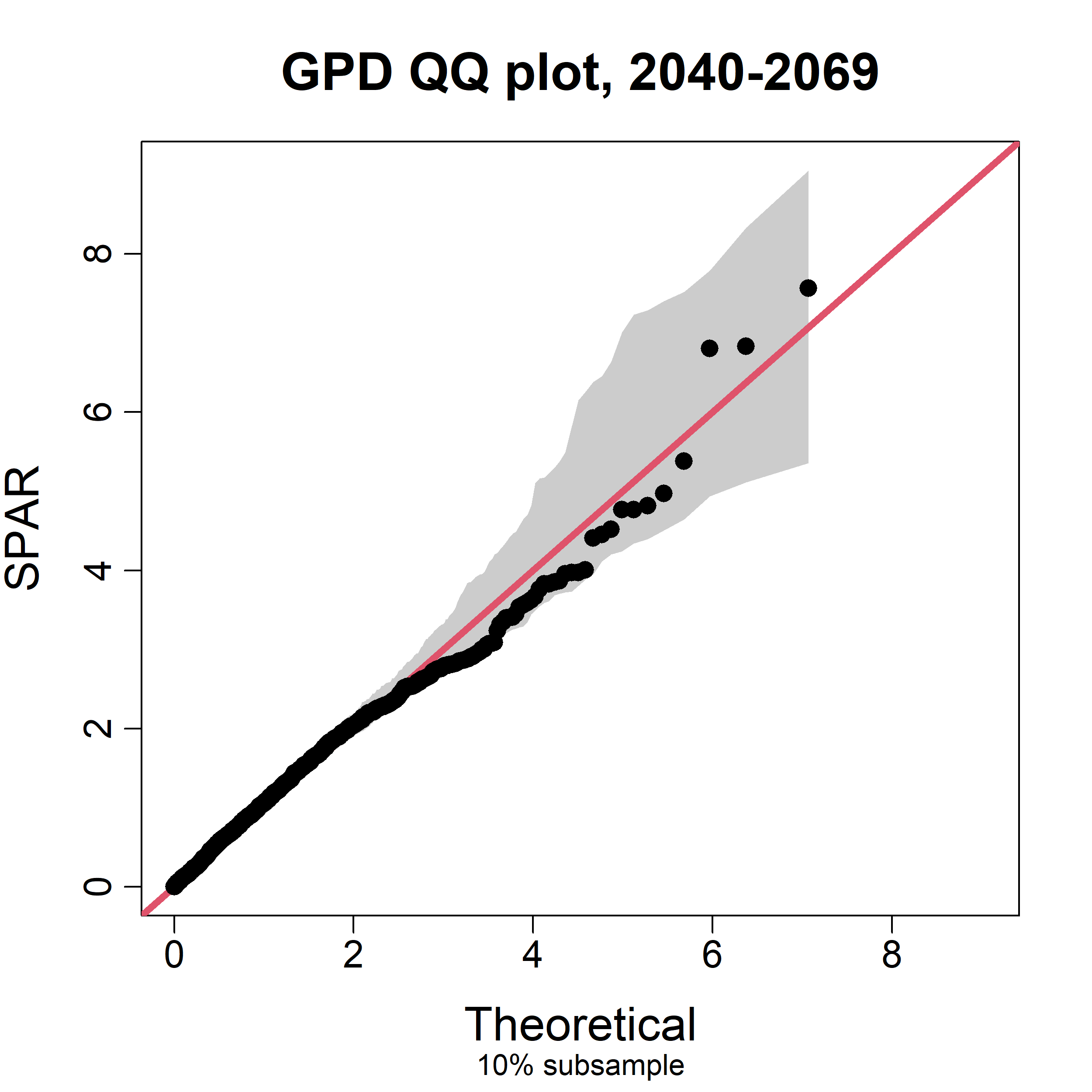}
    \end{minipage}%
    \hfill
    \begin{minipage}{0.24\textwidth} 
        \centering
        \includegraphics[width=\textwidth]{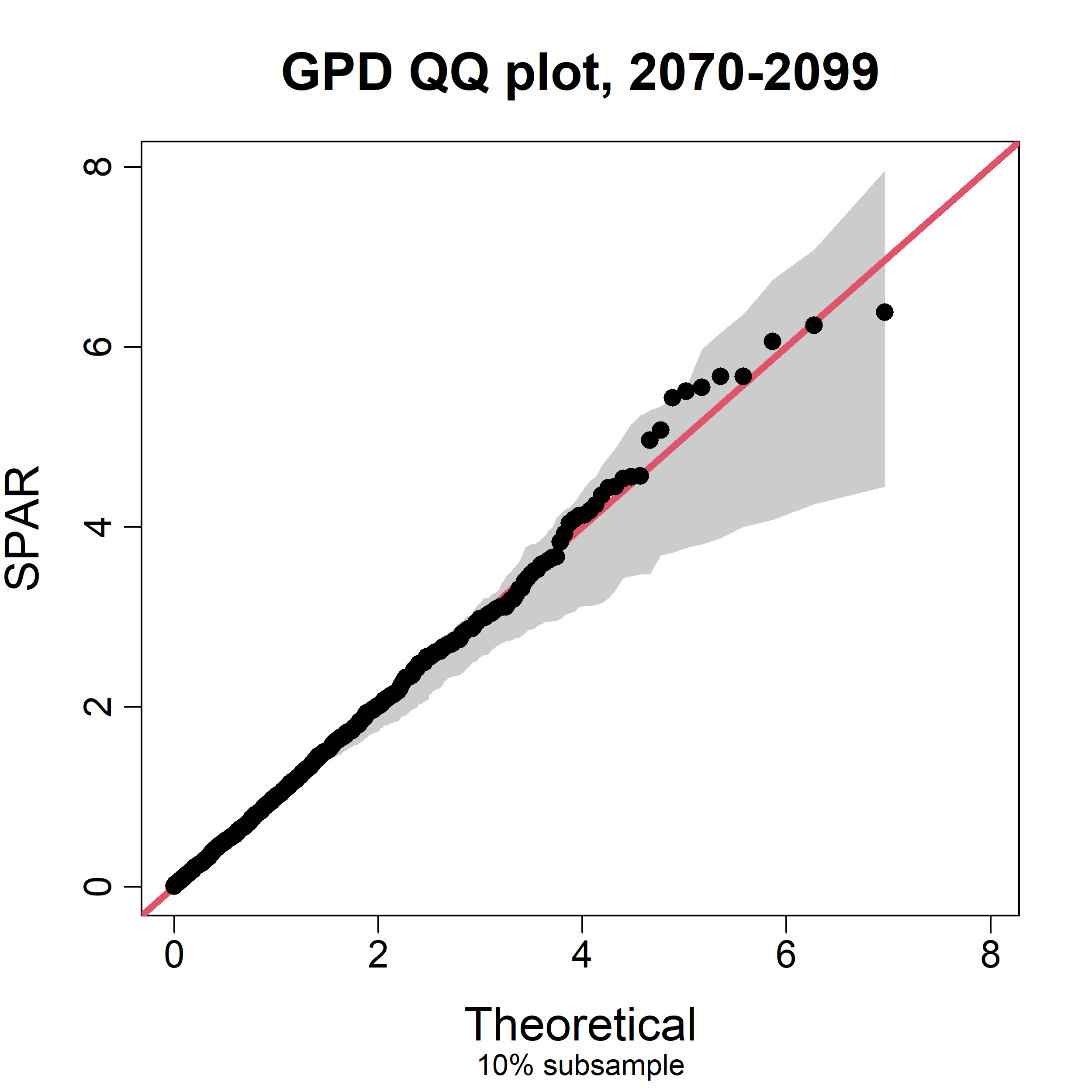}
    \end{minipage}%
    \caption{GPD QQ plots for sequential time windows, ordered chronologically from left to right, for the subsample containing $10\%$ of the full data.}
    \label{fig:gpd_qq_plots_10}
\end{figure}

\begin{figure}[h] 
    \centering
    \begin{minipage}{0.24\textwidth} 
        \centering
        \includegraphics[width=\textwidth]{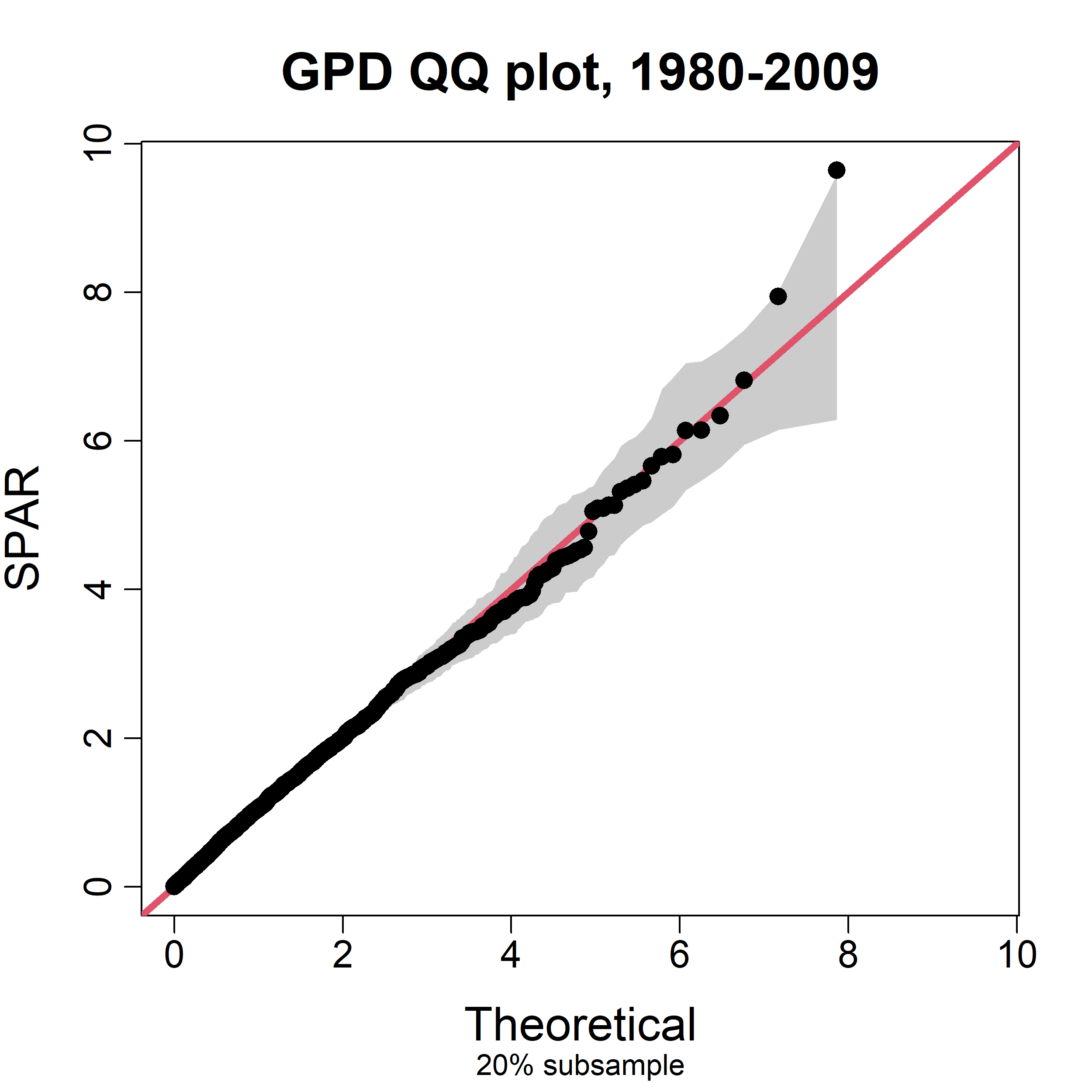}
    \end{minipage}%
    \hfill
   \begin{minipage}{0.24\textwidth} 
        \centering
        \includegraphics[width=\textwidth]{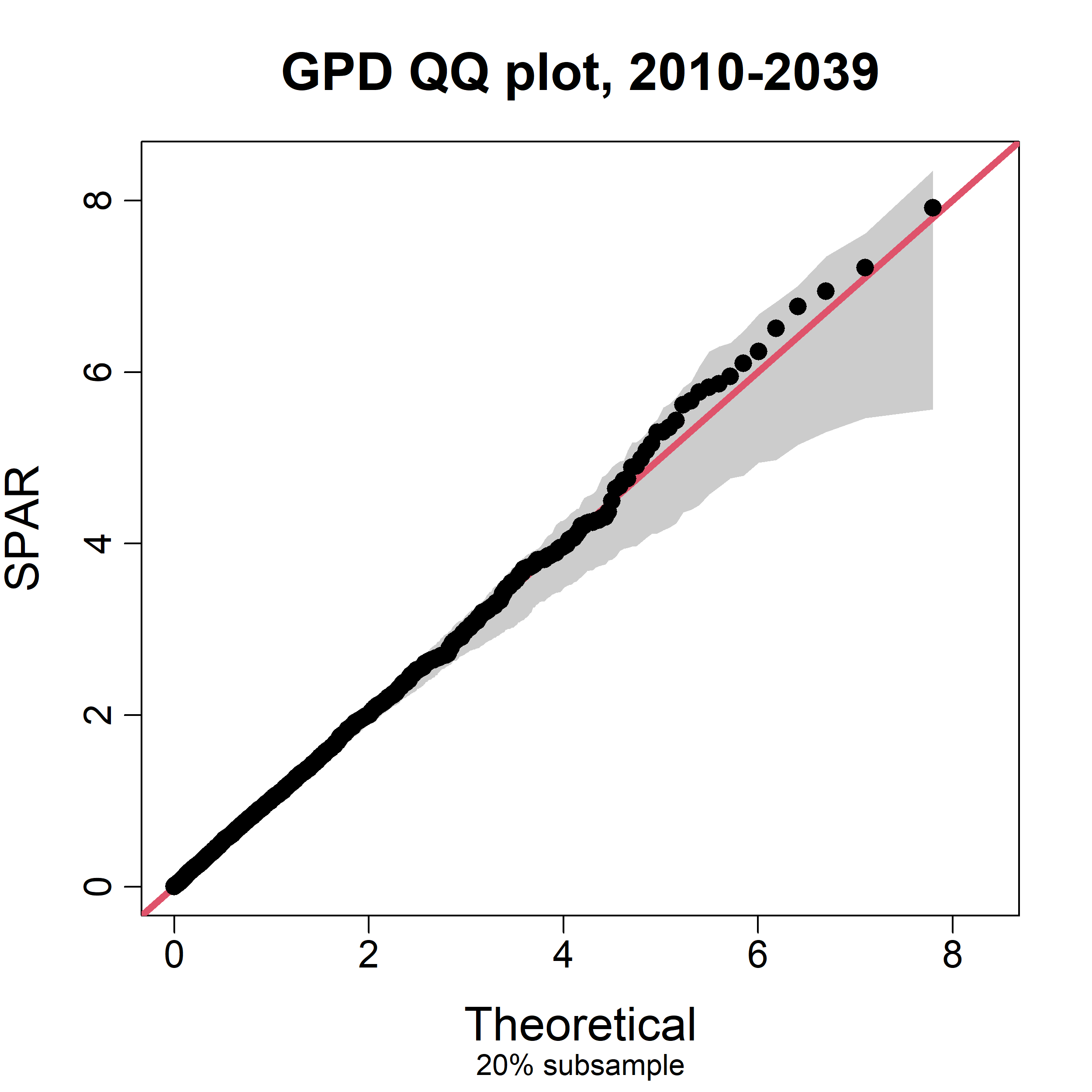}
    \end{minipage}%
    \hfill
    \begin{minipage}{0.24\textwidth} 
        \centering
        \includegraphics[width=\textwidth]{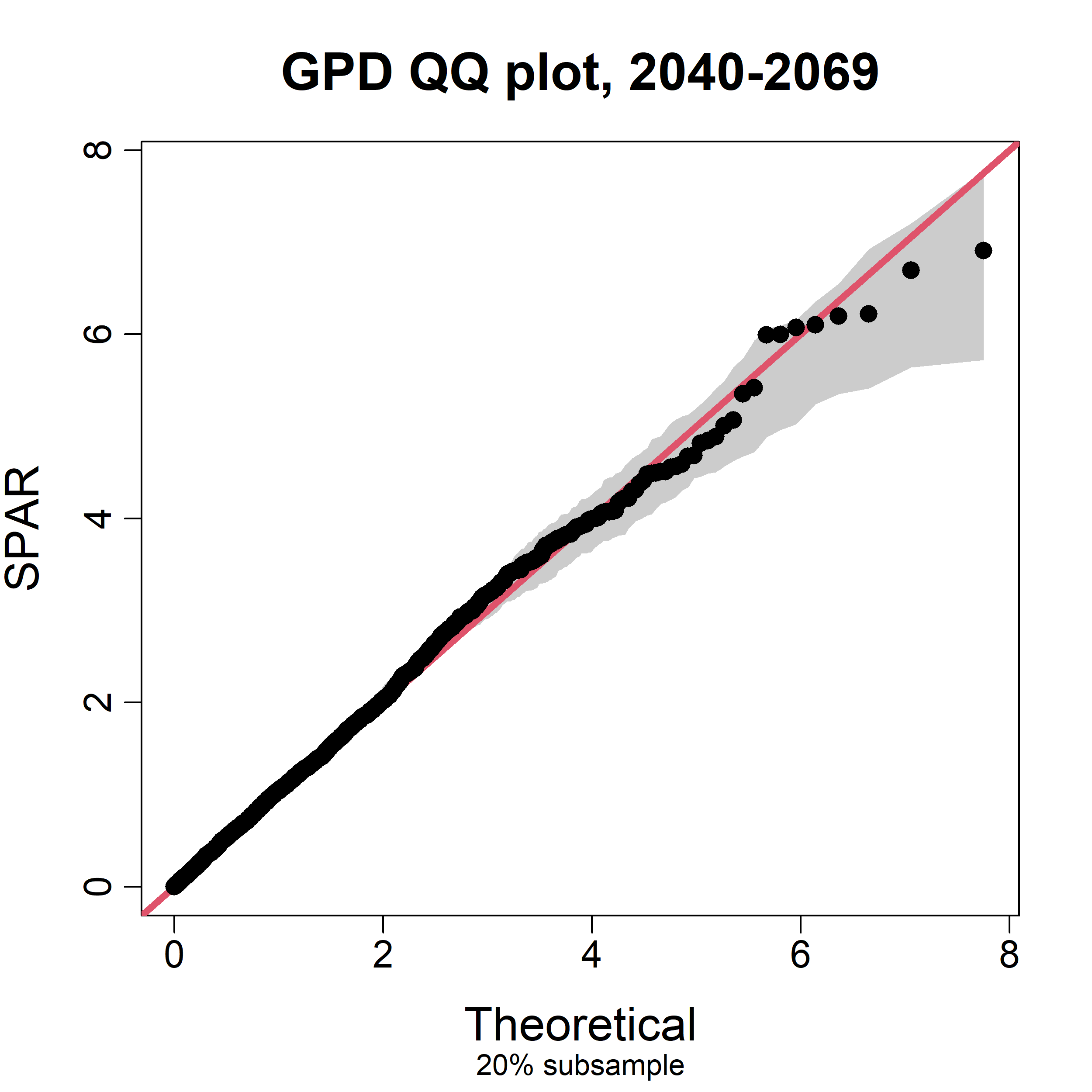}
    \end{minipage}%
    \hfill
    \begin{minipage}{0.24\textwidth} 
        \centering
        \includegraphics[width=\textwidth]{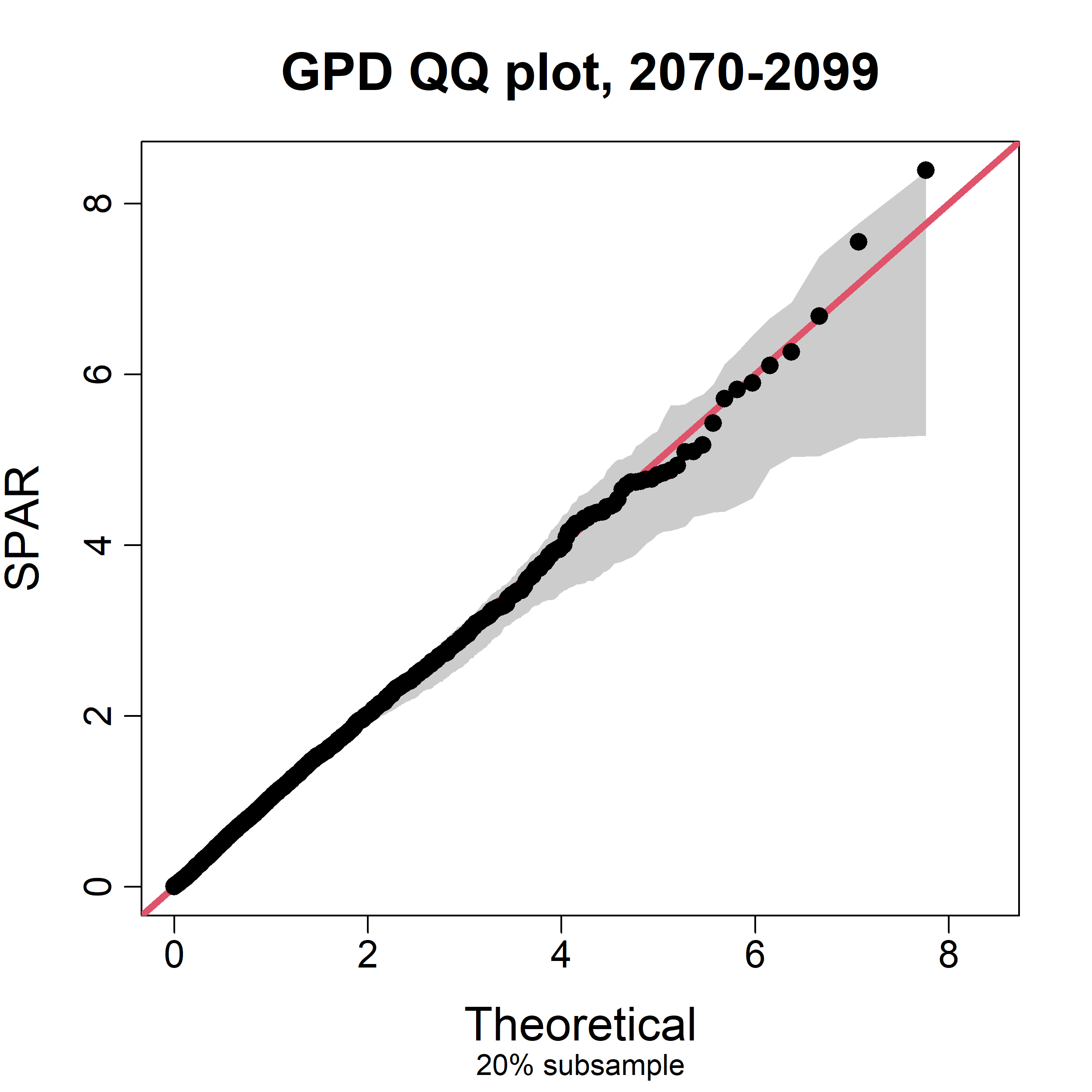}
    \end{minipage}%
    \caption{GPD QQ plots for sequential time windows, ordered chronologically from left to right, for the subsample containing $20\%$ of the full data.}
    \label{fig:gpd_qq_plots_20}
\end{figure}

\begin{figure}[h] 
    \centering
    \begin{minipage}{0.48\textwidth} 
        \centering
        \includegraphics[width=\textwidth]{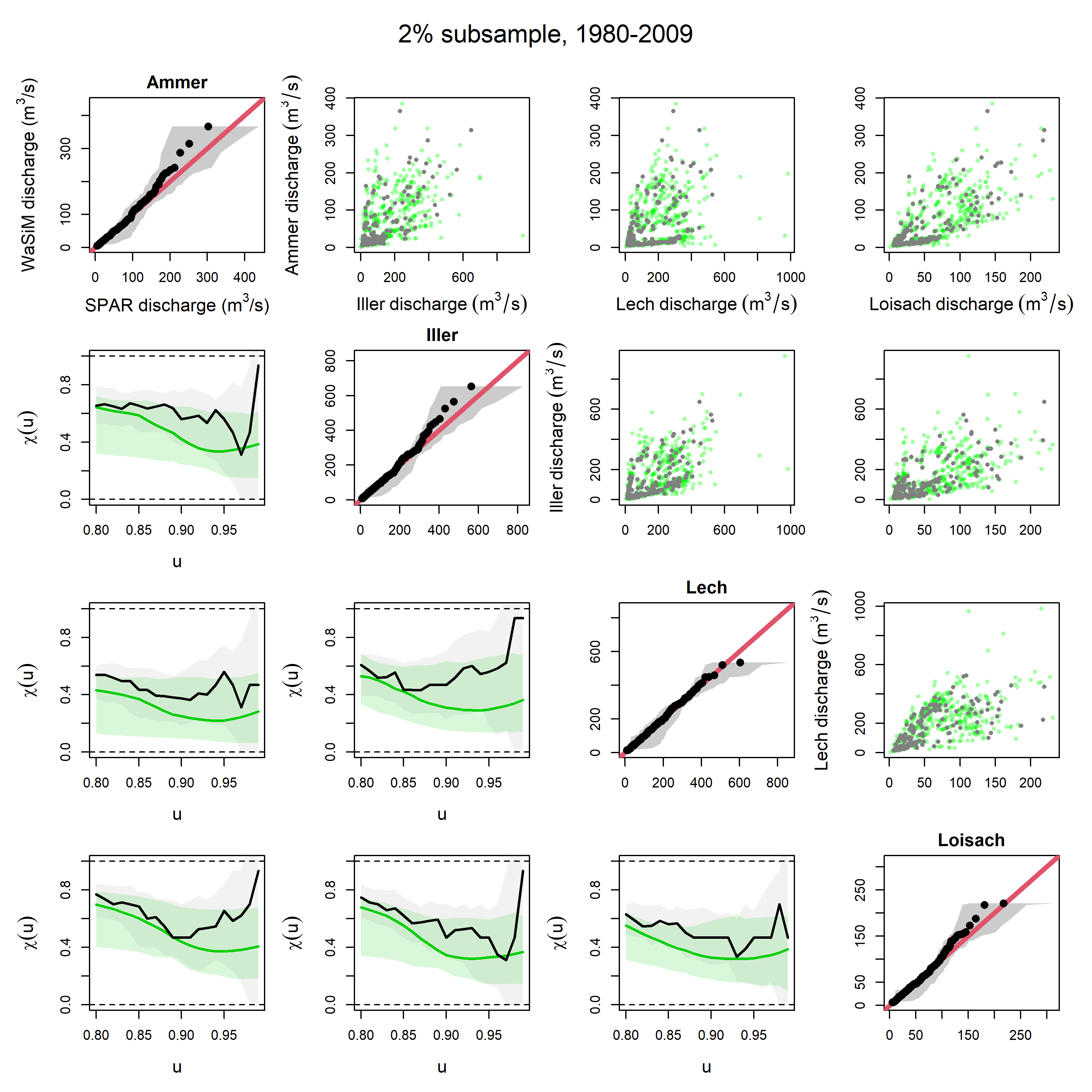}
    \end{minipage}%
    \hfill
   \begin{minipage}{0.48\textwidth} 
        \centering
        \includegraphics[width=\textwidth]{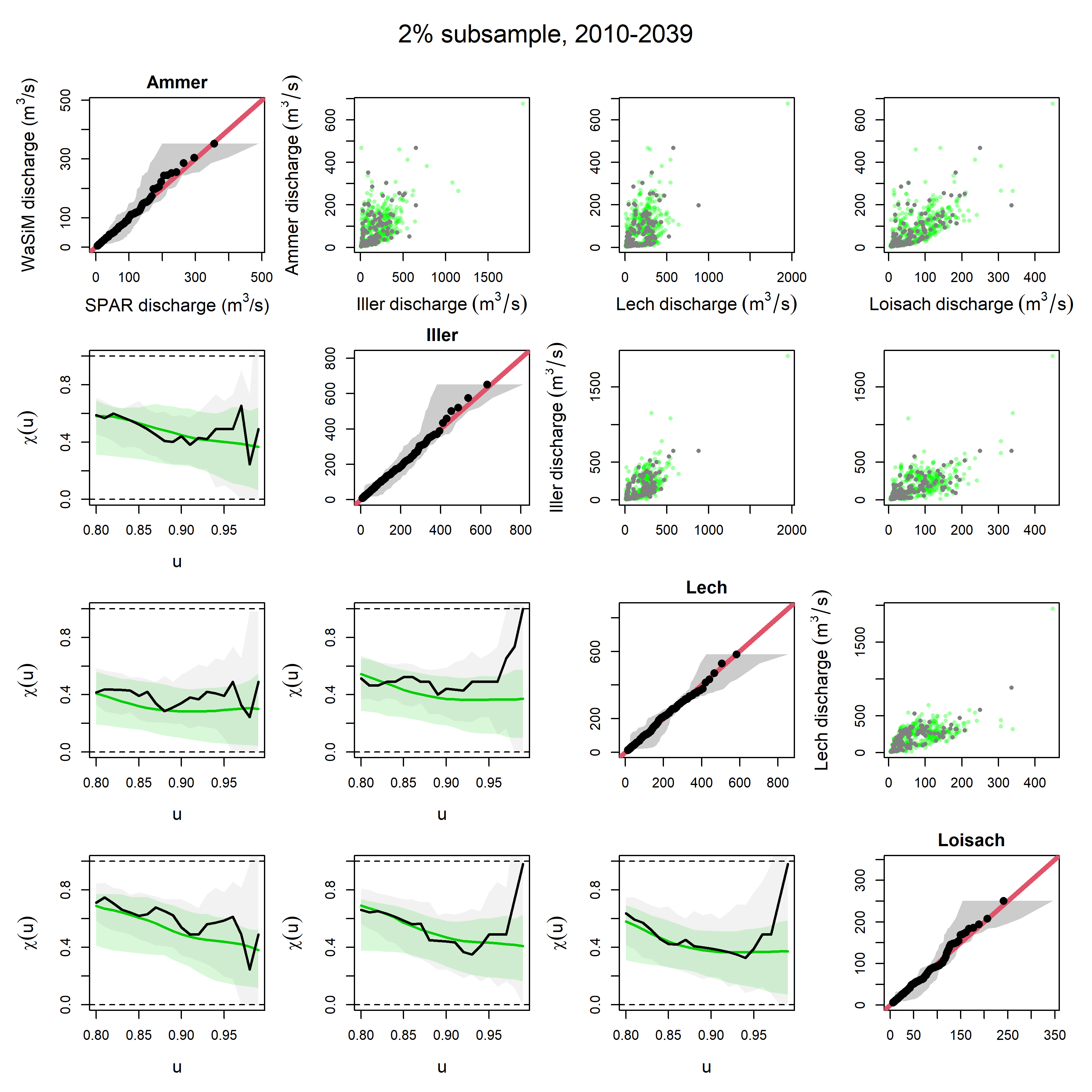}
    \end{minipage}%
     \hfill
   \begin{minipage}{0.48\textwidth} 
        \centering
        \includegraphics[width=\textwidth]{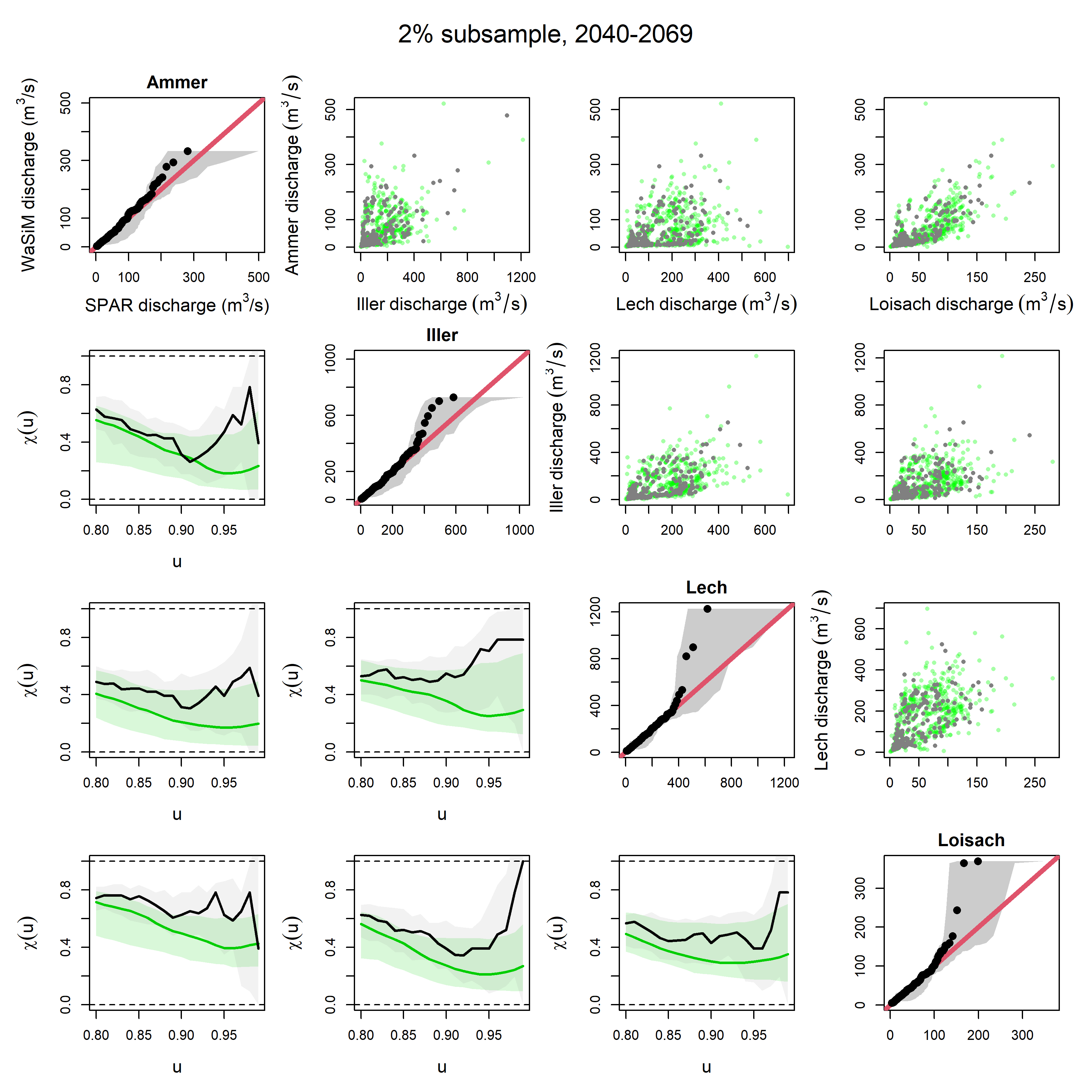}
    \end{minipage}%
    \hfill
   \begin{minipage}{0.48\textwidth} 
        \centering
        \includegraphics[width=\textwidth]{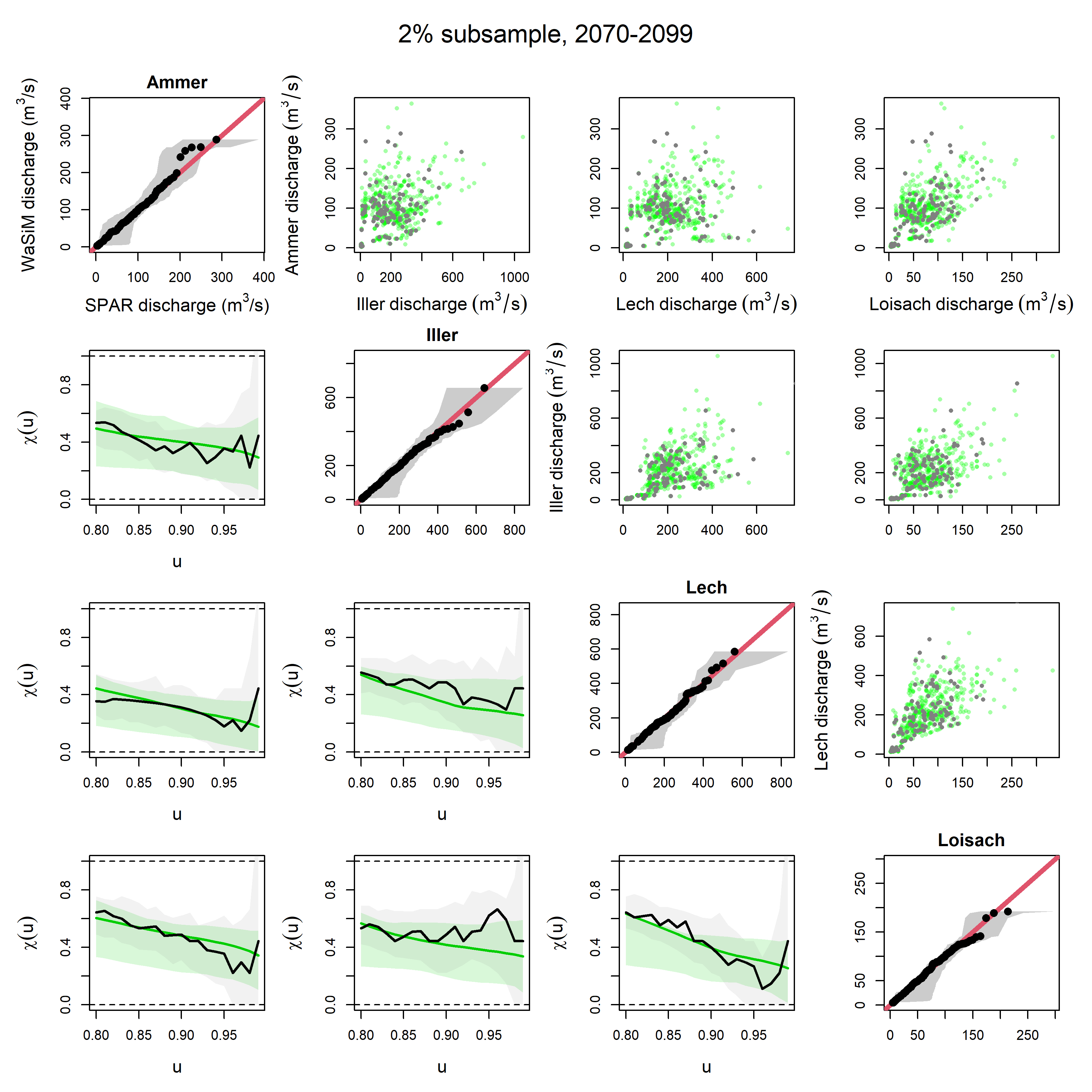}
    \end{minipage}%
    \caption{Model performance in the joint tail region $\mathbb{Q}^c_{u}$ for the subsamples containing 2\% of the full data sample. See Figure~5 in the main manuscript for details on the interpretation of individual panels.}
    \label{fig:scatters_2}
\end{figure}

\begin{figure}[h] 
    \centering
    \begin{minipage}{0.48\textwidth} 
        \centering
        \includegraphics[width=\textwidth]{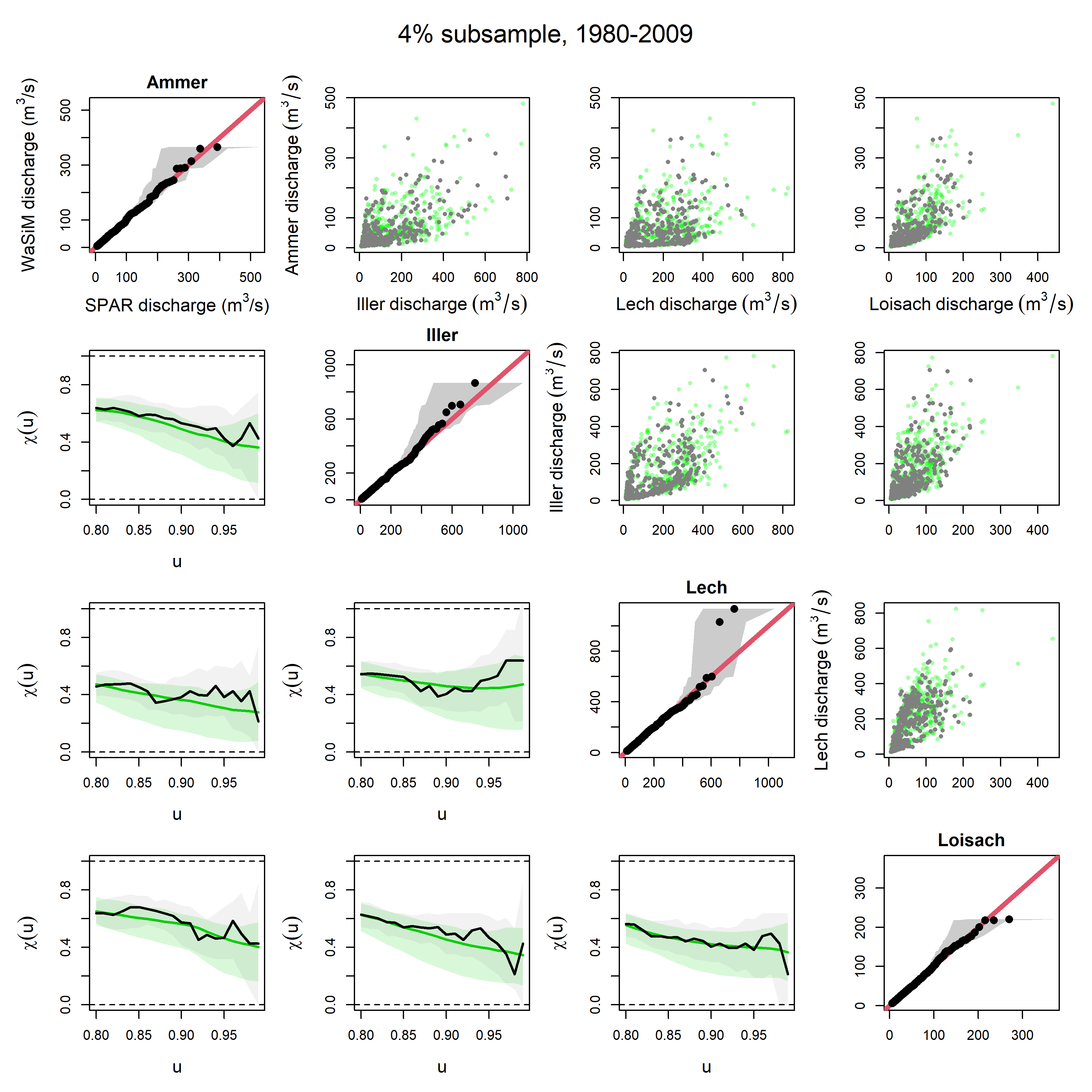}
    \end{minipage}%
    \hfill
   \begin{minipage}{0.48\textwidth} 
        \centering
        \includegraphics[width=\textwidth]{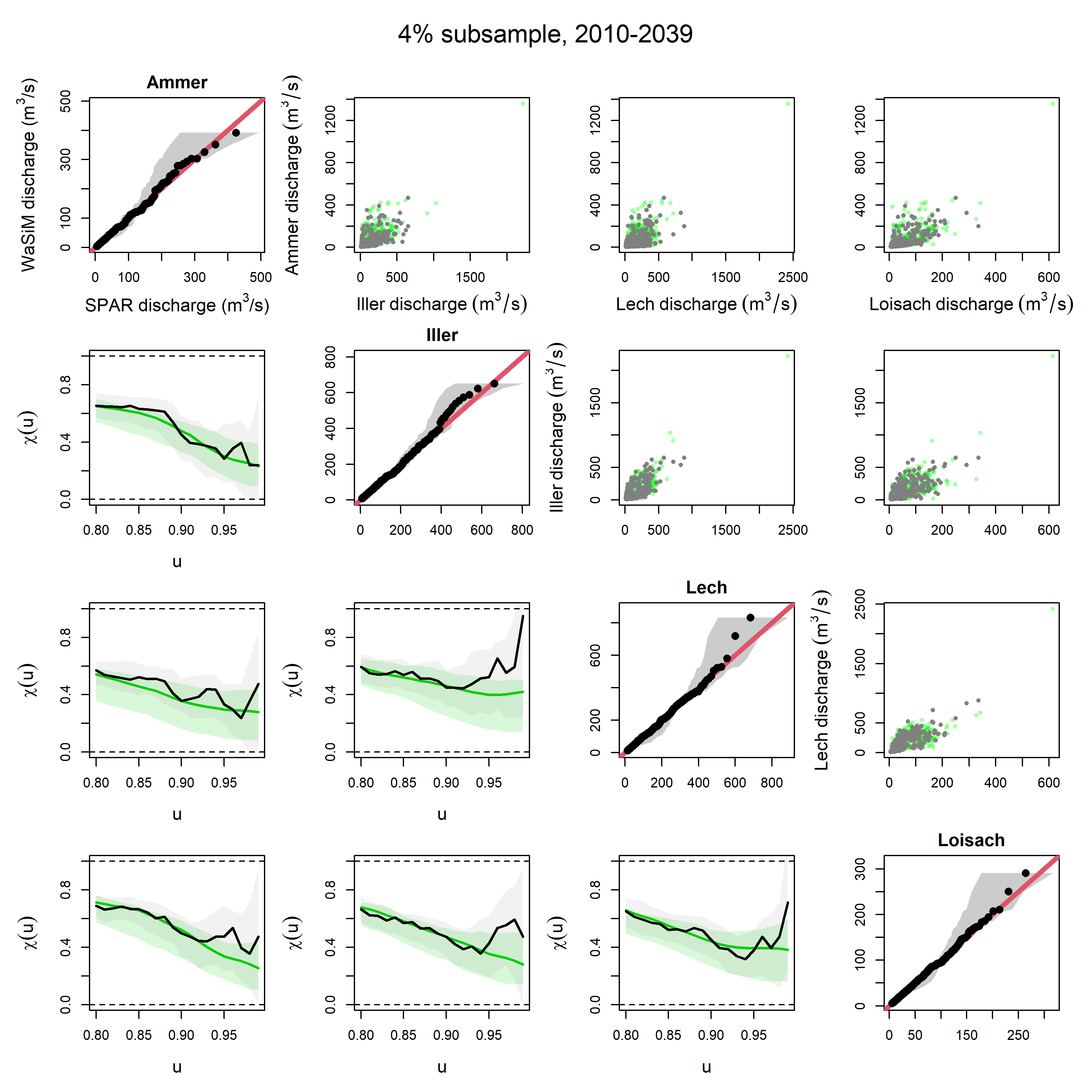}
    \end{minipage}%
     \hfill
   \begin{minipage}{0.48\textwidth} 
        \centering
        \includegraphics[width=\textwidth]{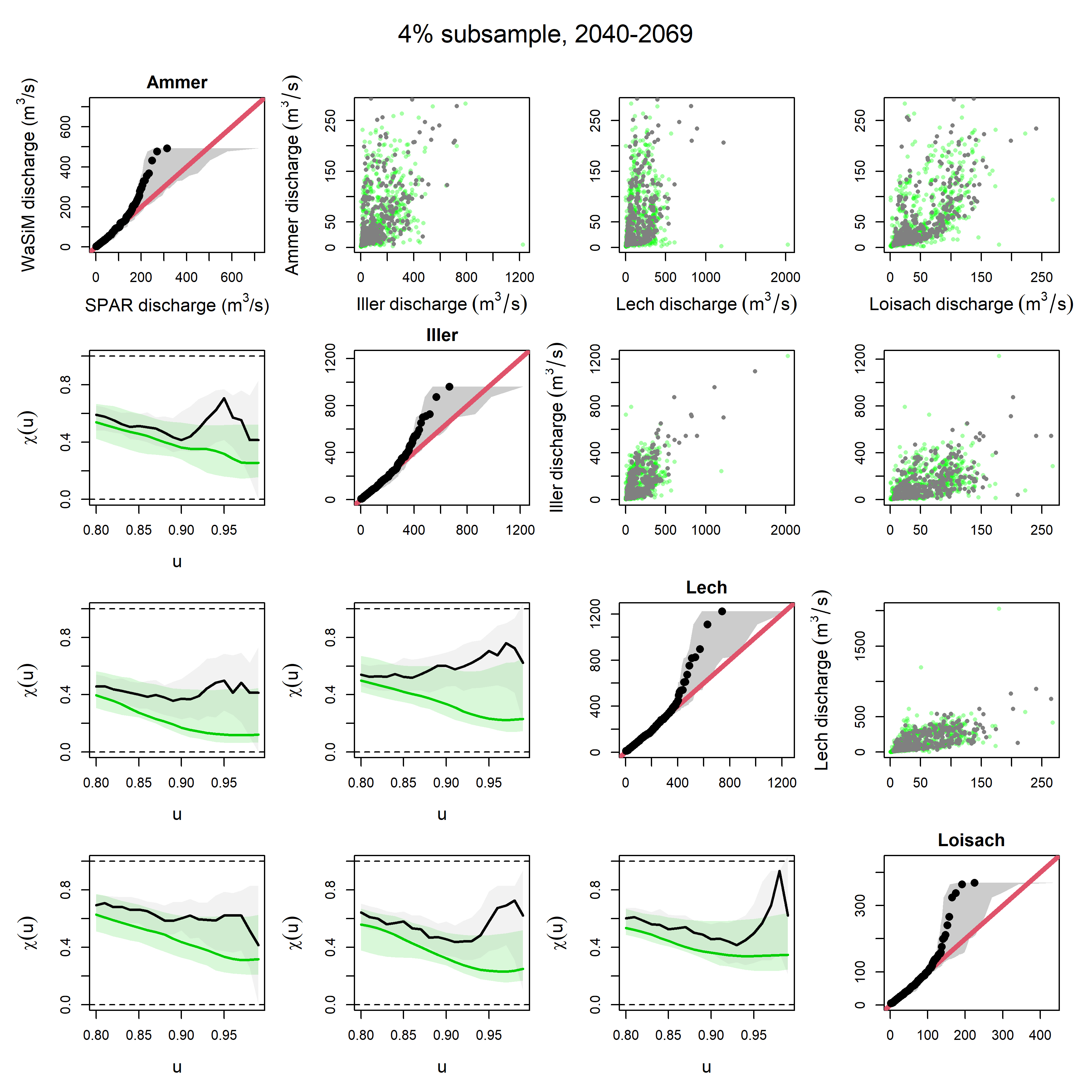}
    \end{minipage}%
    \hfill
   \begin{minipage}{0.48\textwidth} 
        \centering
        \includegraphics[width=\textwidth]{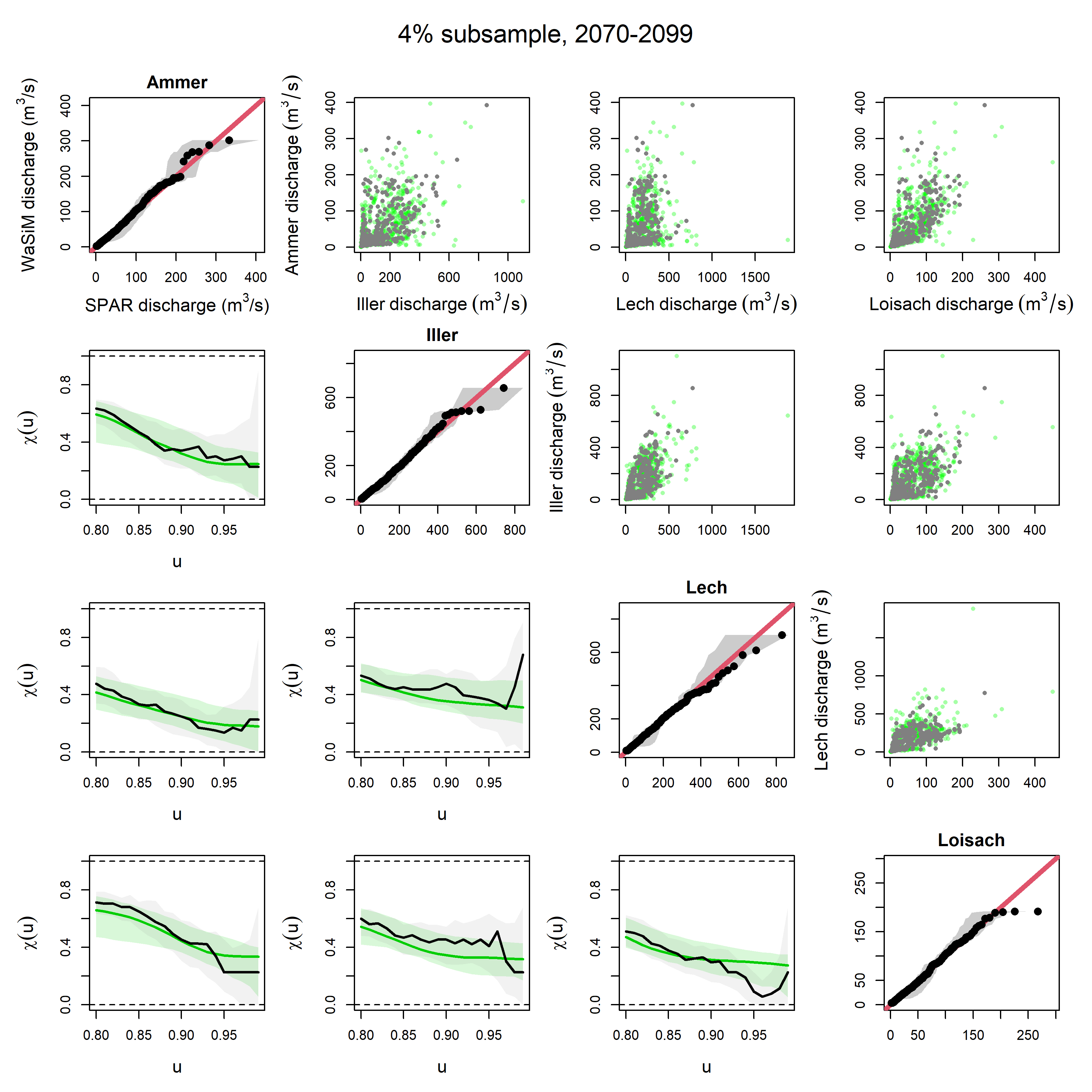}
    \end{minipage}%
    \caption{Model performance in the joint tail region $\mathbb{Q}^c_{u}$ for the subsamples containing 4\% of the full data sample. See Figure~5 in the main manuscript for details on the interpretation of individual panels.}
    \label{fig:scatters_4}
\end{figure}

\begin{figure}[h] 
    \centering
    \begin{minipage}{0.48\textwidth} 
        \centering
        \includegraphics[width=\textwidth]{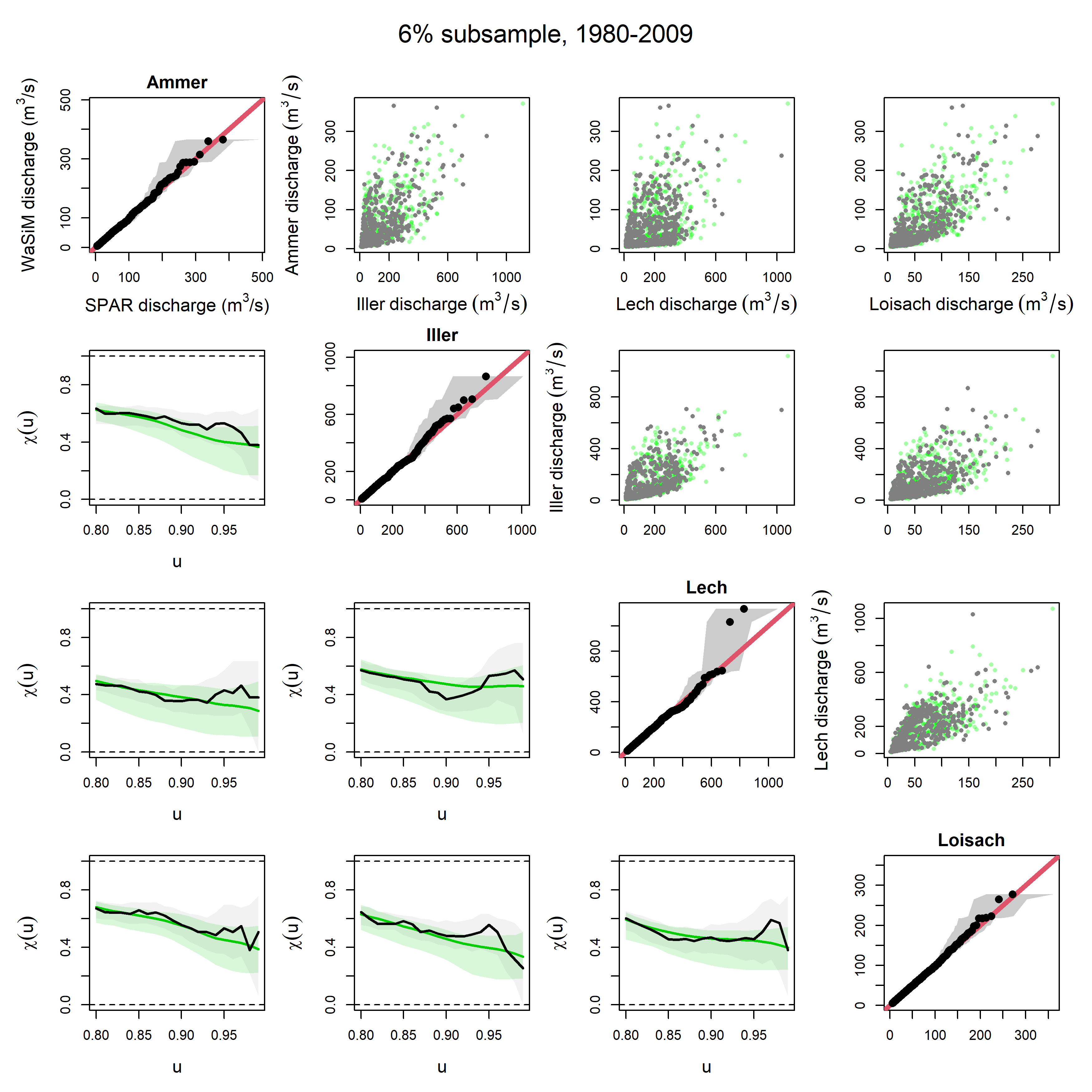}
    \end{minipage}%
    \hfill
   \begin{minipage}{0.48\textwidth} 
        \centering
        \includegraphics[width=\textwidth]{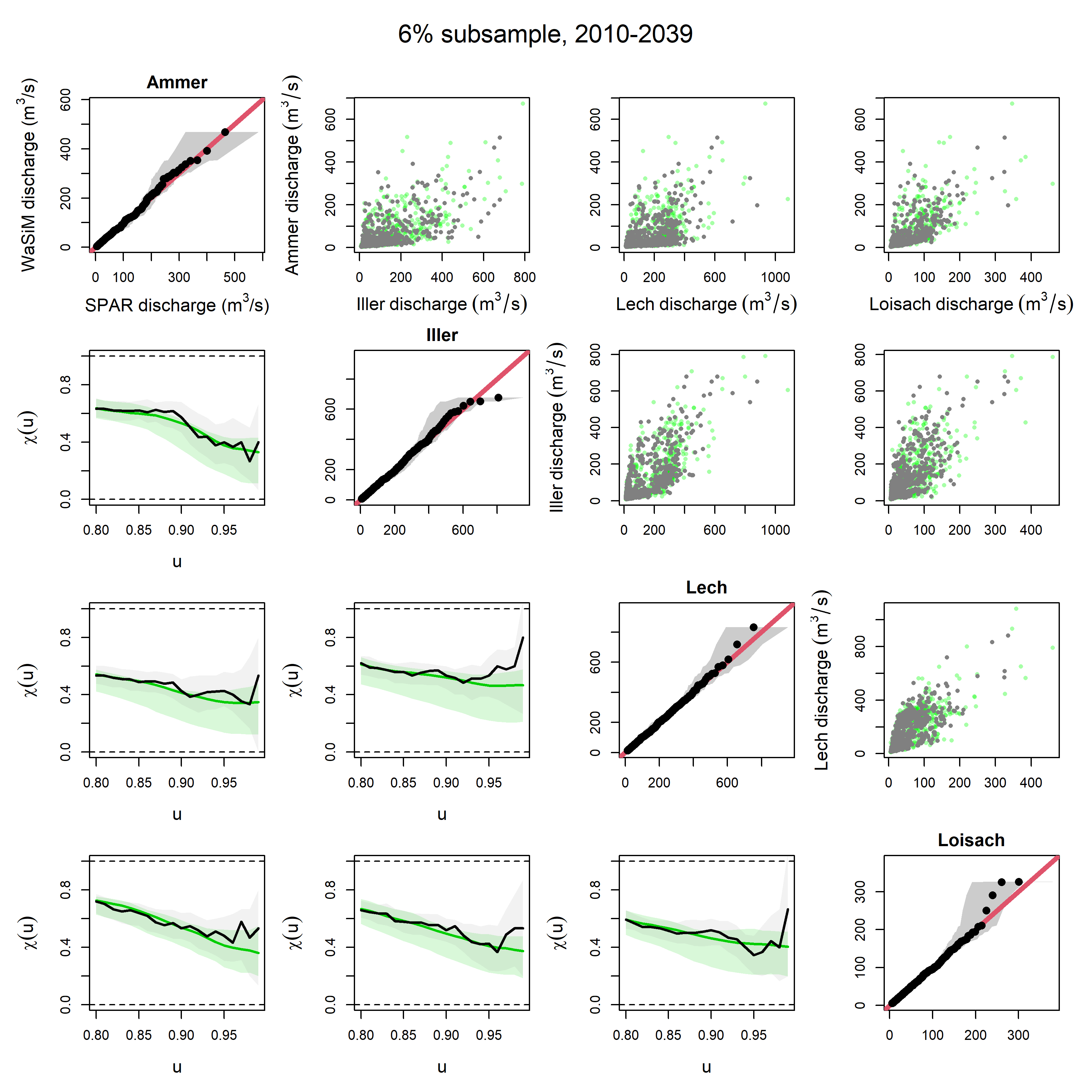}
    \end{minipage}%
     \hfill
   \begin{minipage}{0.48\textwidth} 
        \centering
        \includegraphics[width=\textwidth]{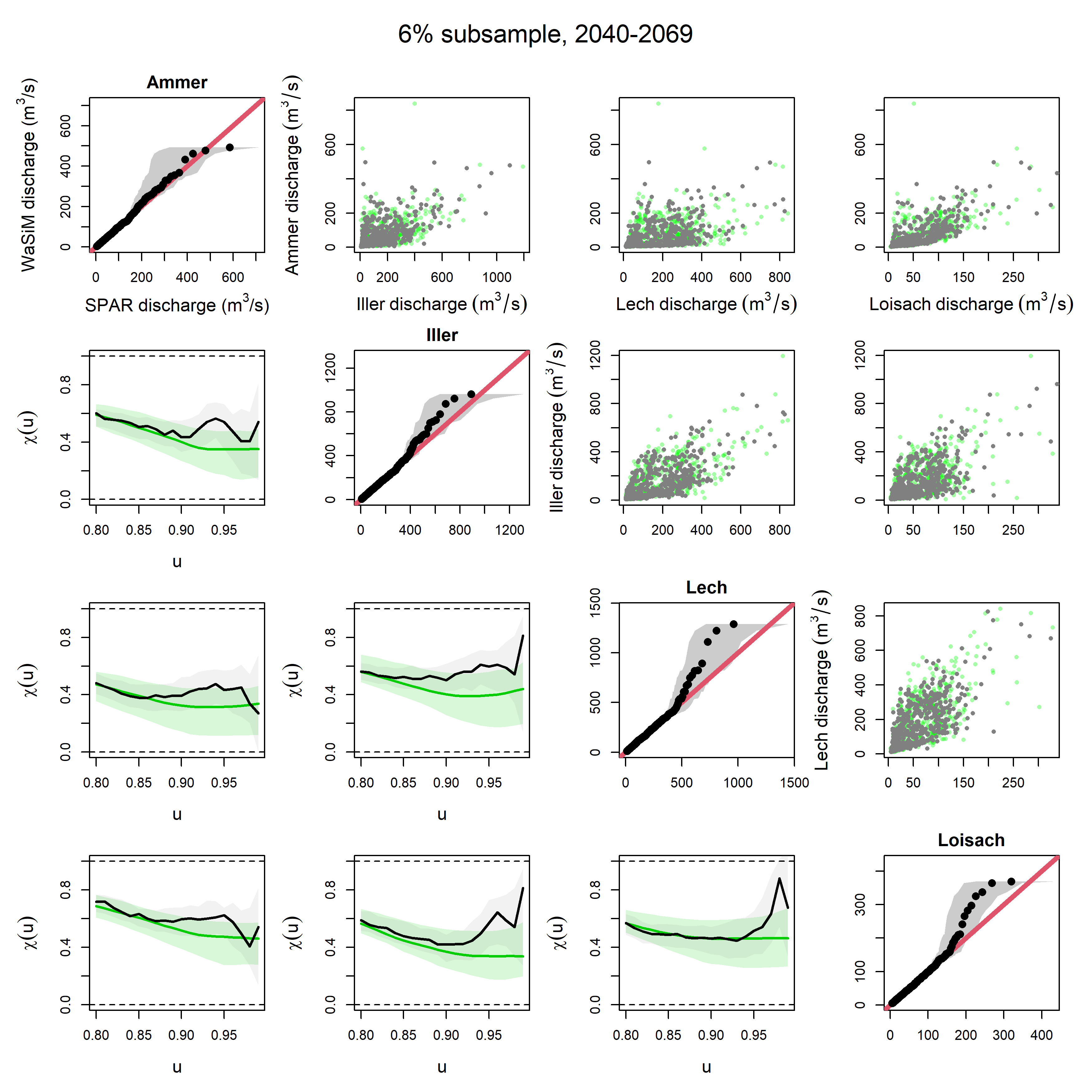}
    \end{minipage}%
    \hfill
   \begin{minipage}{0.48\textwidth} 
        \centering
        \includegraphics[width=\textwidth]{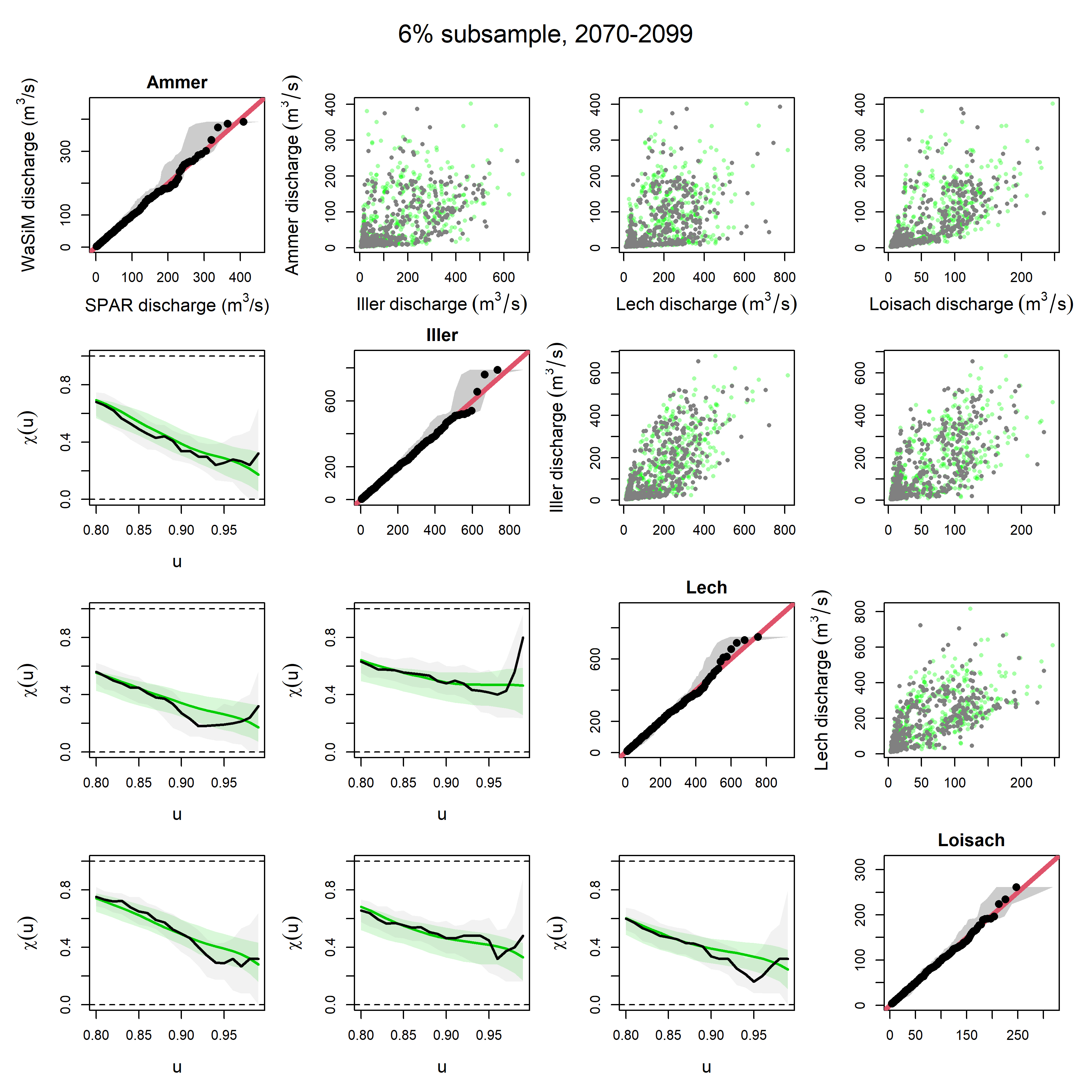}
    \end{minipage}%
    \caption{Model performance in the joint tail region $\mathbb{Q}^c_{u}$ for the subsamples containing 6\% of the full data sample. See Figure~5 in the main manuscript for details on the interpretation of individual panels.}
    \label{fig:scatters_6}
\end{figure}

\begin{figure}[h] 
    \centering
    \begin{minipage}{0.48\textwidth} 
        \centering
        \includegraphics[width=\textwidth]{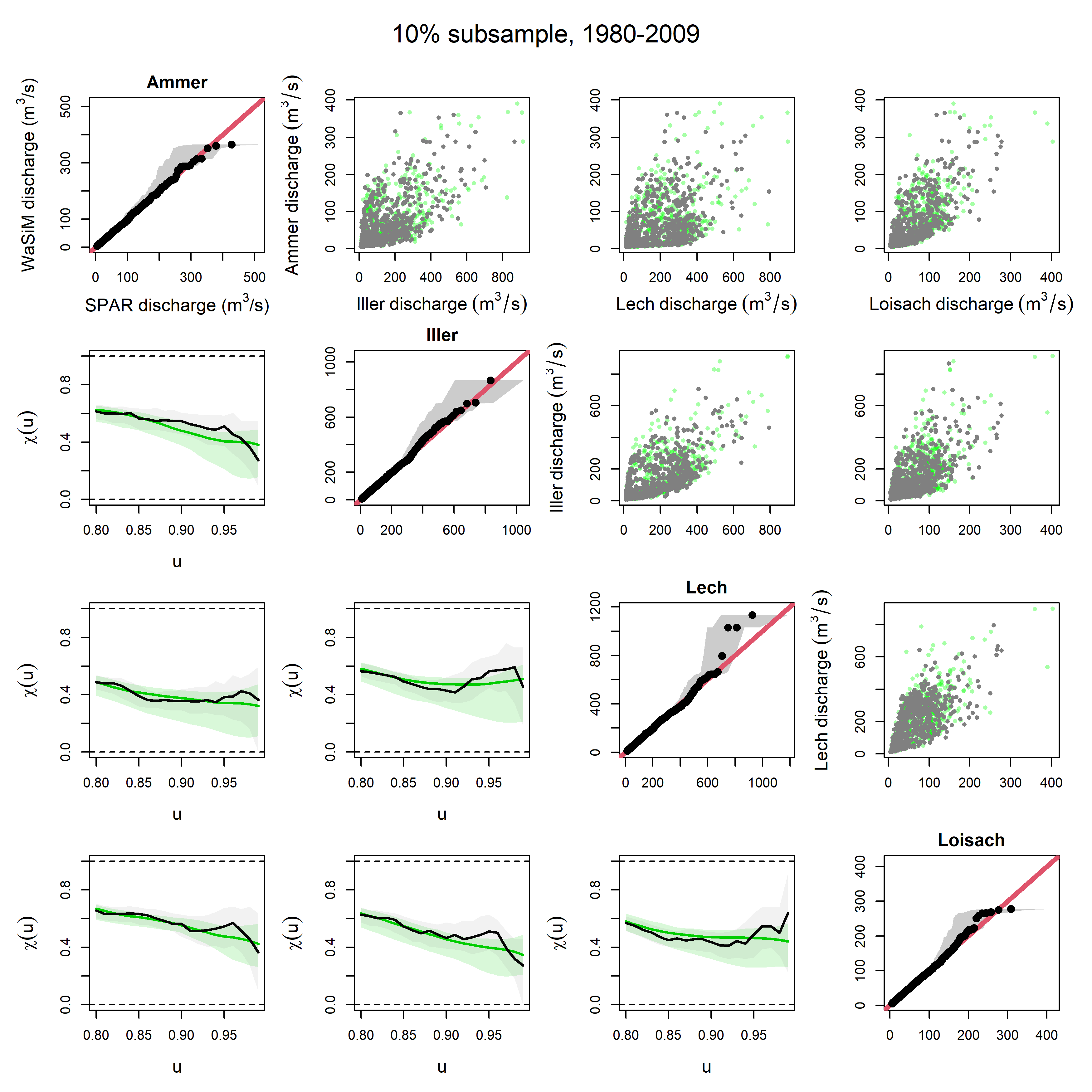}
    \end{minipage}%
    \hfill
   \begin{minipage}{0.48\textwidth} 
        \centering
        \includegraphics[width=\textwidth]{Figs/Scatterplots/scatterplot_tw_2_nmodel_5_cis.png}
    \end{minipage}%
     \hfill
   \begin{minipage}{0.48\textwidth} 
        \centering
        \includegraphics[width=\textwidth]{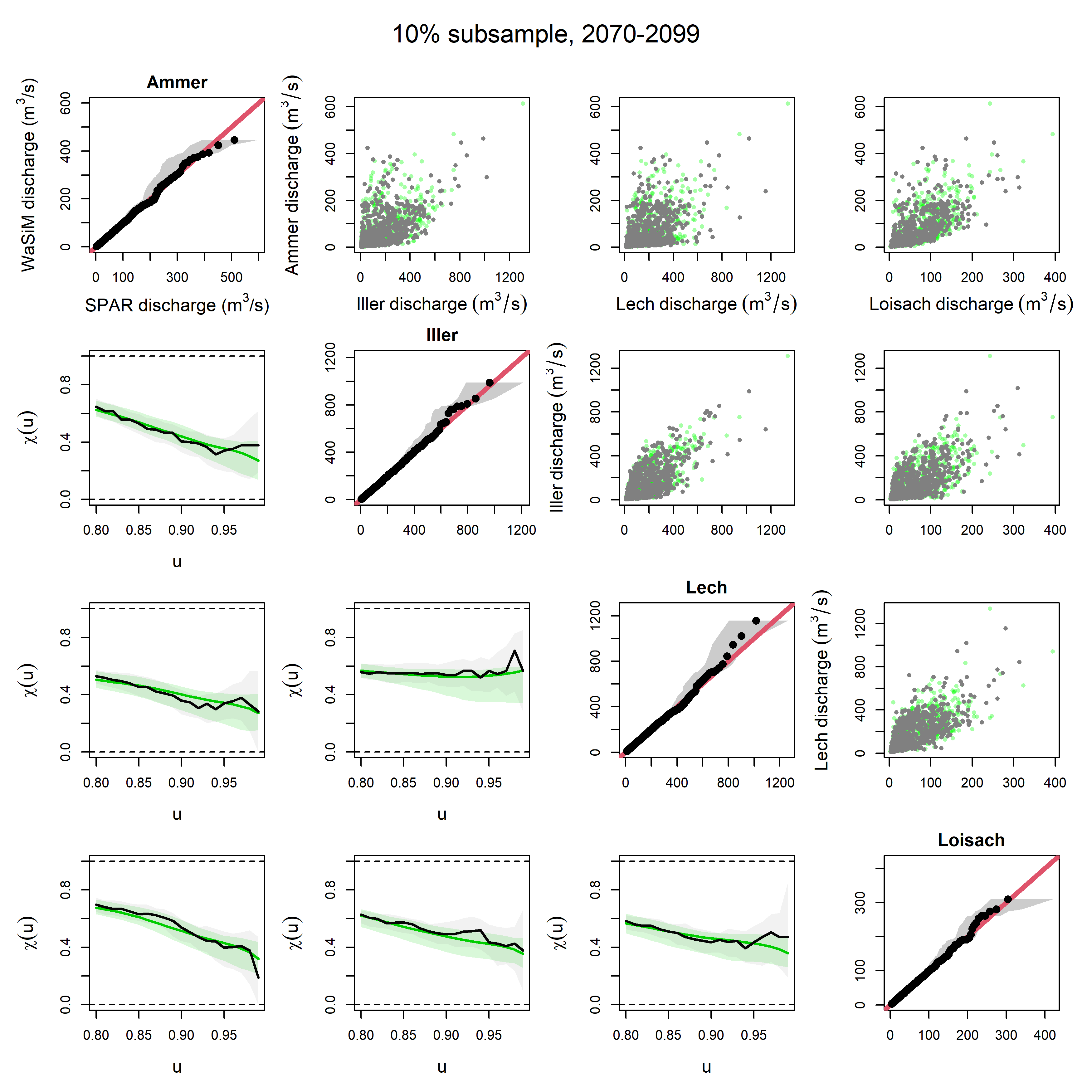}
    \end{minipage}%
    \caption{Model performance in the joint tail region $\mathbb{Q}^c_{u}$ for the subsamples containing 10\% of the full data sample. See Figure~5 in the main manuscript for details on the interpretation of individual panels.}
    \label{fig:scatters_10}
\end{figure}

\begin{figure}[h] 
    \centering
    \begin{minipage}{0.48\textwidth} 
        \centering
        \includegraphics[width=\textwidth]{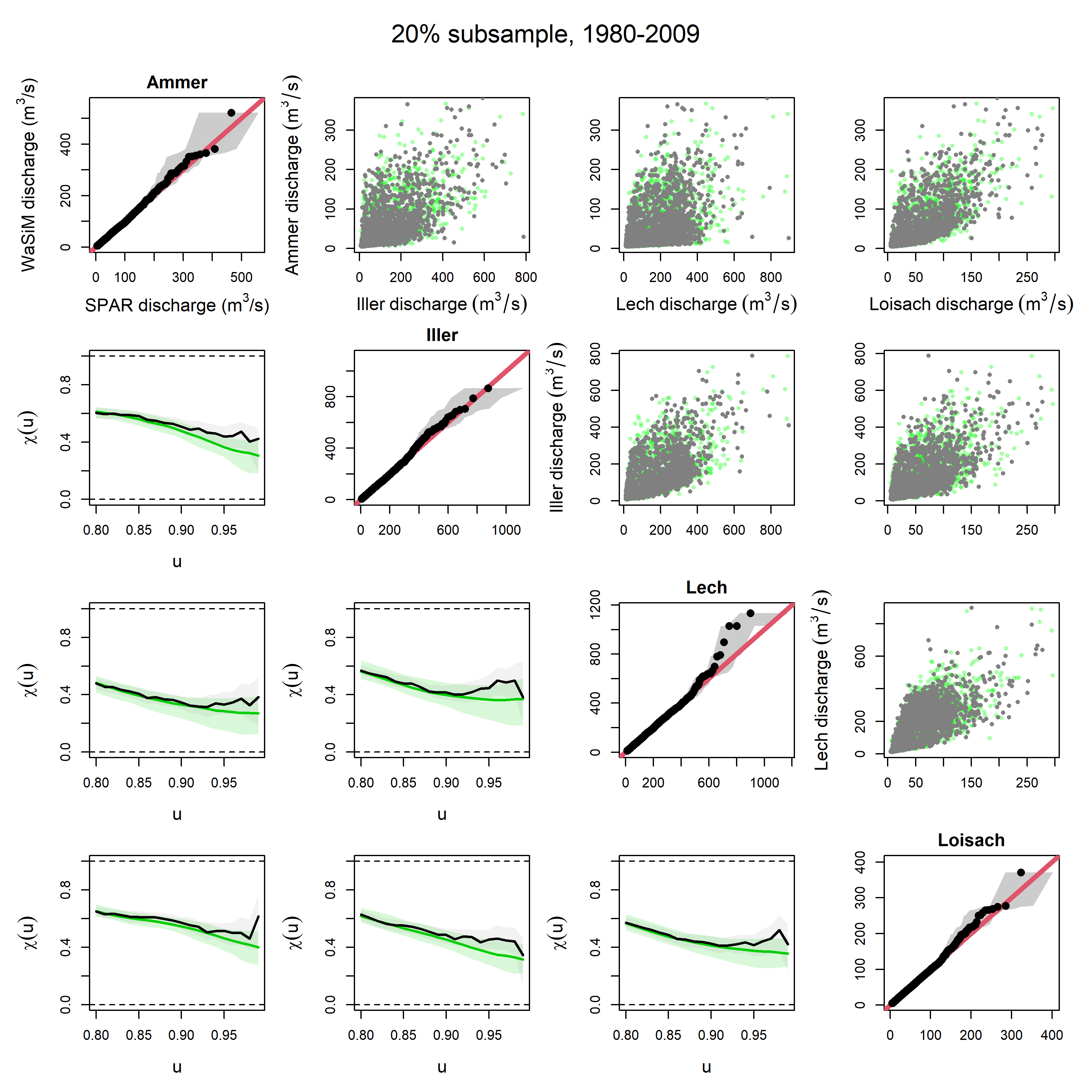}
    \end{minipage}%
    \hfill
   \begin{minipage}{0.48\textwidth} 
        \centering
        \includegraphics[width=\textwidth]{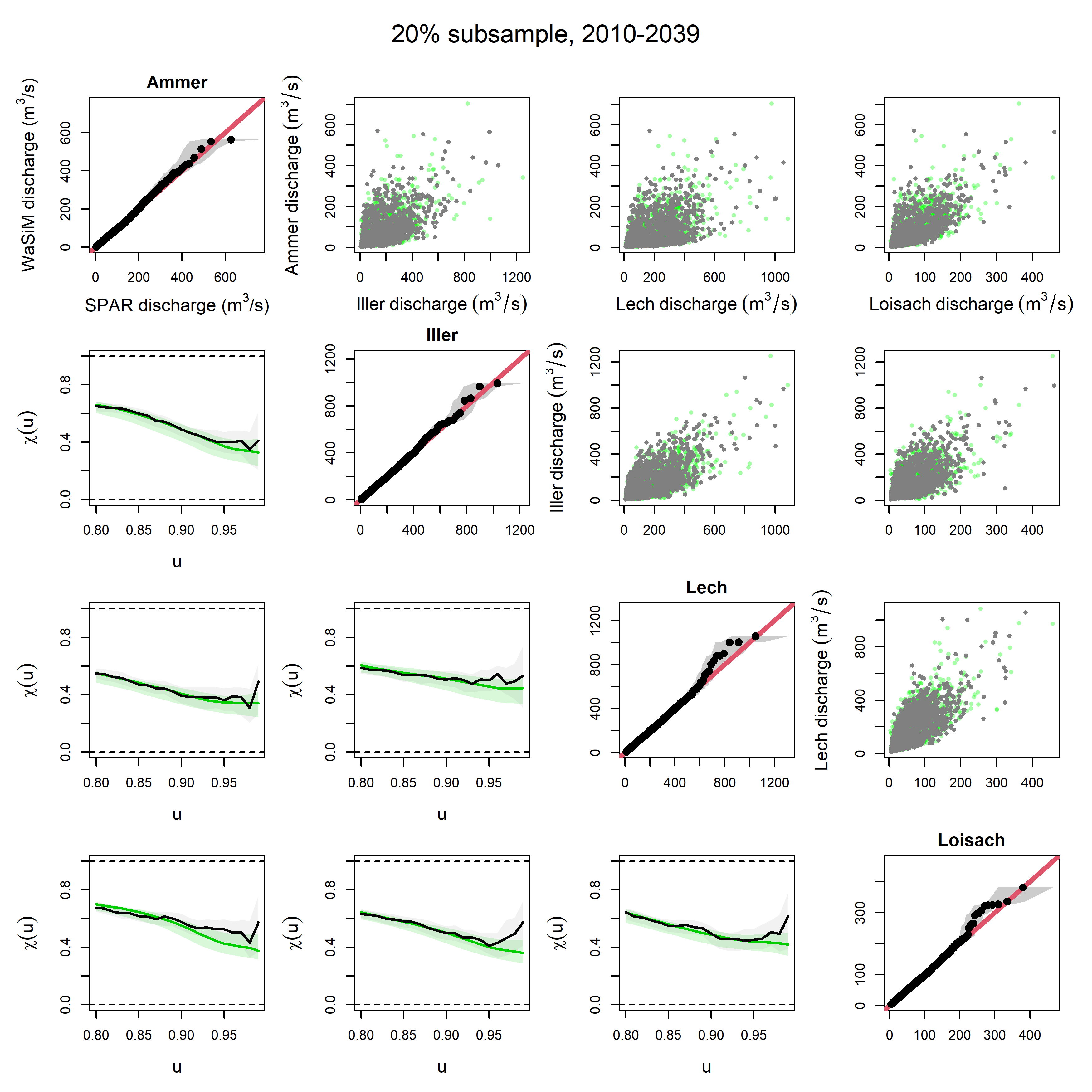}
    \end{minipage}%
     \hfill
   \begin{minipage}{0.48\textwidth} 
        \centering
        \includegraphics[width=\textwidth]{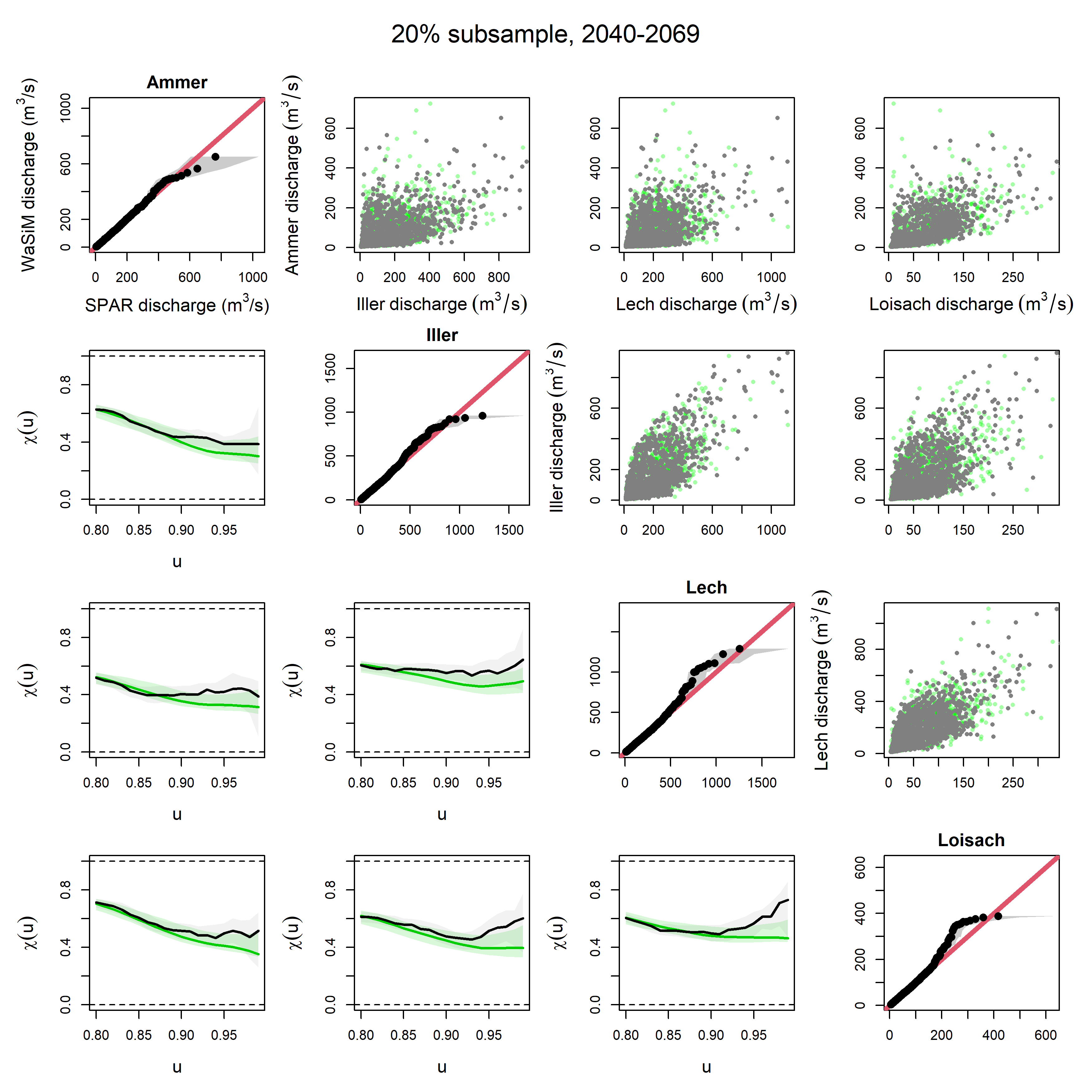}
    \end{minipage}%
    \hfill
   \begin{minipage}{0.48\textwidth} 
        \centering
        \includegraphics[width=\textwidth]{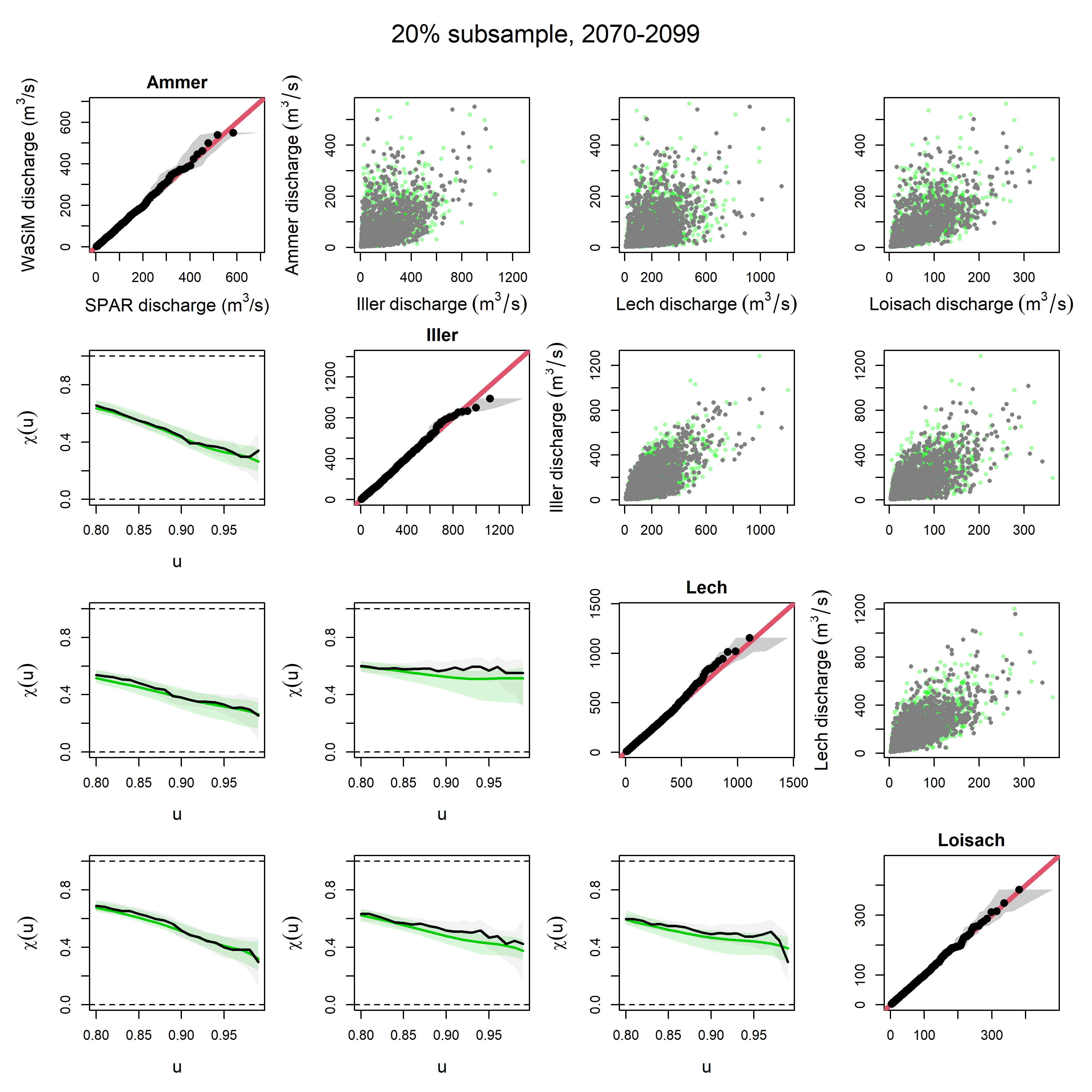}
    \end{minipage}%
    \caption{Model performance in the joint tail region $\mathbb{Q}^c_{u}$ for the subsamples containing 20\% of the full data sample. See Figure~5 in the main manuscript for details on the interpretation of individual panels.}
    \label{fig:scatters_20}
\end{figure}

\begin{figure}[h] 
    \centering
    \begin{minipage}{0.48\textwidth} 
        \centering
        \includegraphics[width=\textwidth]{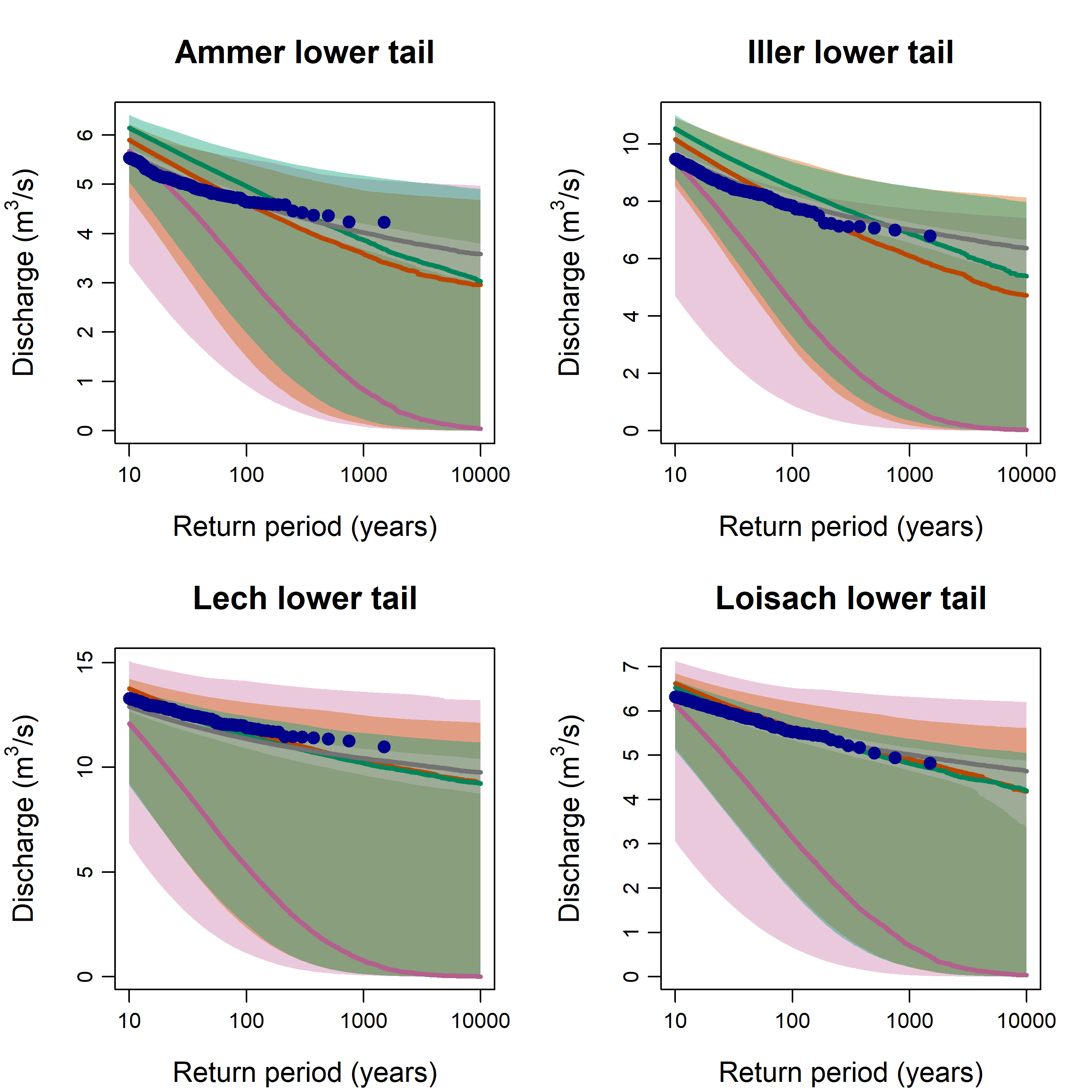}
    \end{minipage}%
    \hfill
   \begin{minipage}{0.48\textwidth} 
        \centering
        \includegraphics[width=\textwidth]{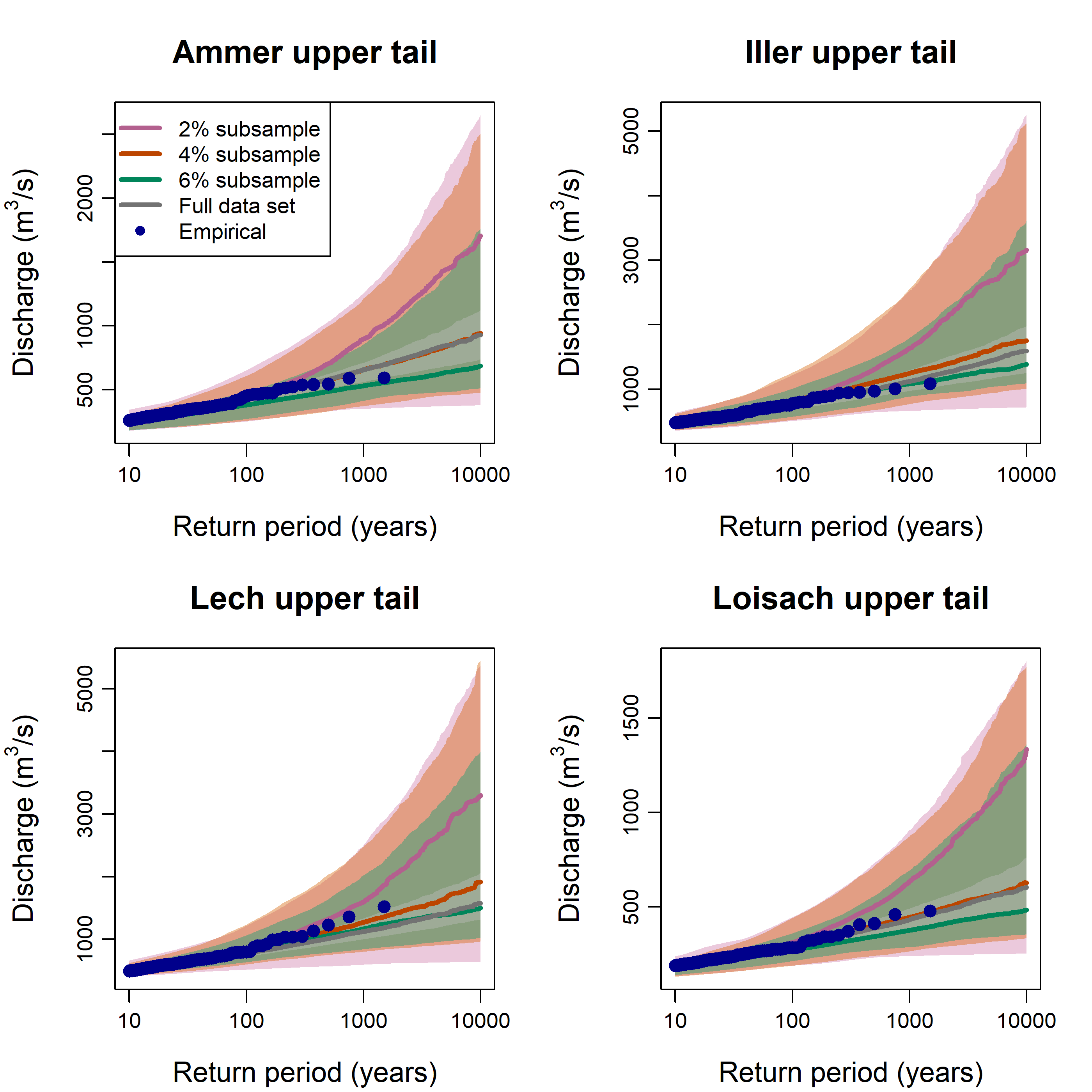}
    \end{minipage}%
    \hfill
    \begin{minipage}{0.48\textwidth} 
        \centering
        \includegraphics[width=\textwidth]{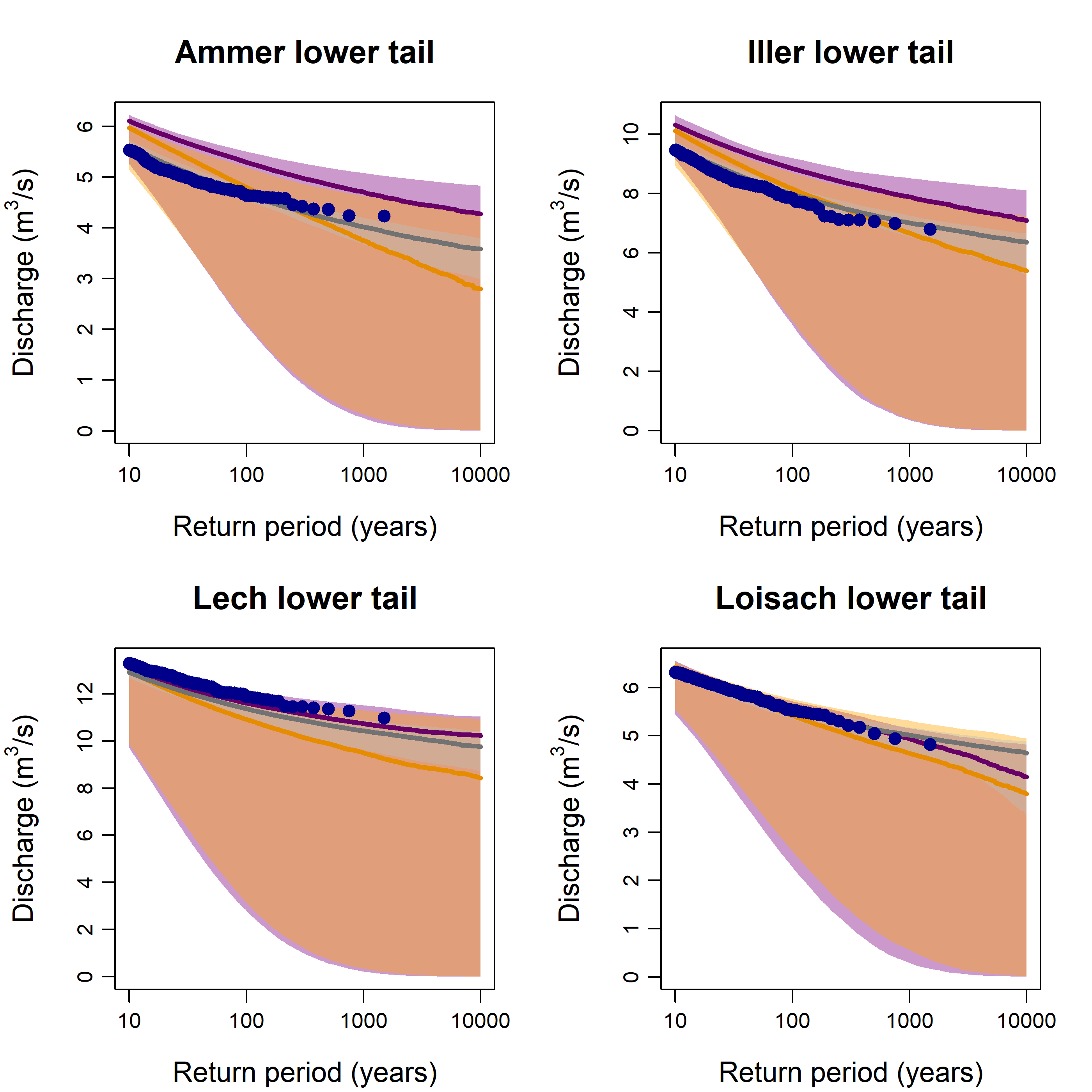}
    \end{minipage}%
    \hfill
   \begin{minipage}{0.48\textwidth} 
        \centering
        \includegraphics[width=\textwidth]{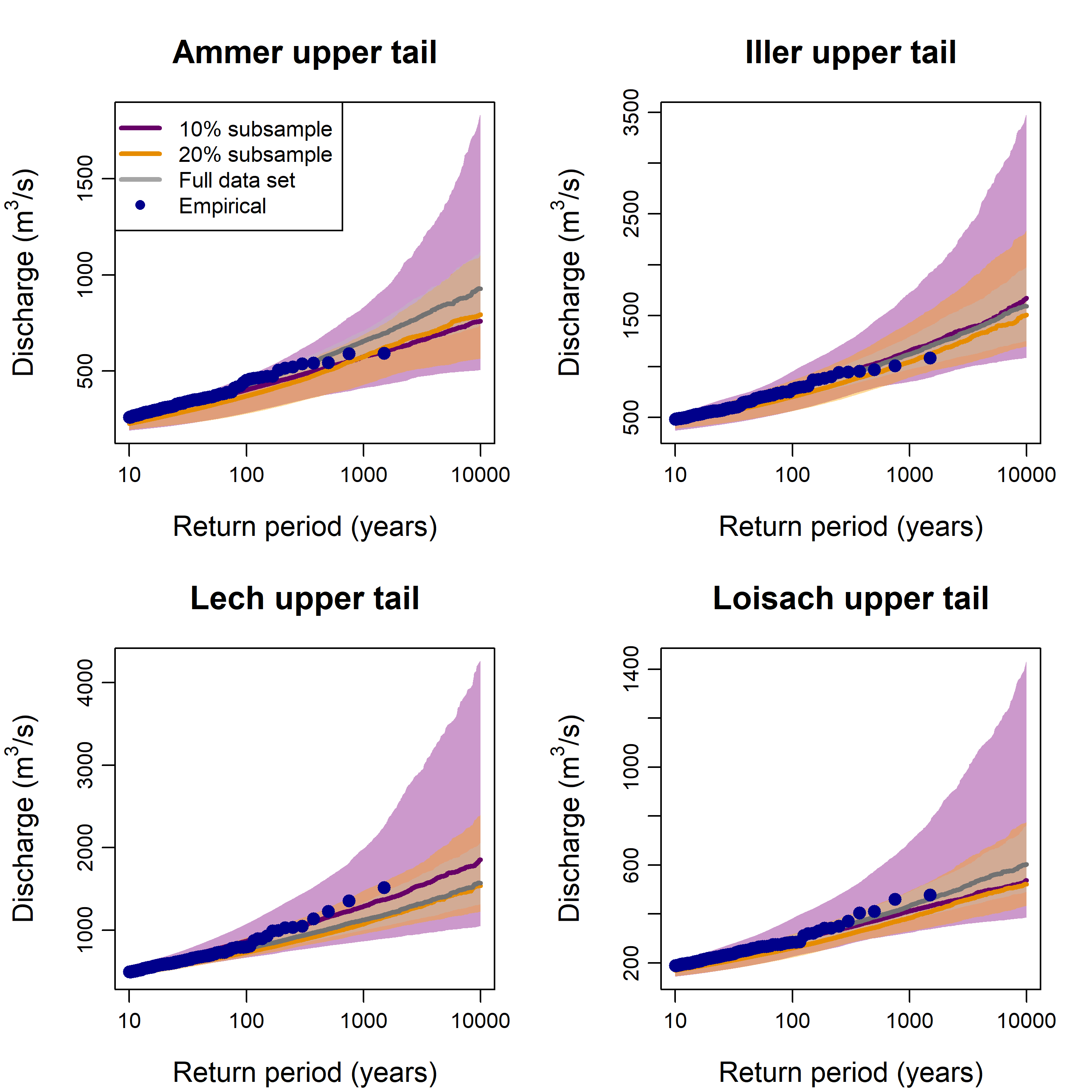}
    \end{minipage}%
    \caption{Comparison of return level estimates from different sample sizes in the lower and upper tails of each marginal variable for the 1980--2009 time period. See Figure~8 in the main manuscript for details on the interpretation of individual panels.}
    \label{fig:ret_level_comp_1}
\end{figure}

\begin{figure}[h] 
    \centering
    \begin{minipage}{0.48\textwidth} 
        \centering
        \includegraphics[width=\textwidth]{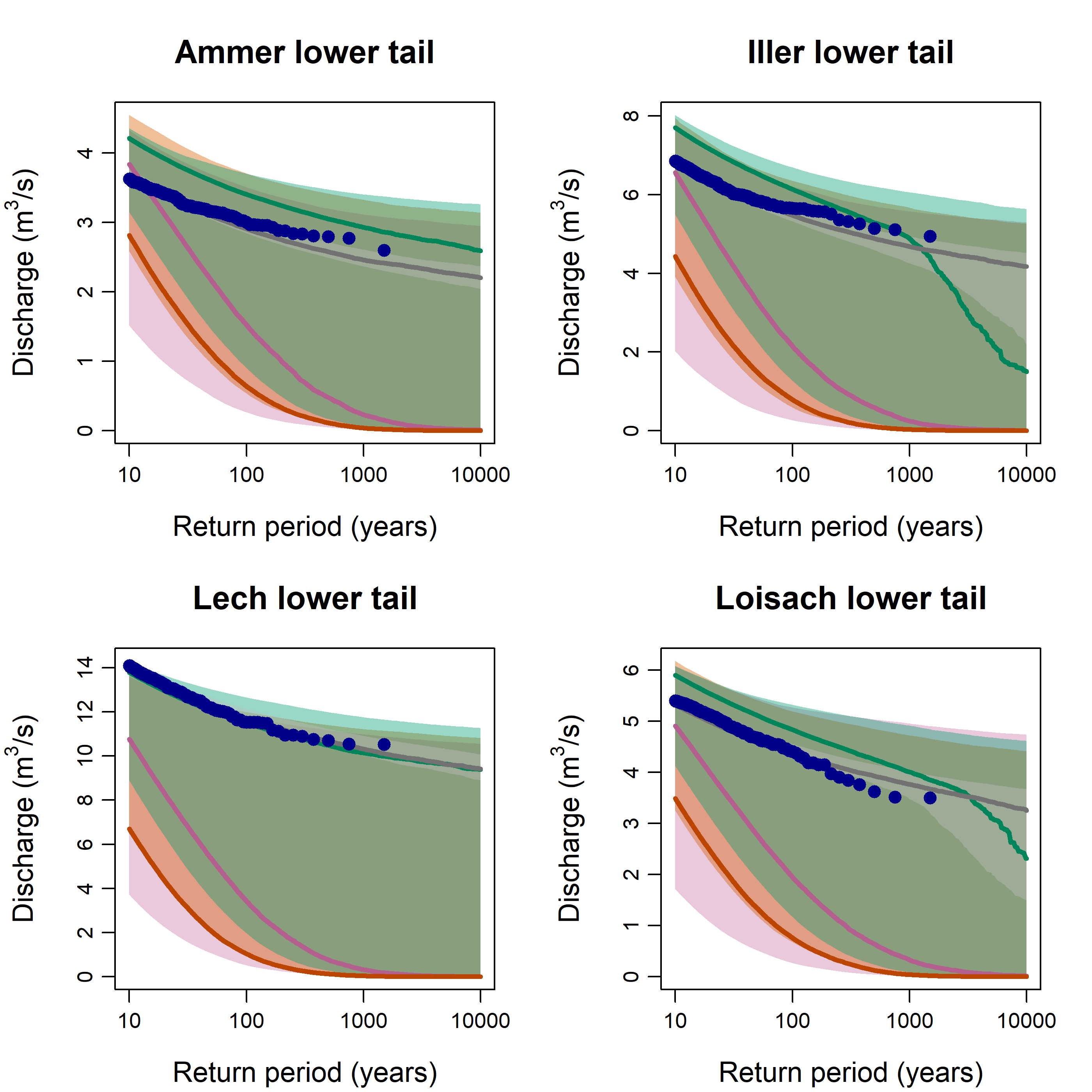}
    \end{minipage}%
    \hfill
   \begin{minipage}{0.48\textwidth} 
        \centering
        \includegraphics[width=\textwidth]{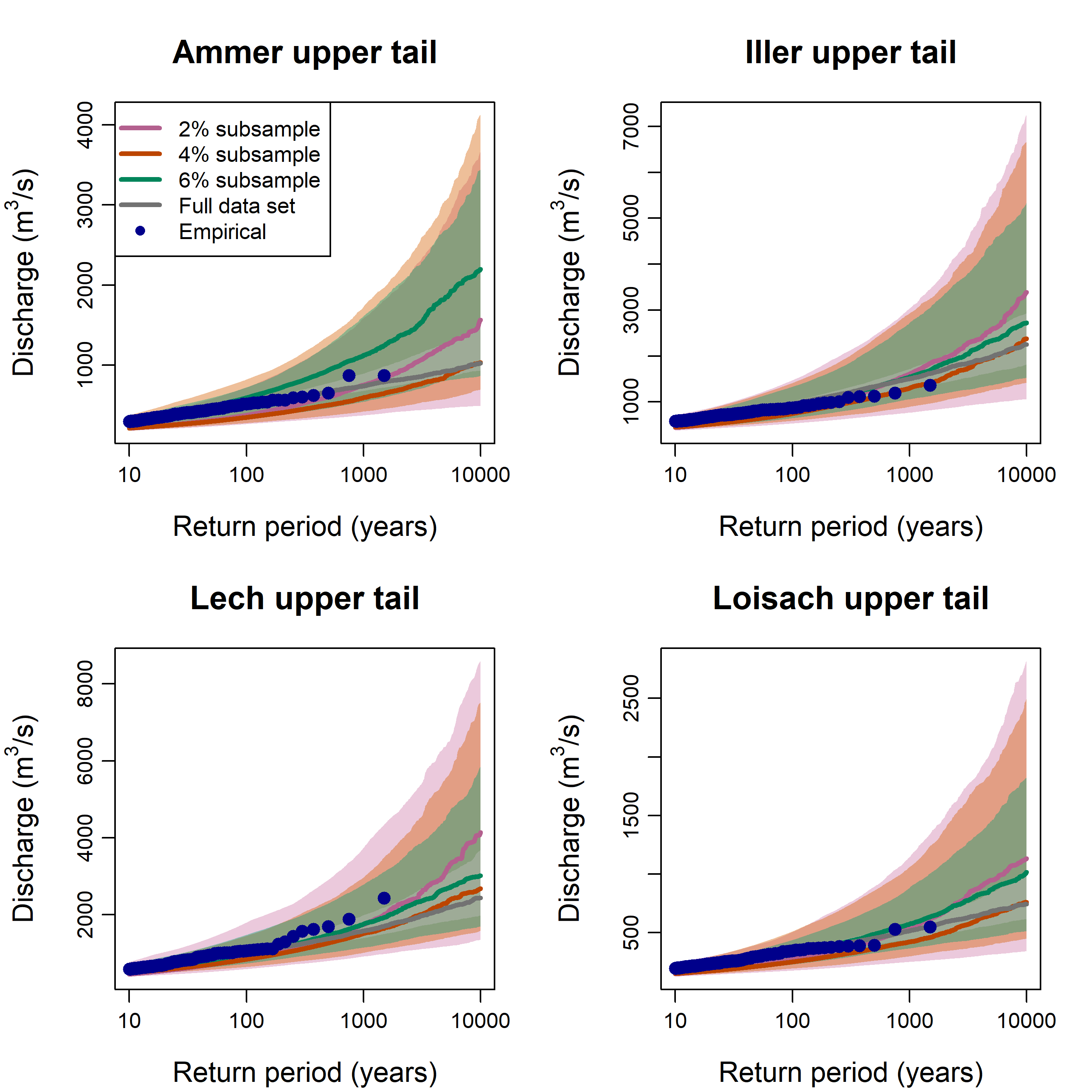}
    \end{minipage}%
    \hfill
    \begin{minipage}{0.48\textwidth} 
        \centering
        \includegraphics[width=\textwidth]{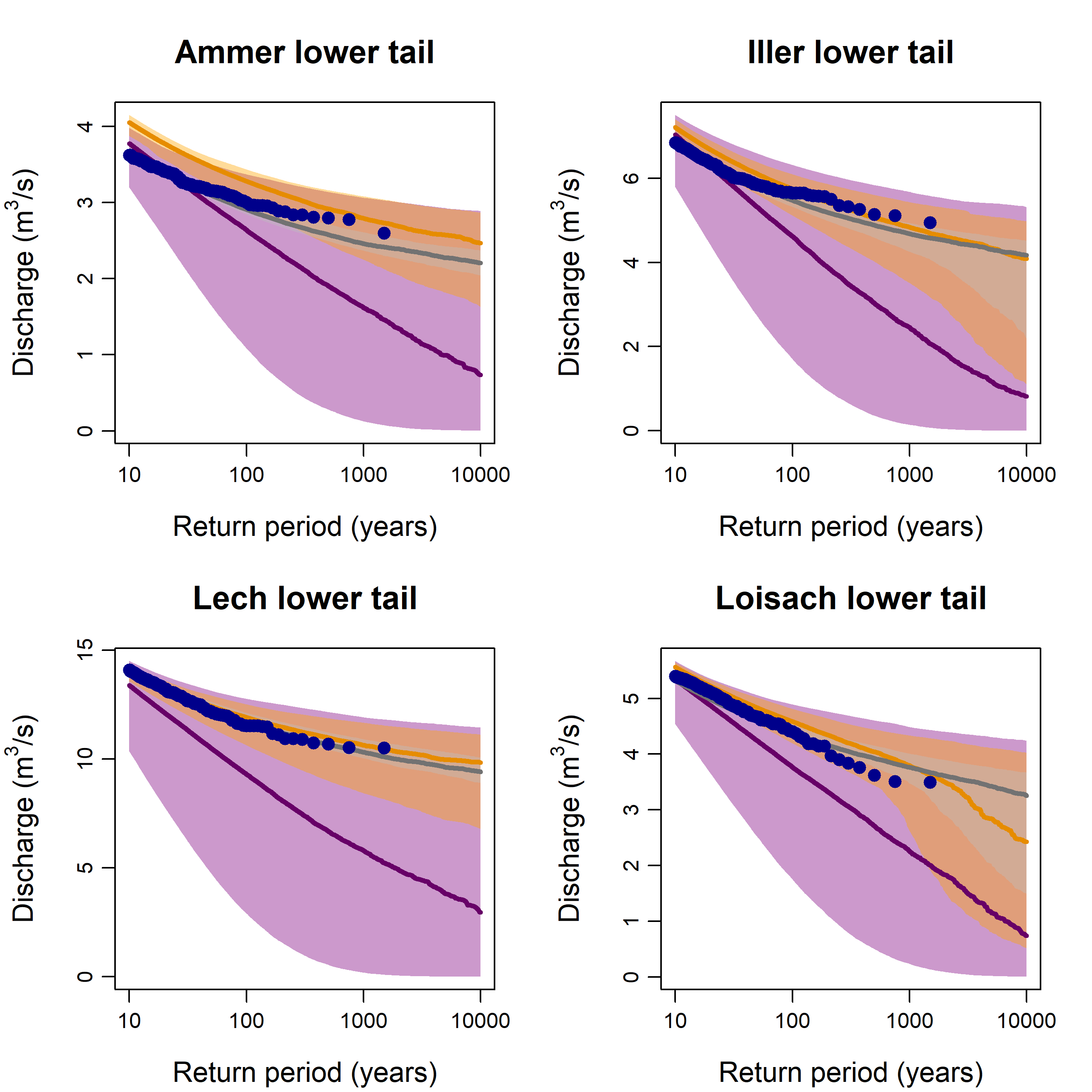}
    \end{minipage}%
    \hfill
   \begin{minipage}{0.48\textwidth} 
        \centering
        \includegraphics[width=\textwidth]{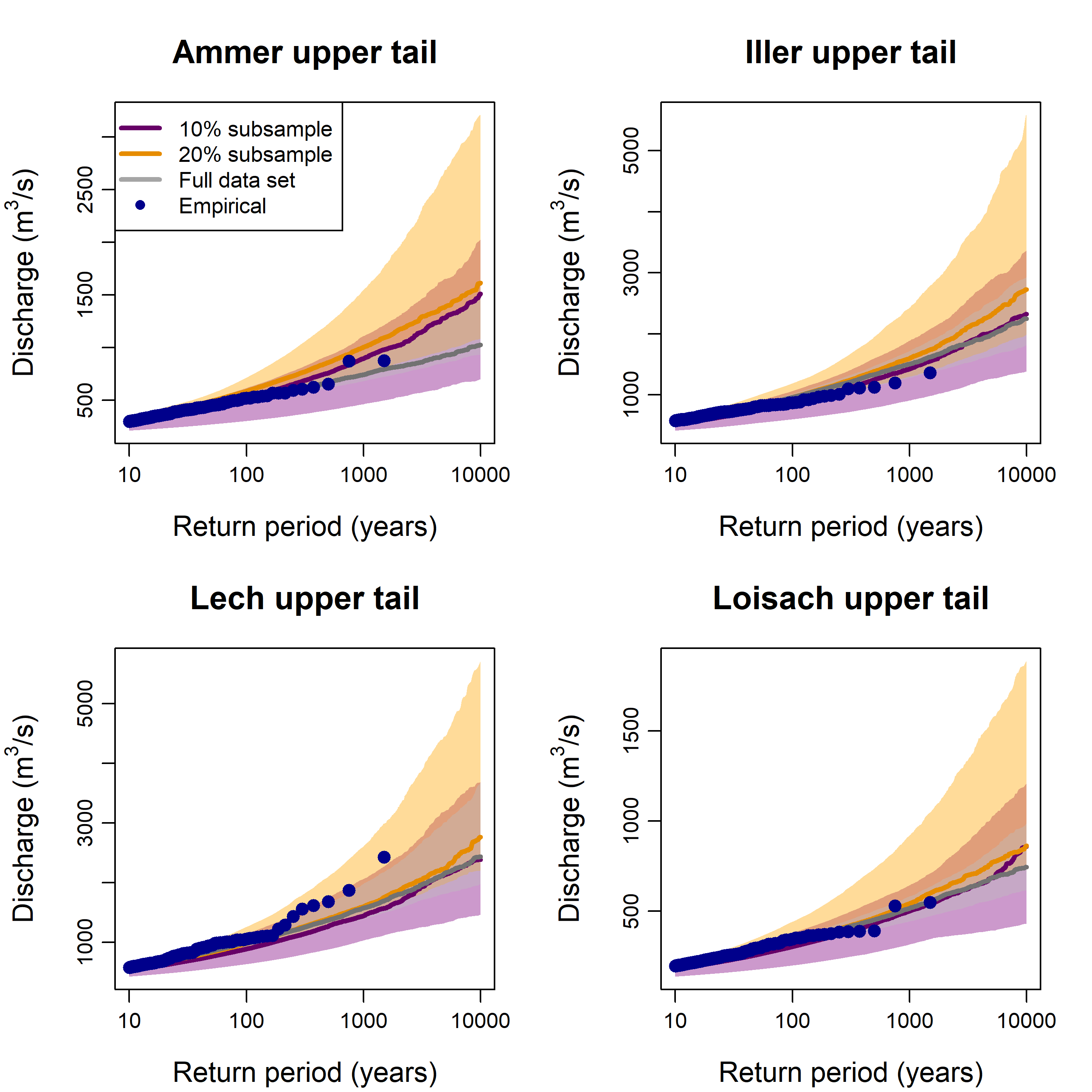}
    \end{minipage}%
    \caption{Comparison of return level estimates from different sample sizes in the lower and upper tails of each marginal variable for the 2040--2069 time period. See Figure~8 in the main manuscript for details on the interpretation of individual panels.}
    \label{fig:ret_level_comp_3}
\end{figure}

\begin{figure}[h] 
    \centering
    \begin{minipage}{0.48\textwidth} 
        \centering
        \includegraphics[width=\textwidth]{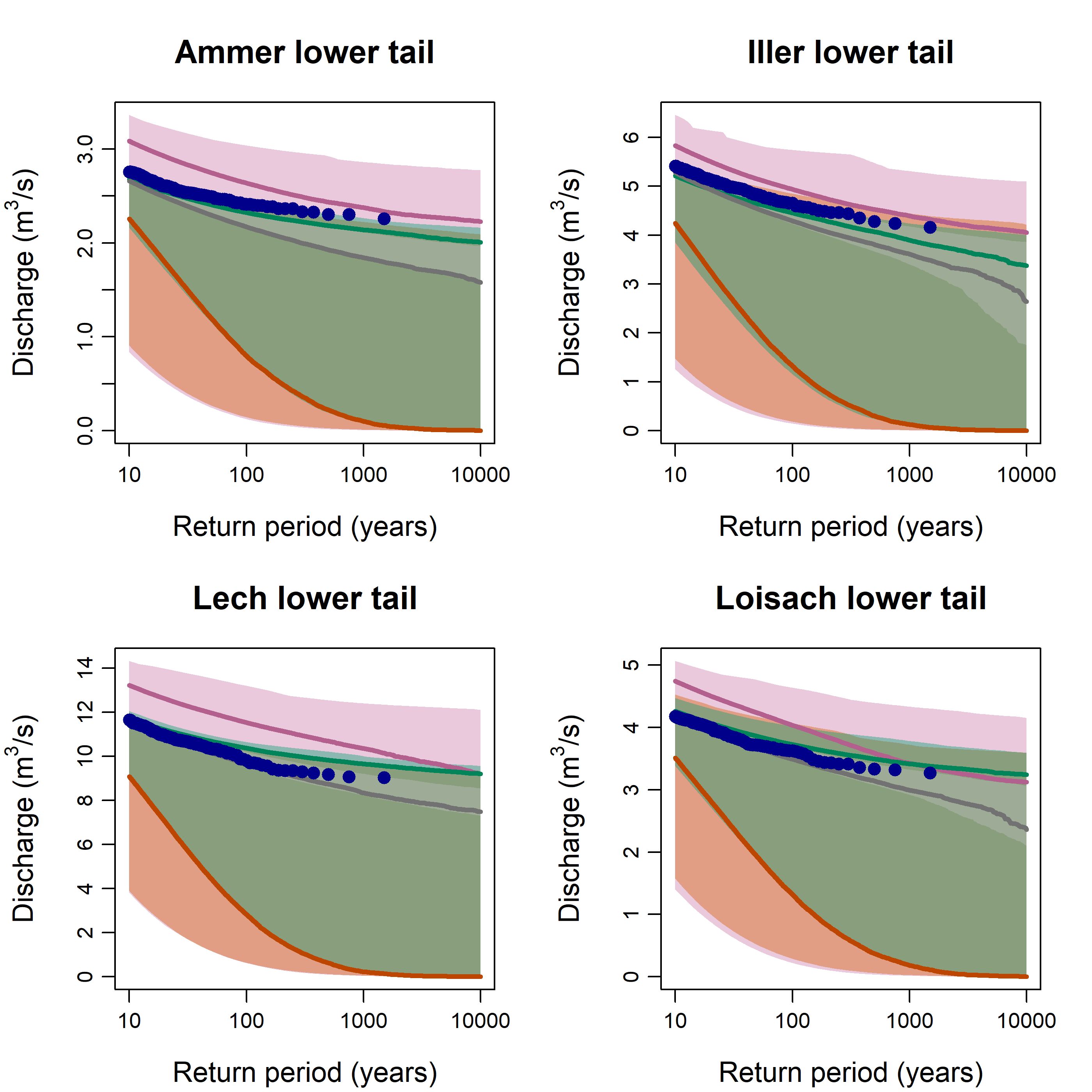}
    \end{minipage}%
    \hfill
   \begin{minipage}{0.48\textwidth} 
        \centering
        \includegraphics[width=\textwidth]{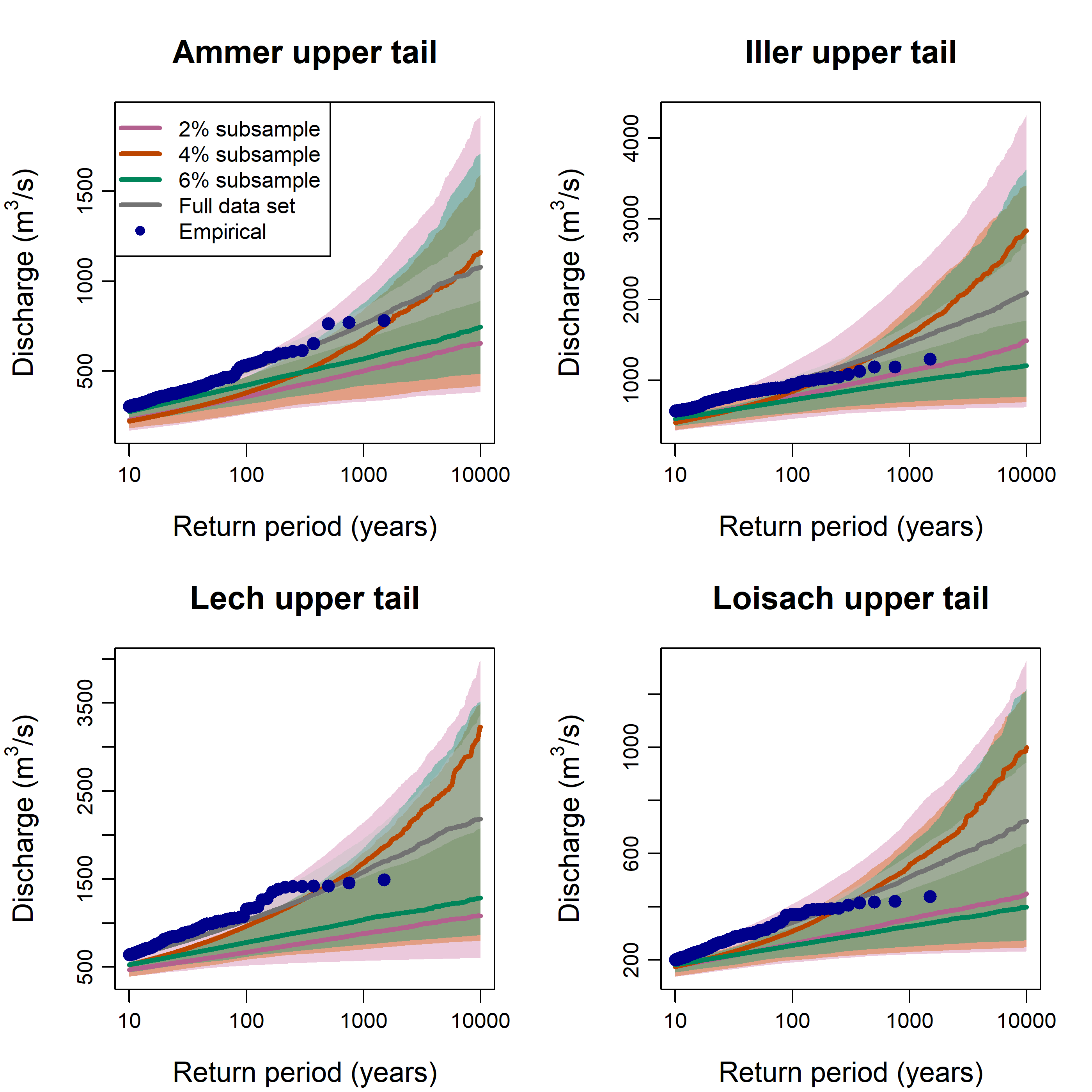}
    \end{minipage}%
    \hfill
     \begin{minipage}{0.48\textwidth} 
        \centering
        \includegraphics[width=\textwidth]{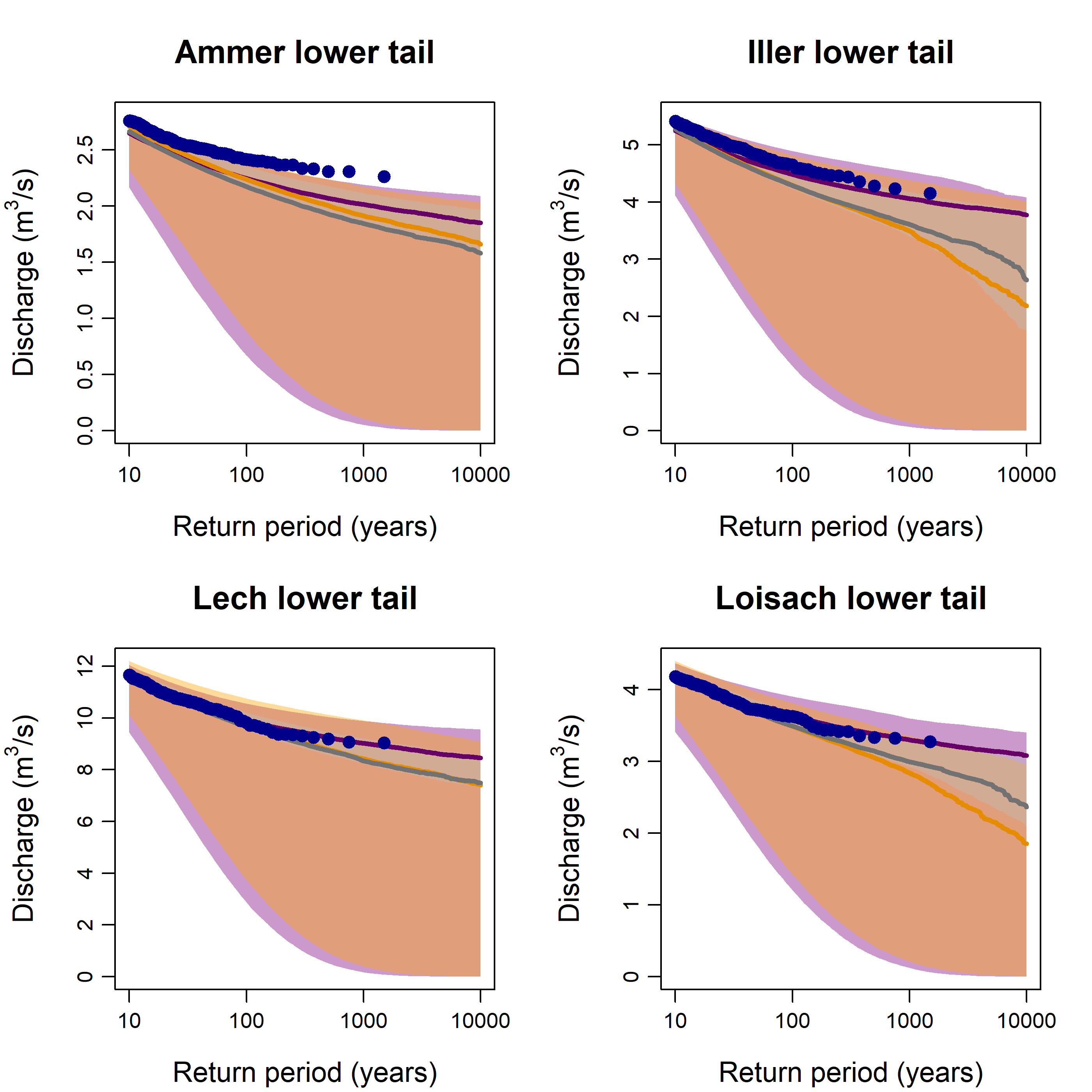}
    \end{minipage}%
    \hfill
   \begin{minipage}{0.48\textwidth} 
        \centering
        \includegraphics[width=\textwidth]{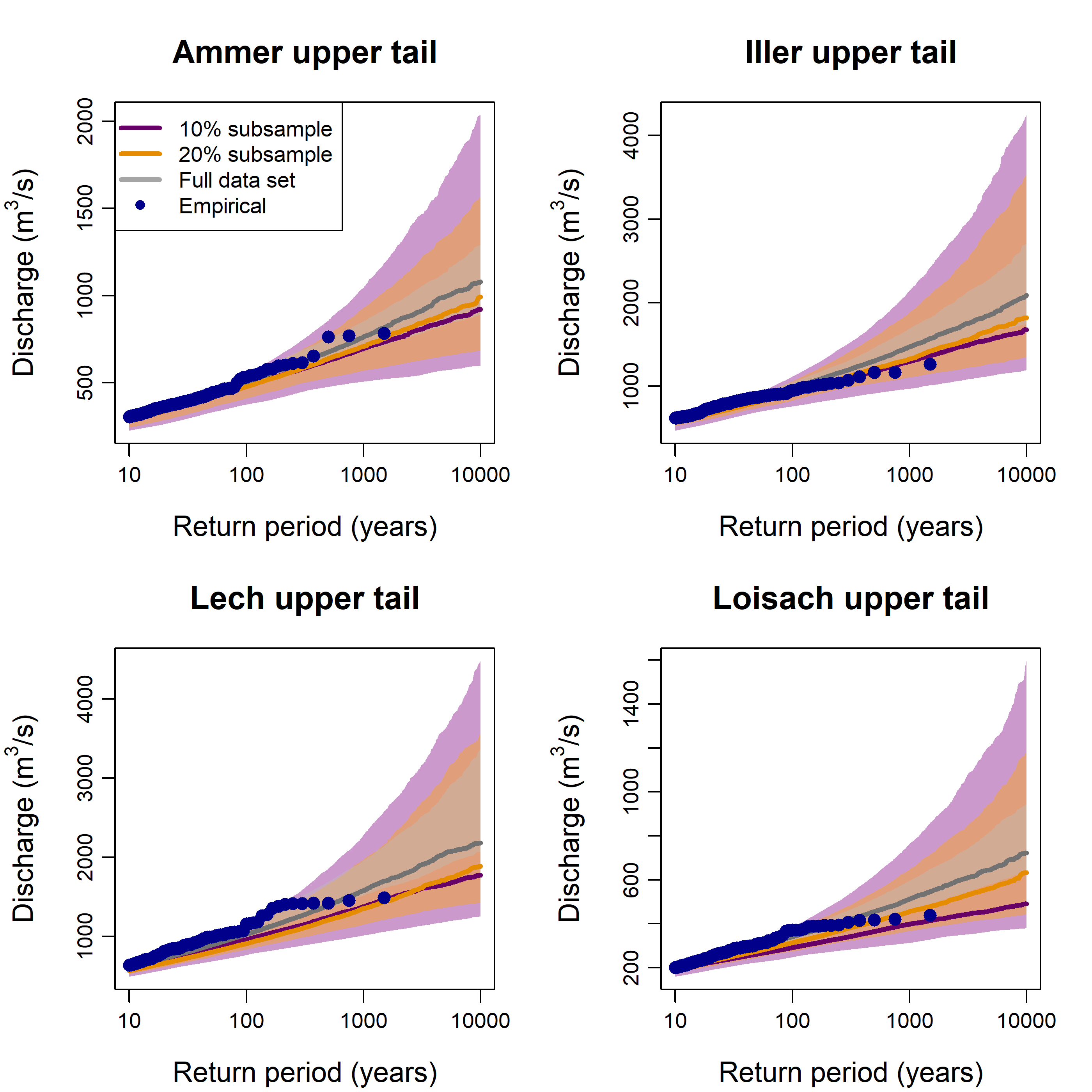}
    \end{minipage}%
    \caption{Comparison of return level estimates from different sample sizes in the lower and upper tails of each marginal variable for the 2070--2099 time period. See Figure~8 in the main manuscript for details on the interpretation of individual panels. }
    \label{fig:ret_level_comp_4}
\end{figure}

Figures~\ref{fig:gpd_qq_plots_2}--\ref{fig:scatters_20} illustrate the diagnostic plots obtained for the model fits on smaller subsamples. Figures~\ref{fig:ret_level_comp_1}--\ref{fig:ret_level_comp_4} compare the return level estimates from each sample size for each time window besides the third.   

\clearpage

\section{Probability plots for smaller sample sizes} \label{appen:probs_smaller_ss}

\begin{figure}[h]
    \centering
    \includegraphics[width=\linewidth]{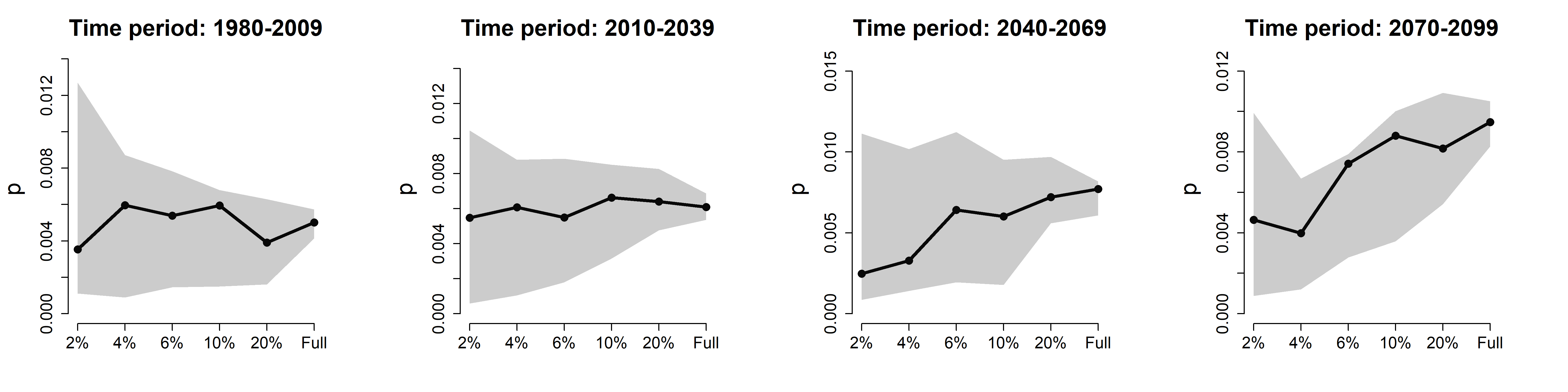}
    \caption{Comparison of $\Pr(S>1,000 (m^3/s))$ estimates over different subsample sizes for each time window, ordered chronologically from left to right.}
    \label{fig:probs_l1_upper_comp}
\end{figure}

\begin{figure}[h]
    \centering
    \includegraphics[width=\linewidth]{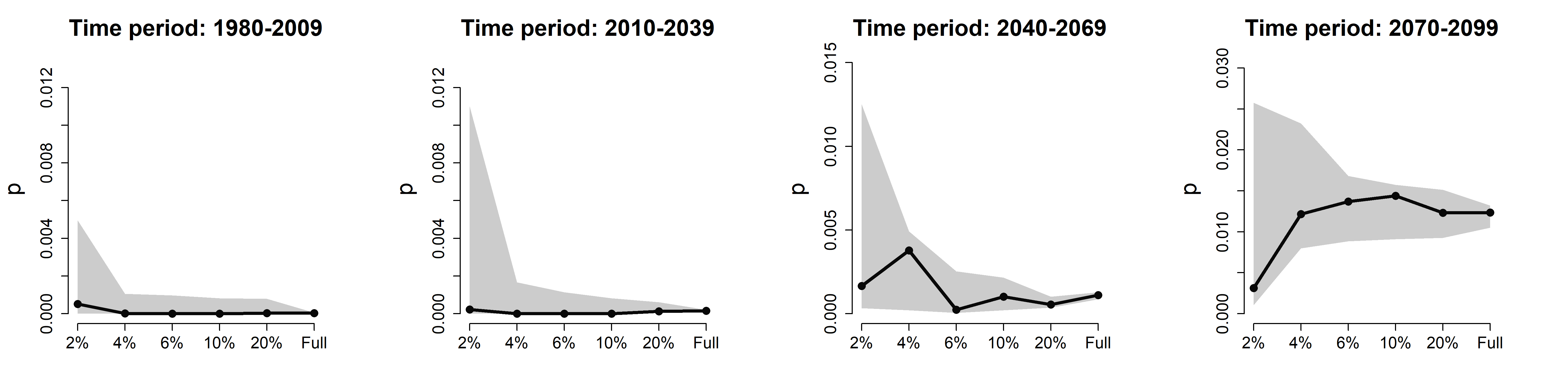}
    \caption{Comparison of $\Pr(S<30 (m^3/s))$ estimates over different subsample sizes for each time window, ordered chronologically from left to right.}
    \label{fig:probs_l1_lower_comp}
\end{figure}

\begin{figure}[h]
    \centering
    \includegraphics[width=\linewidth]{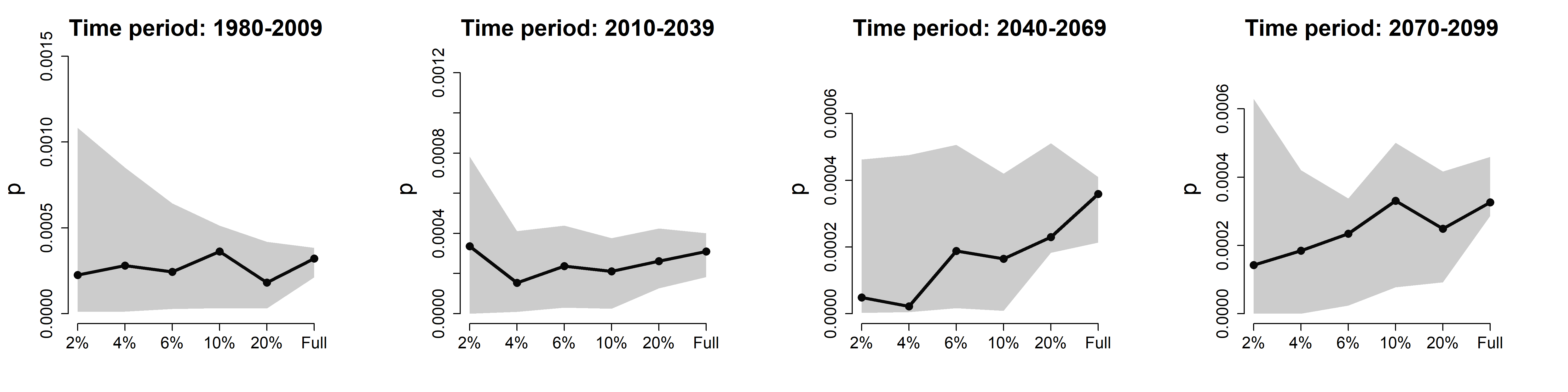}
    \caption{Comparison of $\Pr(X_i > x_i^{1 - 1/N}, \forall i \in \{1,\hdots,4 \} )$ estimates at $N=10$ over different subsample sizes for each time window, ordered chronologically from left to right.}
    \label{fig:probs_surv_comp}
\end{figure}

\begin{figure}[h]
    \centering
    \includegraphics[width=\linewidth]{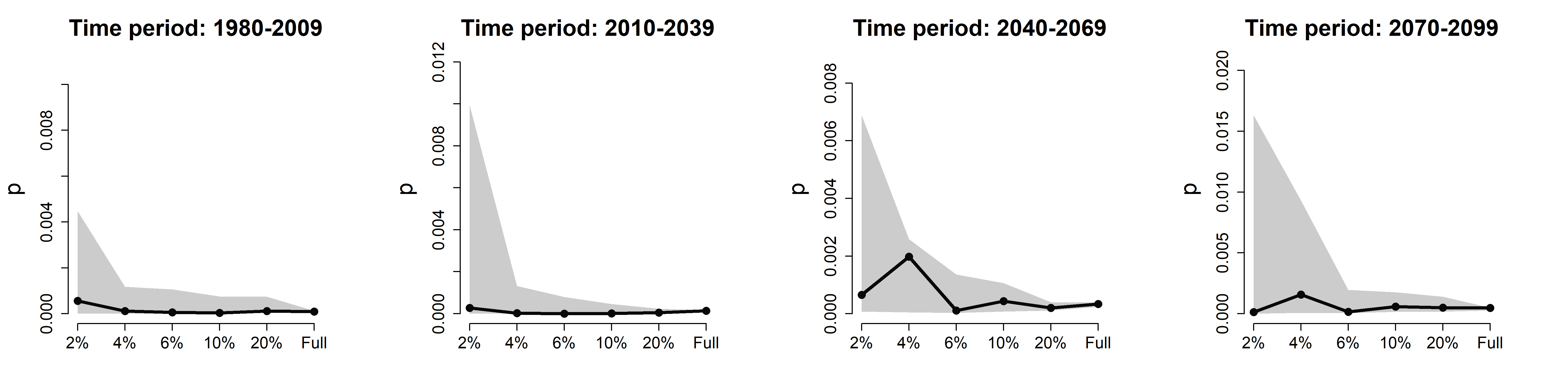}
    \caption{Comparison of $\Pr(X_i < x_i^{1/N}, \forall i \in \{1,\hdots,4 \} )$ estimates at $N=10$ over different subsample sizes for each time window, ordered chronologically from left to right.}
    \label{fig:probs_cdf_comp}
\end{figure}

Figures~\ref{fig:probs_l1_upper_comp}-\ref{fig:probs_cdf_comp} compare the estimated tail probabilities across different subsample sizes and time windows. 

\end{document}